\documentclass[usenatbib]{mnras}
\usepackage[utf8]{inputenc}
\usepackage[english]{babel}
\usepackage{todonotes}

\usepackage{graphicx}	
\usepackage{amsmath}	
\usepackage{amssymb}	
\usepackage{braket}
\usepackage[inline]{enumitem}
\usepackage{color,soul}

\newcommand{\fg}[1]{{Fig.}\,\ref{#1}}
\newcommand{\tb}[1]{{Tab.}\,\ref{#1}}  
\newcommand{\tilk}{\widetilde{\kappa}}
\newcommand{\tilkR}{\widetilde{\kappa}(R)}
\newcommand{\dd}{\text{d}}
 
\title[Curvature for galaxy structural analysis]{Galaxy Structural Analysis with the Curvature of the Brightness Profile}
    
\author[Lucatelli $\&$ Ferrari]{             
	Geferson Lucatelli,$^{1}$\thanks{E-mail: gefersonlucatelli@furg.br}
	Fabricio Ferrari,$^{1}$\thanks{\,E-mail: fabricio.ferrari@furg.br}
	\\                                        
$^{1}$Instituto de Matem\'atica Estat\'istica e F\'isica -- IMEF,   Universidade Federal do Rio Grande -- FURG, Rio Grande, RS 96203-900, Brasil.
}                                          

\date{Accepted 2019 August 1. Received 2019 July  31; in original form 2019 May 24}

\pubyear{2019}                             
 
\begin{document}                           
\label{firstpage}                         
\pagerange{\pageref{firstpage}--\pageref{lastpage}}
\maketitle                                 
\begin{abstract}                           
	In this work we introduce the curvature of a galaxy brightness profile to 
	identify its structural subcomponents in a non-parametrically fashion. 
    Bulges, bars, disks, lens, rings and spiral arms are key to understand the 
    formation and evolution path the galaxy undertook. Identifying them is also 
    crucial for morphological classification of galaxies. We measure and analyse 
    in detail the curvature of $14$ galaxies with varied morphology.
	High (low) steepness profiles show high (low) curvature measures. 
	Transitions between components are identified as local peaks oscillations 
	in the values of the curvature.  We  identify patterns that characterise 
	bulges (pseudo or classic), disks, bars and rings. This method can be 
	automated to identify galaxy components in large datasets or to provide 
	a reliable starting point for traditional multicomponent modelling of galaxy 
	light distribution.
\end{abstract}                             


\section{Introduction}                     
Galaxy formation and evolution are vital to  understand the Universe 
as a whole, for galaxies portray the general structure where they 
emerged and evolved. Galaxy morphology provides us with a framework on 
which we describe galaxy structures that are connected with such 
evolution. The processes that gave birth to bulges, disks, bars, 
rings, arms, halos, for example, are imprinted in the properties or 
absence of these features.

Traditional understanding of galaxy formation and evolution encloses, 
basically, the distinction of the processes that originate elliptical galaxies and 
classical bulges from the formation of disk dominated/spiral galaxies
and pseudobulges. The first mechanism are merger events having violent 
relaxations and hierarchical clustering
\citep{toomre1972,tonini2016,naab2006,hopkins2010}. In the merger scenario, 
it is established that elliptical galaxies are formed by major mergers of 
spiral galaxies \citep{burkert2003}. In the case of spiral, lenticular and 
irregular galaxies, first it was  supposed that they are a result of formative 
evolution where rapid violent processes such as hierarchical clustering and 
merging led to the formation of them 
\citep{white1978,white1991,firmani2003,buta2013}. 

The physical properties of the galaxy components also bring relevant information
regarding the history of the galaxy. Bulges can be formed by two different ways 
and therefore separated in two categories: 
classical originates from violent processes  such as {hierarchical 
accretion} \citep{naab2006,hopkins2010,gadotti2009}, 
are dynamically hot and posses similar properties of elliptical galaxies 
\citep{fisher2008}; 
pseudobulges originates from secular evolution through longer times scales  
where disk material is rearranged by bars and spiral structures in a slow steady 
process
\citep{wyse1997,firmani2003,kormendy_kennicut2004,athanassoula2005,guedes2013,grossi2018}. 
Characteristics of pseudo bulges are not found in elliptical galaxies 
and can be similar to those of disks, which are dynamically 
cold with the kinematics dominated by rotation \citep{fisher2008}. 
Even so, not all bulges can be clearly labelled as classic or pseudo,  
for there are bulges that present a mix of properties of the two types 
\citep{kormendy_kennicut2004}.

{ {More recently, however, with the increase of computational power, numerical simulations
	allowed more profound studies on galaxy formation and evolution} \citep{mo2010,naab2017}. {Examples 
	of such simulations are Millennium }\citep{springel2005}{, EAGLE }\citep{schaye2015}{, Illustris (and TNG)
}\citep{vogelsberger2014,annalisa2017}{, Horizon-AGN }\citep{dubois2014,kaviraj2017}{ (between others)}.}
Therefore the principles of the ideas 
commented above changed slightly and other assumptions raised. 
For example, to cite few: { {pseudobulges are formed by secular and dynamical processes}
\citep{guedes2013} {which can occur together (e.g. through bar dynamics)}
\citep{combes2009,binney2013} {and also from major mergers} \citep{keselman2012,grossi2018}{;}}
dark matter haloes can also evolve in terms of two phases (not only by secular 
evolution), early by major mergers and 
later by minor mergers \citep{zhao2003,diemand2007,ascasibar2008}.

{Besides of how galaxies and its components forms and evolves, galaxy morphological classifications 
	helps to recognise which processes drive galaxy evolution} \citep{buta2013}.
The first ideas were presented by \cite{reynolds1920} and later by 
\cite{hubble1936}, who separated galaxies in classes constituting 
the Hubble sequence - from early to late types galaxies. Their 
classification, and the main procedures adopted in the XX Century,  
were based on visual examination of galaxy images. The classification 
procedure gained in detail along the years, new parameters and schemes 
have been introduced but remained based on the visual inspection of the 
image  by an expert 
\citep{morgan1958,vaucouleurs1959,bergh1960a,bergh1960b,sandage1961,bergh1976, buta2013}.

Within this scenario of components as building blocks of galaxy structure, 
it is relevant to quantify which structures are present in a given galaxy 
and what is their contribution to the total galaxy light compared to other 
components. There are several ways to accomplish that, in general by 
modelling each component in an analytical fashion and then trying a combination 
of them that best describes the galaxy light distribution 
\citep[][among others]{1993caon, peng2002, 
	simard2002,desouza2004, erwin2015}.
This process can describe the galaxy photometry very precisely, 
however there are some drawbacks. In cases of multicomponent, which 
are the majority of galaxies, the minimisation data-model is 
numerically unstable and converges only with the assistance of an 
experienced user, which  limits the range of the model parameters. 
Furthermore, the joint component models fitted to the galaxy 
can be degenerate for different combinations of the parameters
\citep[e.g.][]{2004dejong,andrae2011,sani2011,meert2013}, 
giving the same residuals within 
the photometric errors, and thus are inconclusive. In some cases 
leading to situations where they are not physically reasonable.

These restrictions can be overcome with user inspection, as mentioned. 
Nevertheless, the flood of photometric data that was been made available in 
the last two decades, for example  since the first data release of 
SDSS \citep{sdss_I_2003} (to cite only one big survey) urged us to 
reinvent our basic methods and tools to appropriate from all the 
physical information contained in it. Thus, an automated non-parametric 
method which can infer the basic properties of the components of a given 
galaxy could greatly increase the amount of 
information we could gather from the survey's data. In this way, we 
introduce the \textbf{curvature of the galaxy's brightness profile} 
${\kappa}(R)$ with the purpose of inferring the galaxy's structural 
components. 

The purpose in doing this is that ${\kappa}(R)$ may be used to identify 
if a galaxy is single or multicomponent, and in this case, it would inform 
the radial scale length of each component. This is possible because {$\kappa$}
calculated on the radial profile is a measure of its steepness, and a priori each 
component has its own steepness. We argue that this new approach is more 
physically motivated than previously ones because it is independent of parameters.
In this paper we are firstly introducing the concept  and analysing the results in an 
non-automated way, the automation and improvements are left for a next paper. 

This paper is structured as follows. 
In Section \ref{sec:related_work} briefly exposes approaches  similar to ours. 
In Section \ref{sec:curvature} we introduce our approach for 
galaxy structural analysis and morphometry using the curvature.
In \ref{sec:data}  we discuss the data sample used. 
 The application of the technique  to the data 
is carried out in Section \ref{sec:results_and_discussions}. 
In Sections \ref{sec:discussion} and \ref{sec:conlcusions} we discuss
and summarize our results, respectively.
Also, supplementary material is found in the Appendix Sections. 
The Appendix \ref{app:kurvature_sersic} deal with the curvature of a Sérsic 
profile and Appendix \ref{app:k_plots} displays the galaxy images and curvature plots 
of our analysis related to Section \ref{sec:results_and_discussions}.

\section{Related Work}
\label{sec:related_work}
Several techniques to distinguish  galaxy components have been developed over
the years. Basically they are classified  whether they perform the 
modelling in the extracted brightness profile of the galaxy ($1$D modelling) or  
directly in the galaxy image ($2$D modelling). 
The most widespread $1$D approach is to  fit  ellipses to a set of isophotes  -- 
an evolution of aperture photometry done at the telescope 
\citep{jedrzejewski1987,jungwiert1997,erwin2003,lavers2004,erwin2004,laurikainen2005, gadotti2007,perez2009}. 

For each isophote there is a collection of geometric and physical 
parameters that describe it, including the brightness profile 
(energy flux as a function of distance) which is later used to 
compare to  a model of the components of the galaxy. As a side 
effect, this technique is often used to identify behaviour in the 
parameters (ellipticity, position angle, Fourier coefficients of the 
ellipse expansion) that may indicate the domain of different 
components or, in another context, to give hints of ancient 
interactions, mergers or cannibalism. Also with a similar purpose, 
other non-parametric approaches like unsharp  masking 
\citep{malin1977,erwin2003,kim2012} and structure
maps \citep{pogge2002,kim2012} are used.

 Unidimensional 
techniques are better suited to extract geometric information 
of individual isophotes, but incorporating the instrument response 
function (the PSF -- point spread function) is not trivial. 
Moreover, there is no consensus on which distance coordinate 
to use to extract the profile -- major or minor axis or different 
combination of both -- whose choice impacts  the leading parameters 
\citep{ferrari2004}.
Two-dimensional algorithms operate directly in the galaxy image, 
so there is no ambiguity in extracting the brightness profile. Real PSFs can be 
incorporated in the models and the process (convolution) in this conserves energy. 
Their disadvantage are the high computational cost -- not so critical nowadays -- 
and the instability to initial conditions. This instability increases wildly with the number 
of components and  free parameters in the models. In some situations (e.g. a 
combination of bulge, bar and disc), the algorithm only converges if the initial 
parameters are very close to the final ones, which makes the algorithm itself 
unnecessary \citep[see for example ][]{haussler2007}.

Regardless the method, $1$D or $2$D, the galaxy is then modelled with a 
combination of components, usually described  with a Sérsic function 
\citep{Sersic1968} with different $n$, scale lengths and intensity to 
resemble a bulge, a disk, a bar and so on.
\citep[See for example][and references therein]{1993caon,peng2002,desouza2004,laurikainen2005,gadotti2008,simard2011,kormendy2012,bruce2014,argyle2018,erwin2015}.

The curvature itself, as far as we know, has not been used in the 
context of galaxy structural analysis, nevertheless it has been used 
in different fields of science in a similar manner.
In medicine, for example, it is used to recognize the existence of breast
tumours \citep{lee2015}, where it is applied on the imaging data 
of the photograph of each patient's breast in order to classify the tumours. 
\cite{luders2006}  used the curvature 
to study the process of brain gyrification according genre, 
a process present on the cerebral cortex 
which forms folds, composed by peaks and valleys
(see their methodology for details). They have used 
the mean curvature as a external measure 
of how the normal vector of the surface of the brain changes across it. 
The idea behind it is similar to ours, because here
the curvature $\kappa(R)$ specifies how the normal vector of $I(R)$ 
changes along radius, giving information of what kind of component 
is present in each specific region of the profile. 
Other works using curvature were found on dental studies
\citep{zhang2015,destrez2018} and in general medicine
\citep{preim2014}.

{In summary, we intend that the curvature may be used as a non-parametric 
	tool to identify different galactic components and constrain their scale 
	lengths.  This  might support the structural analysis made 	with standard 
	procedures commented above. For example:  instead of force 
	arbitrarily multiple  S\'ersic functions in structural decompositions, the results from 
	curvature might provide an initial estimate about the morphology of each component and their scale lengths; 
	also it can be comparable with the results obtained from the ellipse fitting technique, where properties  
	are also extracted under the variation of position angle and ellipticity along radius,
	since these may change between different components.  }

\section{Curvature of the brightness profile}
\label{sec:curvature}                      
\label{sec_definition}
\subsection{Definitions}      
With the purpose to identify galaxy substructures non-parametrically  we 
introduce  the \textbf{curvature of the brightness profile} ${\kappa}(R)$. 
The concept of curvature comes from differential geometry  
\citep[e.g.][]{tenenblat2008}: given a function $f = f(x)$ it is possible to 
measure how it deviates from a straight line (the same reasoning 
can be extended to higher dimensional spaces) by means  of a curvature 
measure on $f$, that is $\kappa[f(x)]\equiv \kappa(x)$.To clarify these 
concepts of curvature, consider \fg{fig:kurcircles1}. Along the path $S$, 
at each different point $P$ we can draw a circle with radius $\mathcal{R}$ 
passing through it and its neighbour points. This circle is called of 
{\bf osculating circle}. The measure of curvature of the path at point $P$ 
is inversely proportional to the osculating circle of radius $\mathcal{R}$ 
bypassing through that same point whose surface normal vector points towards 
to the centre of the circle, that is $\kappa \propto 1/\mathcal{R}$ 
\citep[see also additional schematics in][]{crane2013,lee2015}. Along the path, 
we can understand how changes the curvature of different points from $P$ 
to $P'$ drawing many osculating circles with radii $\mathcal{R}$ and 
$\mathcal{R}'$ as we move along the path. In \fg{fig:kurcircles1}, at $P$ 
the curvature is negative and small because $\mathcal{R}$ is large; at $P'$ 
the curvature $\kappa'$ is positive and larger than $\kappa$ because 
$\mathcal{R}'$ is small; at $P''$ the curvature tends to zero because 
$\mathcal{R}''$ goes infinite (we have not draw the corresponding circle).

\begin{figure}
	\centering
	\includegraphics[width=0.85\linewidth]{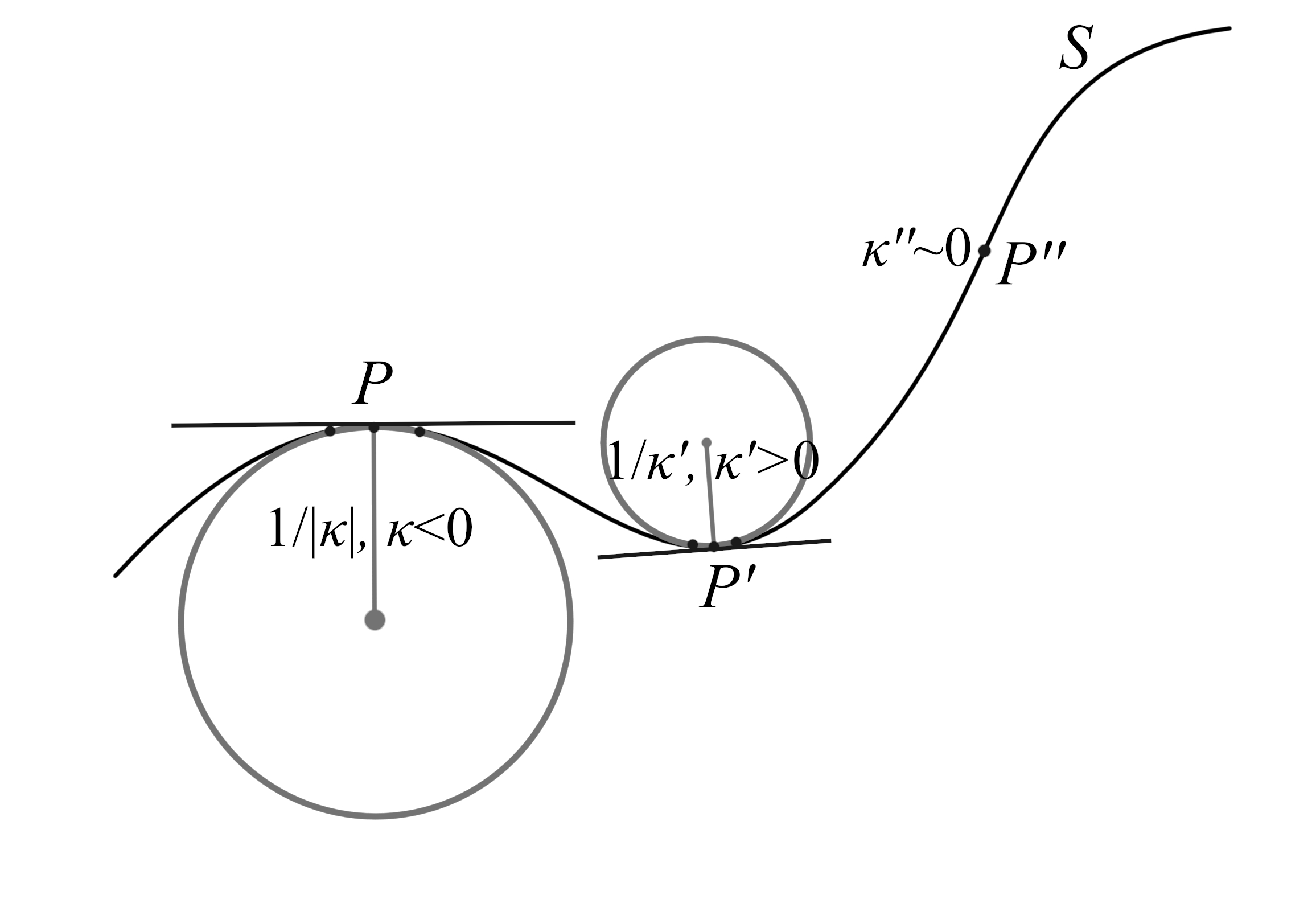}
	\caption{Idealized graphical representation of how the curvature is
		measured along a path $S$ at different points $P$, $P'$ and $P''$. 
		At each point an osculating circle with radius $\mathcal{R}$ can be 
		drawn so that $\kappa \propto 1/\mathcal{R}$.}
	\label{fig:kurcircles1}
\end{figure}

In this work we take $f(x)$ to be the radial brightness profile $I(R)$ 
of the galaxy (more details below). From the curvature's definition of a 
path \citep[e.g][]{tenenblat2008} we have
\begin{align}                              
\kappa(R) =  \dfrac{\dfrac{d^2 I(R)}{dR^2}}{\left[1 + \left(\dfrac{d I(R)}{dR}\right)^2\right]^{3/2}}.
\label{kR}                                 
\end{align}

Equation (\ref{kR}) is a particular case when we have a one-dimensional  
continuous function. However, in this work we will deal with non-continuous 
quantities, which are the discrete data of light profiles of the galaxies.
Therefore, we seek for a discrete curvature measure on our data which 
involves numerical operations. For two discrete vectors representing two 
quantities, say $I$ and $R$, the curvature is defined as 
\begin{align}
\label{numerical_K}
    \kappa = \frac{\delta R_i \delta^2I_i-\delta I_i \delta^2R_i}
    {(\delta R_i^2 + \delta I_i^2)^{3/2}}
\end{align}
where $\delta$ represents the numerical differentials (derivatives) between 
the discrete $i$-values of $R$ and $I$. 

As commented previously, the curvature is inversely proportional to the osculating circle 
respective to a point $P$ in a path $S$. This means that in any arbitrary scale, both 
$x$-axis and $y$-axis variables must have the same scale for our interpretation. 
In other words, the variation of $x$ and $y$ must be represented in a space whose metric 
preserves its components or properties. Therefore, for practical purposes, it 
is convenient to use an orthonormal referential system, consequently we can trace an 
osculating circle for a set of points in relation to a centred point $P$ in the path $S$
represented in this orthonormal system. Otherwise 
we would have an ellipse instead of a circle, and the interpretation will be different
since it is not possible to draw an osculating circle in the $x-y$ space with different 
scales.

Therefore, having two quantities which does not have  the same physical meaning 
and with different scales, such as intensity $I$ and radius $R$, it is necessary to transform them into a space 
having equal metric. The orthonormal system is obtained with a normalization of equations
(\ref{kR}) and (\ref{numerical_K}), that is to confine both $I(R)$ and $R$ to vary in the unitary interval $[0,1]$.

Before we proceed to the normalization, there is a particular  result from the 
curvature for we should take into account: it is zero for a straight line ($\dd f/\dd x = 0$)
and non zero for other general cases. A galactic disk generally follow an exponential profile 
\citep{freeman1970} (see Appendix \ref{sec:sersic_law}). Then in log space, 
this profile is a straight line and we can conclude that the curvature of a disk in 
log space is close to zero. With this in mind we argue that is more useful to normalize 
the logarithm of $I(R)$,  $\log[I(R)]$, instead of $I(R)$. 
Our point is that we will be able to easily distinguish disks
from non disks using {$\kappa$}.
Hence a normalization $\nu(R)$ for $\log[I(R)]$ in the range $[0,1]$ 
can be written as
\begin{align}
\nu(R)\equiv \frac{\log[I(R)] -\min(\log I) }{\max(\log I) - \min(\log I)}.
\label{normalization_IR}
\end{align}

The next step is the normalization for  $R$. We need to change the variation 
of the dimensional variable $R$ to another quantity that is dimensionless and
confine it in the range $[0,1]$. Generally, observations do not reach the faintest 
parts of a galaxy, so it is usual to define a galaxy size. 
One common choice is the  Petrosian radius $R_p$  \citep{petrosian1976}, defined by the radius where the Petrosian function 
\begin{align}
\eta(R) = \frac{\Braket{I}(R)}{I(R)} 
\label{petro_function}
\end{align}
has a definite value, i.e.
\begin{align}
\eta(R_p) = \eta_0.
\label{petro_radius}
\end{align}
Here $\Braket{I}(R)$ is the mean intensity inside $R$ and $I(R)$ the intensity at $R$. Following  
\citep{bershady2000A,blanton2001,ferrari2015} we use $\eta_0=5$ 
with $2R_p$ as the size of galaxy. The normalization of $R$ now reads
\begin{align}
\chi =\frac{R}{2R_p}.
\end{align} 
{With this normalization at $2R_p$, we also set that the value of $\min(\log I)$ in equation }(\ref{normalization_IR})
{becomes $\min(\log I)\sim \log I(2R_p)$ under the assumption that the profiles generally always 
	decreases as $R$ increases.}

Now, the curvature in terms of normalized variables  measures the rate of 
change of $\nu(R)$ in terms of the new variable $\chi$. So the complete 
normalized curvature is obtained taking the derivative of $\nu(R)$ 
with respect to $\chi$. The relationship between the differentials of $\dd R$ 
and $\dd\chi$ is 
\begin{align}
\dd \chi = \frac{\dd R}{2R_p}.
\end{align}
Taking the first and second derivatives,
\begin{align}
\frac{\dd\nu}{\dd\chi} = 2R_p \frac{\dd\nu}{dR} \qquad {\rm and} \qquad  
\frac{\dd^2\nu}{\dd\chi^2} = 4R_p^2 \frac{\dd^2\nu}{\dd R^2}
\end{align}
we obtain  the {\bf normalized curvature} as 
\begin{align}
\label{final_normalized_curvature}
\tilkR =  
4R_p^2\dfrac{\dd^2 \nu}{\dd R^2}
\left[1 + 4R_p^2\left(\dfrac{\dd \nu}{\dd R}\right)^2\right]^{-3/2}
.
\end{align}

\begin{figure}                             
	\centering                                
	\includegraphics[width=0.8\linewidth]{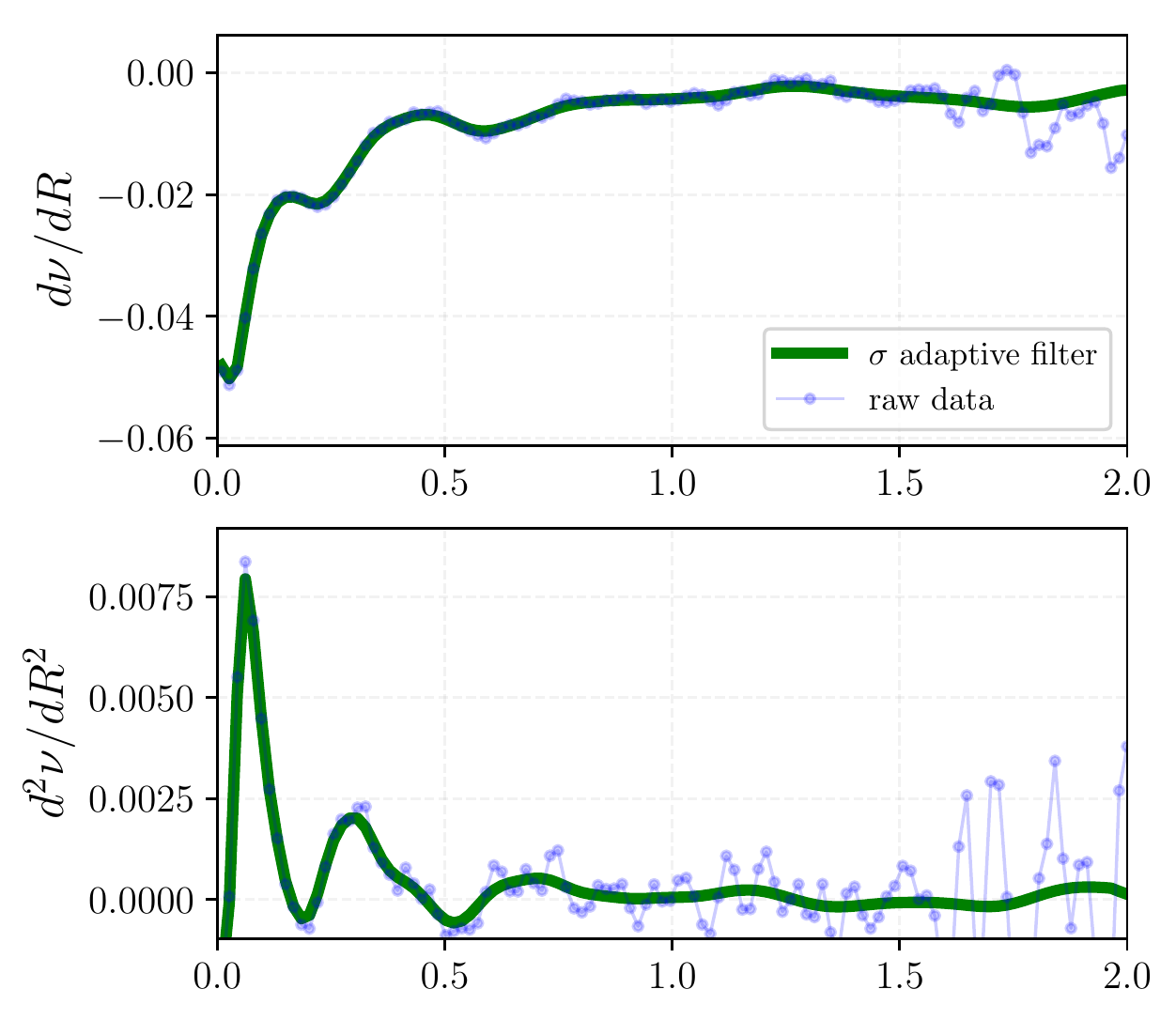}
	\includegraphics[width=0.8\linewidth]{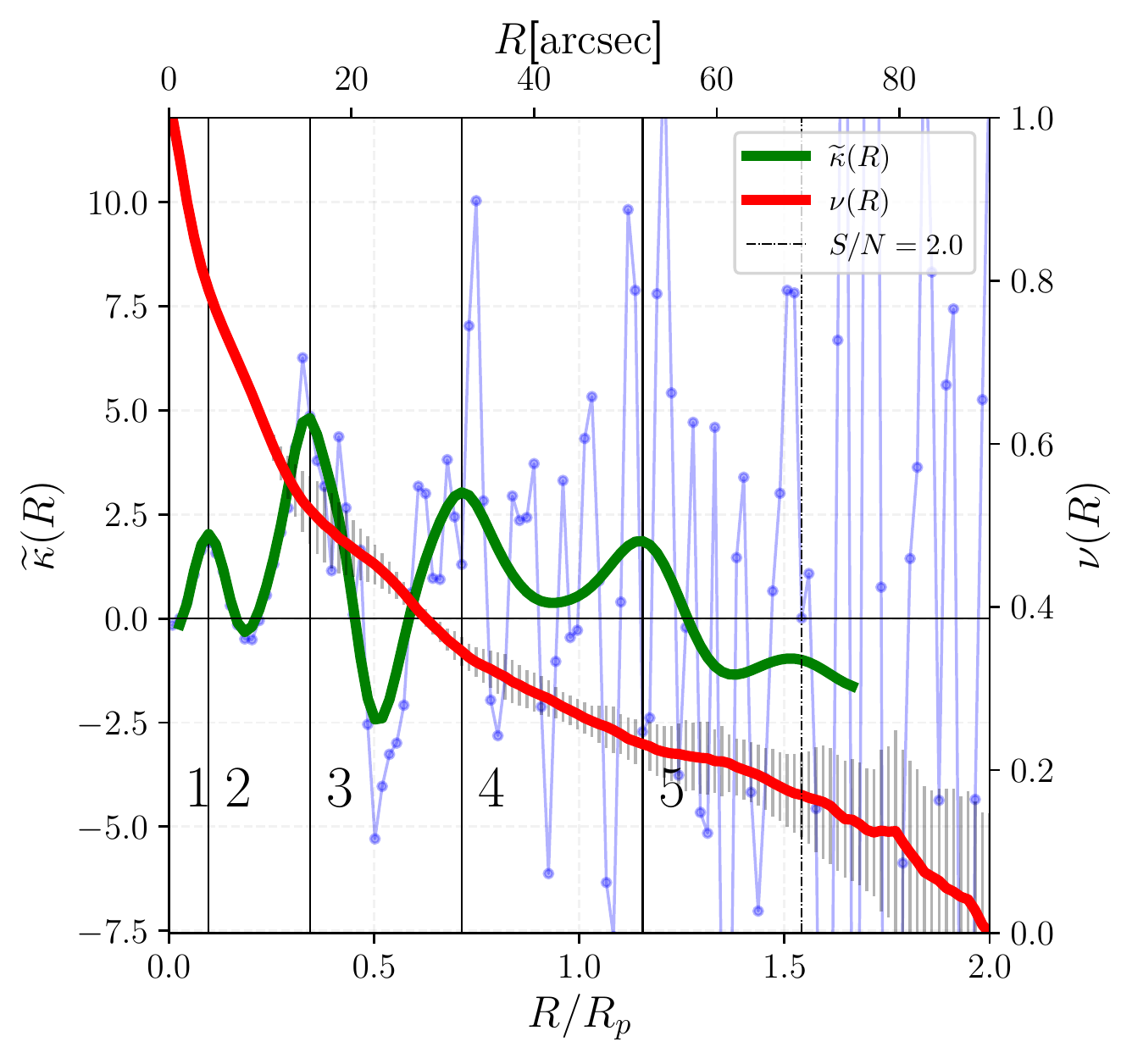}
	\caption{
		Example of the procedure to calculate $\tilkR$ for the EFIGI/SDSS 
		data of galaxy NGC 1211/PGC 11670. \textbf{Top:} first derivative of  
		$I(R)$, raw (blue) and filtered (green). \textbf{Middle:} the same but for 
		the second derivative of $I(R)$. \textbf{Bottom:} log-normalized brightness profile 
		$\nu(R)$ (red) and the raw (blue) and filtered (green)  curvature. The 
		filter is discussed in Section \ref{sec:filter}. The solid
		vertical lines represent the possible transition between regions dominated by 
		different components, and will be explained latter on Section \ref{example_kur_ngc1211}. The vertical dashed  line delimit 
		the regions where the signal-to-noise $S/N$ goes below $2$ and we 
		avoid to interpret profiles outside that region.}                   
	\label{fig:with_and_without_filter_k}     
\end{figure}   

Returning to the discrete case in Eq. (\ref{numerical_K}), 
the discrete normalized curvature in terms of the new variables is, according to 
the new normalized variables, simply
\begin{equation}
\label{normalized_discrete_kur}
\tilk  = \frac{\delta\chi_i\; \delta^2 \nu_i - \delta\nu_i \; \delta^2\chi_i}
{\left(\delta\chi_i^2 + \delta\nu_i^2  \right)^{3/2}}.
\end{equation}

As an example, \fg{fig:with_and_without_filter_k} shows  $\tilk$ measured  
for the EFIGI  galaxy NGC 1211/PGC 11670,  together with the normalized 
profile $\nu(R)$ and the related derivatives of $\nu(R)$ (raw and filtered, 
see Section \ref{sec:filter}) used to calculate the curvature. 

The rationale behind using the brightness profile curvature 
$\tilk$ to identify structural subcomponents in galaxy 
light is based on the fact that the curvature of disks will be 
null whilst that of bulges will be positive and dependent on its 
S\'ersic index; regions highly affected by the PSF would tend 
to be negative in $\tilk$; the transition between different 
components would be manifested in curvature changes; galaxies 
with disks (either inner or outer) will present $\tilk$\  that 
are zero over the region dominated by the disk (see next section). 
In Appendix \ref{app:kurvature_sersic} we use equation 
(\ref{final_normalized_curvature}) to work out the expressions for 
the curvature of Sérsic profile models. In Section \ref{sec:results}
we develop this concepts in practice examining  how $\tilk$\ 
behaves for different combinations of galaxy subcomponents.

\subsection{Measuring $I(R)$ with {\sc Morfometryka}}
\label{sec:morfometryka}

\begin{figure*}
	\centering
	\includegraphics[width=0.90\linewidth]{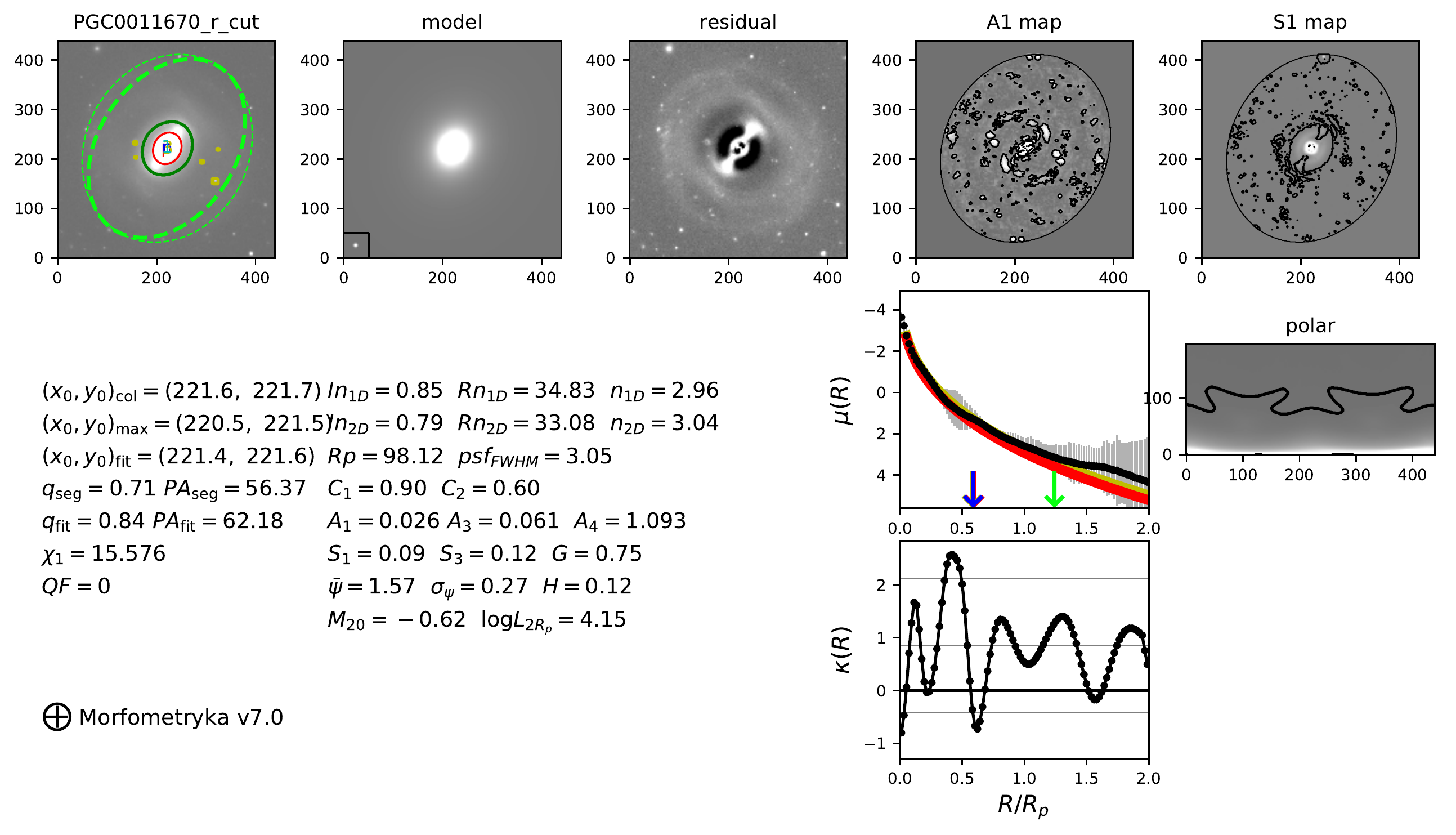}
	\caption{Output from {\sc mfmtk} for the SDSS  $r$ image of NGC1211. 
		The top panel shows (from left to right) the galaxy image, a single 
		Sérsic fits model, residual image, and the asymmetry $A_1$ and smoothness 
		$S_1$ maps. The numerical quantities are related to photometry and
		morphometry. The Sérsic parameters used to the model are the ones with the $2D$ 
		index,	performed in a two dimensional modelling \citep[see][]{ferrari2015}.
		The bottom right panel displays the surface brightness profile $\mu(R)$, the
		curvature $\widetilde{\kappa}(R)$ and the polar plot used for $\sigma_\psi$.
	}
	\label{fig:mfmtk_output_example}
\end{figure*}

The brightness profile is measured as part of the processing done by 
\textsc{Morfometryka} \citep{ferrari2015}.
\textsc{\mdseries Morfometryka} is an algorithm designed to perform
several photometric and morphometric measurements on a galaxy image in
an automated way with no user interaction. The inputs
are the galaxy image stamp and respective point spread function (PSF)
image. It then measures the background in an iterative way,
segmentates the image -- separating galaxy, other objects and
background -- and measure basic geometric parameters of the 
segmented region, like the centre, the position angle, major and minor
axis (see \fg{fig:mfmtk_output_example} for an example of an output result from {\sc Morfometryka}).
Based on this information, it performs aperture photometry on
similar ellipses from the centre up to 2$R_p$ (the Petrosian
radius), spaced 2 pixel apart and having the width of 1 pixel. 
The brightness profile $I(R)$ is the azimuthally averaged value of 
the ellipses aforementioned; the error is the standard deviation of the 
same pixel set.

\subsection{The Filters for  $\widetilde{\kappa}(R)$}
\label{sec:smoothing}
\label{sec:filter}
Measuring the curvature, according to Eq. (\ref{kR}), would be just a matter 
of  evaluating the first and second derivatives of the brightness 
profile. In practice,  with discrete points for $I(R)$ contaminated 
by noise, the direct estimation of $\dd\nu/\dd R$ and $\dd^2\nu/\dd R^2$ is 
worthless because the noise in $I(R)$ is amplified by the derivative operator (a 
high pass filter).  \fg{fig:with_and_without_filter_k} shows 
this noise magnification  that yield a very scattered curvature (blue points). 

One way to overcome the limitations imposed by the noise is to use a filter to 
enhance the signal-to-noise of the data. Many linear filters attenuate the 
signal as well as the noise. For our purpose, we need a filter that 
attenuates the noise but keeps the overall structure present in the data. 
In general, the signal-to-noise ratio is higher in inner regions of 
the galaxy data and decreases to outermost regions. Therefore, the best 
solution to use in the filtering task is an adaptive filter, 
which takes into account the level of the dispersion and adapts the  level of the 
smoothing accordingly. 

We adopt a simple adaptive Gaussian filter $G$, which changes the Gaussian 
dispersion   $\sigma$ according to the local signal to noise ratio. We   
write a linear relation between $\sigma$  and the radial distance 
$R$, assuming  that inner regions are less noisy and constituted by 
small structures, and outer regions, more noisy and large structures, which  is 
true for the majority of galaxies.  We then write 
\begin{align}
\sigma(R) = \left(\sigma_{\max}-\sigma_{\min}\right)\frac{R}{2R_p} + \sigma_{\min}
\end{align}
where $\sigma_{\min}$ is the standard deviation of the filter at the centre 
of the data and $\sigma_{\max}$ in the outermost region. Generally 
$\sigma_{\min}\lesssim 2$ and $\sigma_{\max} \sim 0.1 \times (2R_p)$.
We performed tests with this design  and verified  that a single step 
of filtering in $\nu(R)$ is enough to remove effectively the noise from the data; 
to overcome edge effects caused by the filter when smoothing points at 
the edges, we discard the points corresponding to $2\sigma_{\rm min}$ at the beginning 
and $2\sigma_{\rm max}$ and the end of $\tilk$ -- note  that the 
green line of $\tilkR$ in \fg{fig:with_and_without_filter_k} ends
at $R/R_p \sim 1.7$.

\section{Data Sample}
\label{sec:data_applications}
\label{sec:data}
The data in the present study is constituted by galaxies that already have
been studied in terms of structural decompositions and multicomponent
analysis. We have selected galaxies contained in the following works 
\citep{wozniak1991a,prieto2001b,lavers2004,lauer2007,gadotti2007,
	gadotti2008,gadotti2009,compere2014,salo2015,gao2017,yildrim2017}
(individual references bellow). The data were extracted according 
to their availability in NED and other databases. These are: 
EFIGI sample \citep{bailard2011}, Hubble Space Telescope Archive (HLA\footnote{
	Based on observations made with the NASA/ESA Hubble Space Telescope,  
	and obtained from the Hubble Legacy Archive, which is a collaboration 
	between the Space Telescope Science Institute (STScI/NASA), the Space 
	Telescope European Coordinating Facility (ST-ECF/ESA) and the         
	Canadian Astronomy Data Centre (CADC/NRC/CSA).}), 
SPITZER Telescope \citep{dale2009}, 
Pan-STARRS-1 telescope\footnote{The Pan-STARRS1 Surveys (PS1) and the PS1 public science 
	archive have been made possible through contributions by the Institute for 
	Astronomy, the University of Hawaii, the Pan-STARRS Project Office, the Max-Planck 
	Society and its participating institutes, the Max Planck Institute for Astronomy, 
	Heidelberg and the Max Planck Institute for Extraterrestrial Physics, Garching, 
	The Johns Hopkins University, Durham University, the University of Edinburgh, 
	the Queen's University Belfast, the Harvard-Smithsonian Center for Astrophysics, 
	the Las Cumbres Observatory Global Telescope Network Incorporated, the 
	National Central University of Taiwan, the Space Telescope Science Institute, 
	the National Aeronautics and Space Administration under Grant No. NNX08AR22G 
	issued through the Planetary Science Division of the NASA Science Mission 
	Directorate, the National Science Foundation Grant No. AST-1238877, the 
	University of Maryland, Eotvos Lorand University (ELTE), the Los Alamos National 
	Laboratory, and the Gordon and Betty Moore Foundation.
\url{https://outerspace.stsci.edu/display/PANSTARRS/Pan-STARRS1+data+archive+home+page}}
\citep{chambers2016}
and Cerro Tololo Inter-American Observatory 1.5m (CTIO$1.5$m) 
telescope\footnote{\url{http://www.ctio.noao.edu/noao/}.}. 

\tb{tab:data_sample} introduces the following information for each galaxy: 
the complete references, where the data was taken, the filter used and the 
morphological classification found in the literature. 
Most of galaxies are in the $r$ band, corresponding to a wavelength around $6250$ angstroms. For the HST images 
we have the filter f160w, which corresponds to the $H$ band with a wavelength 
peak of $1.545$
microns\footnote{
	\url{http://www.stsci.edu/hst/wfc3/ins_performance/ground/components/filters}.}.
For the Spitzer galaxy NGC\,1357, the filter correspond to the IRAC3.6 band with a wavelength 
around of $3.6$ microns.
We have limited our study 
to a small set of 14 galaxies in order to perform a careful analysis with the curvature.
We selected galaxies in each general morphology: ellipticals, 
lenticular, disk/spiral (with and without bar). 

The sample cover the following diversity: 
\begin{enumerate}                          
	\item galaxies with complex structure, e.g. bars, rings, arms, disks, which are notably seem: 
	NGC\,1211 \citep{buta2013}, 
	NGC\,0936 \citep{wozniak1991a}, 
	NGC\,1512 \citep{laurikainen2006,compere2014}, 
	NGC\,7723 \citep{prieto2001b};
	\item galaxies with fine structures, e.g. core, smooth bars or arms 
	+ S0 galaxies with smooth disks which barely can be identified: 
	NGC\,0384 \citep{yildrim2017}, 
	NGC\,1052 \citep{lauer2007}, 
	NGC\,2767 \citep{nair2010,yildrim2017}, 
	NGC\,4267 \citep{wozniak1991a}, 
	NGC\,4417 \citep{kormendy2012},
	NGC\,1357 \citep{gao2017}, 
	NGC\,2273 \citep{lavers2004}, 
	NGC\,6056 \citep{tarenghi1994,prieto2001b};
	\item galaxies with well known structure in shape, 
	for example ellipticals: 
	NGC\,472, NGC\,1270 \citep{yildrim2017}.
\end{enumerate}     

With this set we investigate the behaviour of $\widetilde{\kappa}(R)$ 
for each representative morphology, recognizing 
how the curvature differs between each type. 

\begin{table*}                              
	\centering                                
	\caption{Data sample used in this work.}  
	\label{tab:data_sample}                   
	\begin{tabular}{llccc} 
		\hline                                   
		$i$ & Galaxy Name + filter& Morphology   	 			& Origin  & Reference of the data			\\ \hline                                   
		1  & NGC0384-f160w  	   & S0$^{3}$/E$^{1,2}$		 	& HST	  & \citep{yildrim2017}				\\
		2  & NGC0472-f160w  	   & VY CMPT$^{2}$     	 		& HST     & \citep{yildrim2017}				\\
		3  & NGC0936-$r$    	   & SB0$^{4}$	     	 		& EFIGI   & \citep{bailard2011}				\\
		4  & NGC1052-$r$    	   & E$^{1,5}$(+core$^{6}$)/S0$^{5}$&EFIGI & \citep{bailard2011}			\\
		5  & NGC1211-$r$	       & SB0/a(r)$^{7}$ 	 	 	& EFIGI   & \citep{bailard2011}				\\
		6  & NGC1270-f160w         & E$^{1}$/VY CMPT$^{2}$	 	& HST	  & \citep{yildrim2017}				\\
		7  & NGC1357-3.6$\mu$m     & SA(s)ab$^{1}$	 	 	    & SPITZER & \citep{dale2009,sheth2010}		\\
		8  & NGC1512-$r$  	       & SB(r)a$^{1}$   			& CTIO	  & \citep{meurer2006}				\\
		9  & NGC2273-$r$  	       & SB(r)a$^{1}$   	 	 	& Pan-STARRS	  & \citep{chambers2016}				\\
		10 & NGC2767-$r$  	       & E$^{1}$/S0$^{8}$			& HST	  & \citep{yildrim2017}				\\
		11 & NGC4267-$r$  	       & S0$^{3}$/SB0$^{4,5}$		& EFIGI   & \citep{bailard2011}				\\
		12 & NGC4417-$r$  	       & SA0a$^{9}$ 	     		& EFIGI   & \citep{bailard2011}				\\
		13 & NGC6056-$r$  	       & SB(s)0$^{1,10}$/S0/a$^{11}$& EFIGI   & \citep{bailard2011}				\\
		14 & NGC7723-$r$     & SB(r)b$^{1,10}$			& Pan-STARRS & \citep{chambers2016}		\\
		\hline
		\multicolumn{3}{l}{$^1$ \citep{RC3_1991}. }\\
		\multicolumn{3}{l}{$^2$ \citep{yildrim2017}. }\\
		\multicolumn{3}{l}{$^3$ \citep{nilson1973}. }\\
		\multicolumn{3}{l}{$^4$ \citep{wozniak1991a}. }\\
		\multicolumn{3}{l}{$^5$ \citep{sandage1981}. }\\
		\multicolumn{3}{l}{$^6$ \citep{lauer2007}. }\\
		\multicolumn{3}{l}{$^7$ \citep{gadotti2007}. }\\
		\multicolumn{3}{l}{$^8$ \citep{nair2010}. }\\
		\multicolumn{3}{l}{$^9$ \citep{kormendy2012}. }\\
		\multicolumn{3}{l}{$^{10}$ \citep{prieto2001b}. }\\
		\multicolumn{3}{l}{$^{11}$ \citep{tarenghi1994}. }\\
	\end{tabular}
\end{table*}

\section{Applications to The Data Sample}
\label{sec:results_and_discussions}
\label{sec:results}

\begin{figure}
	\centering                                
	\includegraphics[width=0.85\linewidth]{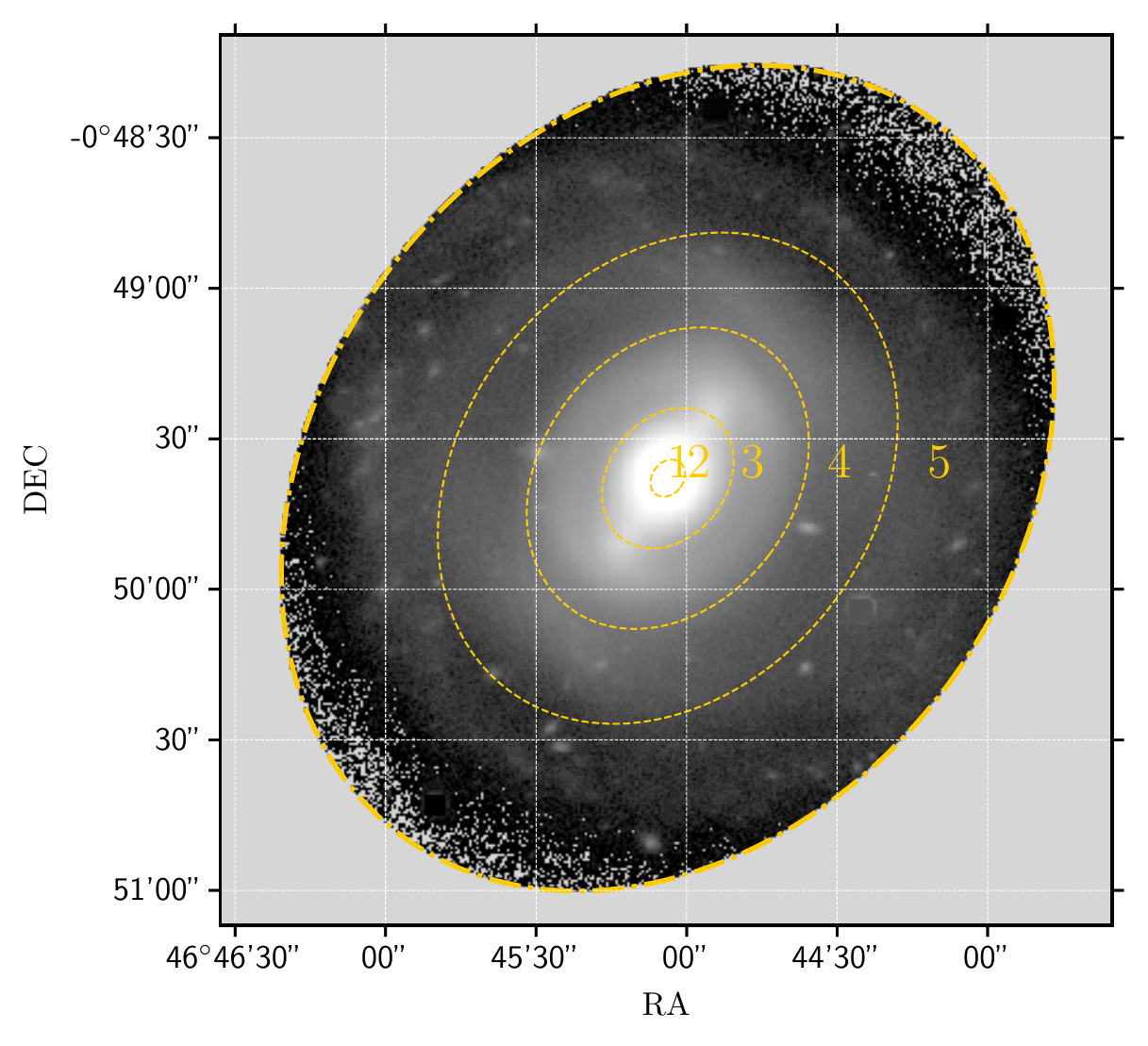}
	\includegraphics[width=0.85\linewidth]{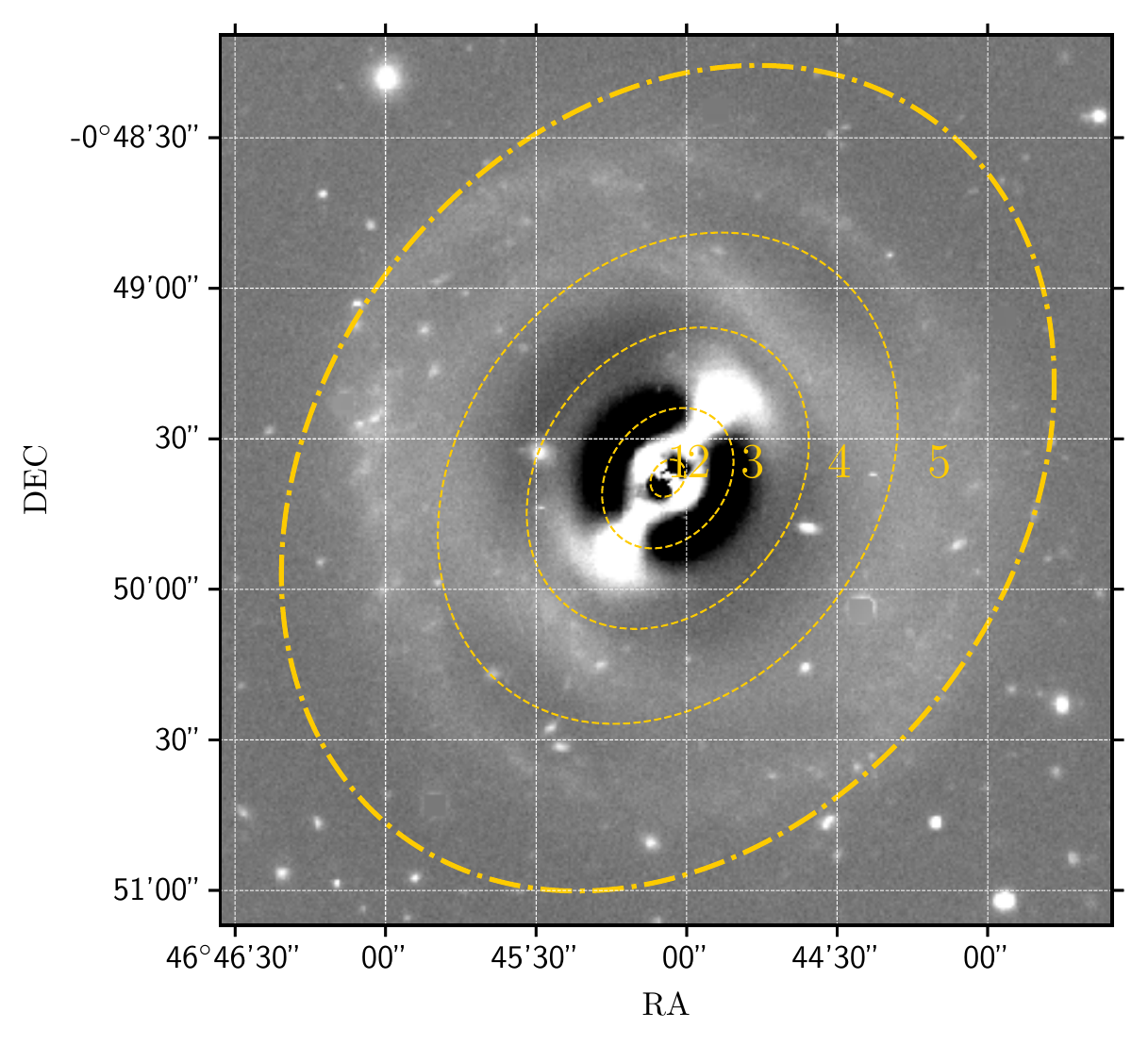}
	\includegraphics[width=0.85\linewidth]{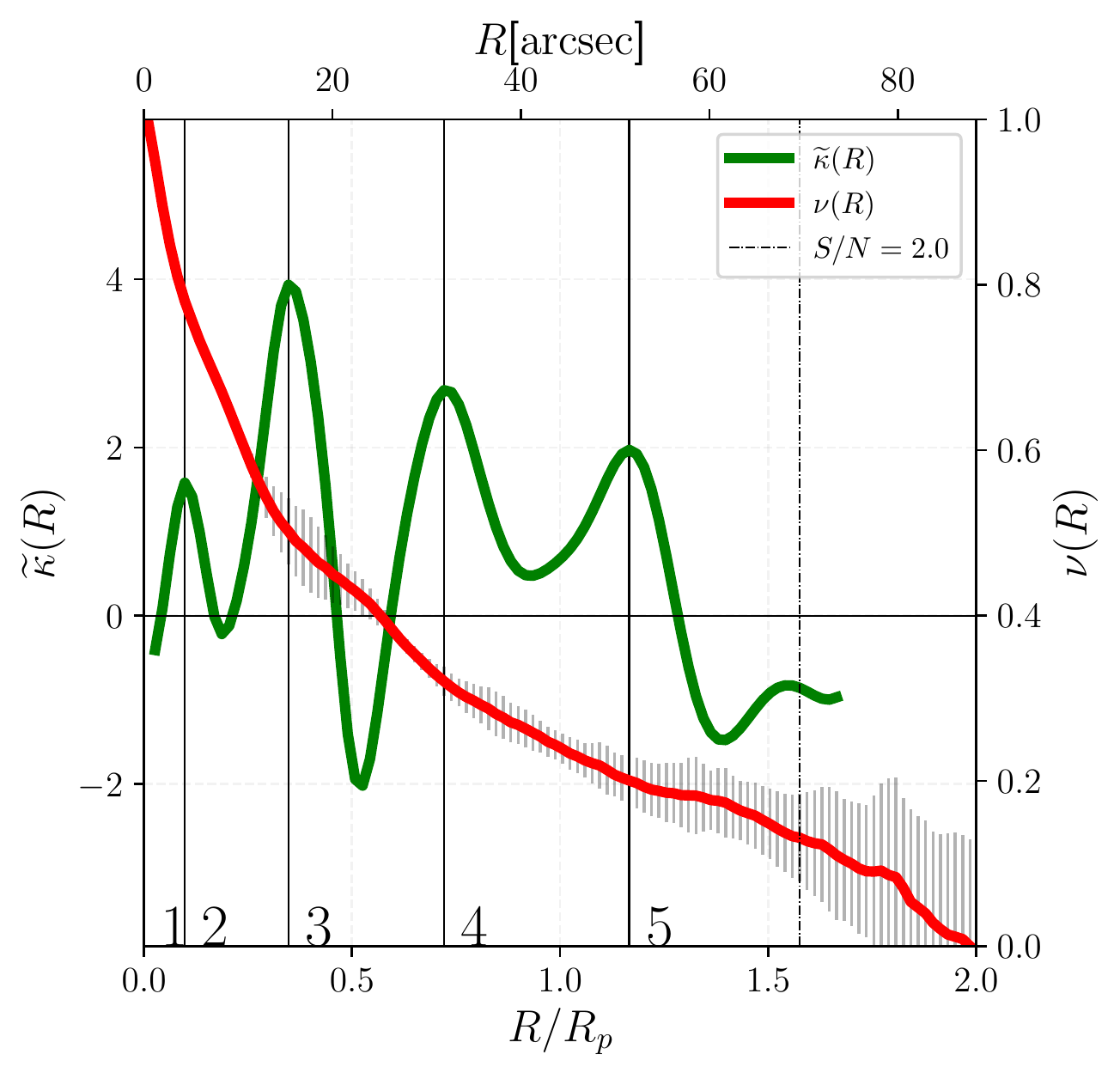}\\
	\caption{
		\textbf{Top:} SDSS $r$ image for  NGC1211 from the EFIGI sample. 
		The yellow lines represent  each 
		component of the galaxy. These regions  were drawn  from the local peaks
		in the green line of $\tilkR$ (see bellow). The numbers in the yellow 
		ellipses and in the $\tilk$ plots refer to the same regions. 
		\textbf{Middle}: Residual from a single Sérsic fit to the image above made 
		with \textsc{Morfometryka}. For both, yellow dotted lines correspond to 
		regions identified in the curvature plot below; yellow dot-dashed line 
		is the $2R_p$ region. \textbf{Bottom:} Normalized log-brightness profile 
		(red curve) and curvature $\tilk$\ calculated from it (green curve). 
		The solid vertical lines delimit regions of different components 
		(see text for details). The vertical dashed line is the limit of confidence 
		of $\mbox{SNR=2}$.}
	\label{fig:PGC0011670_r_kur}                  
\end{figure} 
  
\begin{figure}
	\centering
	\includegraphics[width=0.9\linewidth]{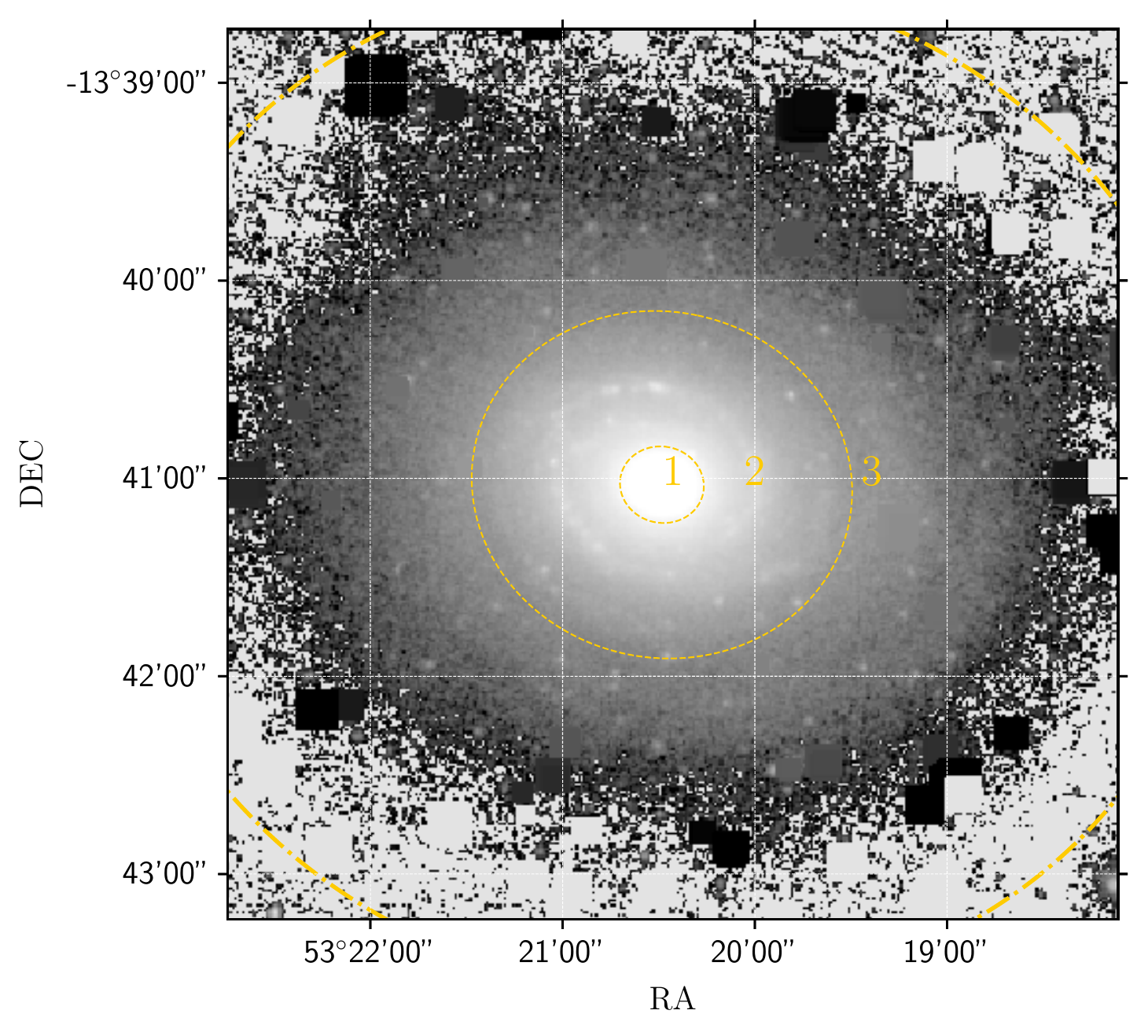}
	\includegraphics[width=0.9\linewidth]{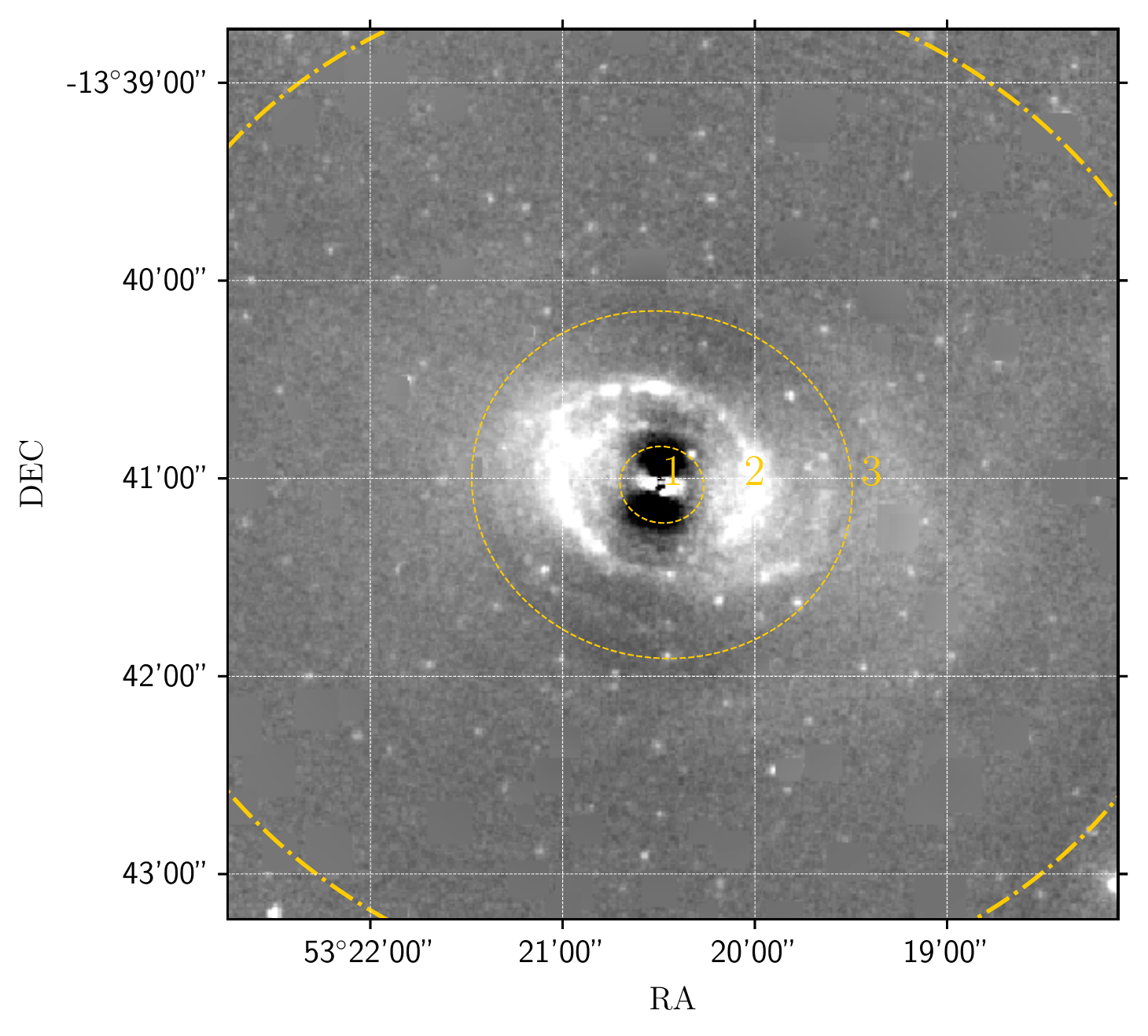}
	\includegraphics[width=0.96\linewidth]{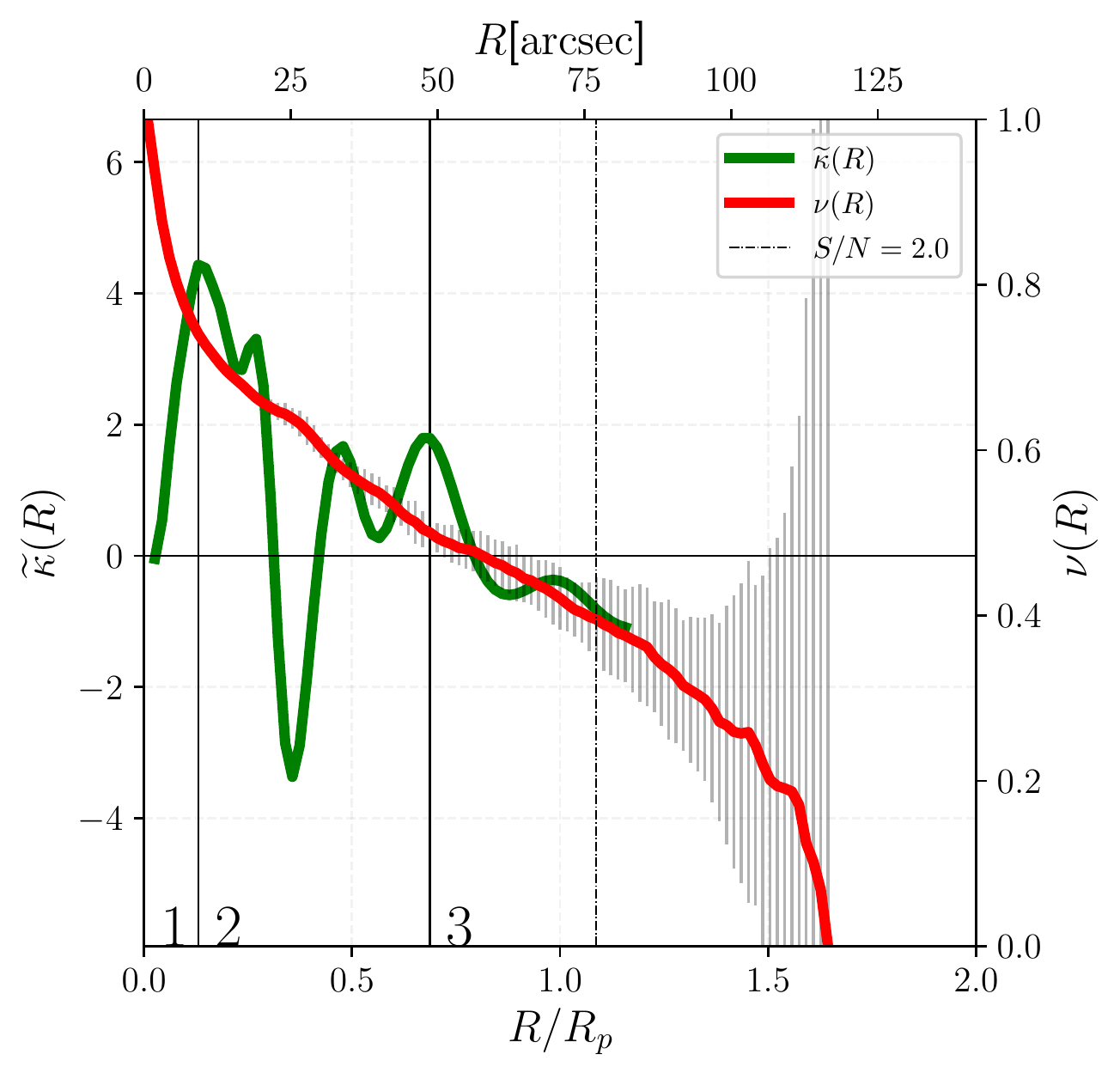}
	\caption{Near infrared band ($3.6\mu$m) of NGC 1357 from Spitzer telescope. 
		Labels and graphic positions are the same as \fg{fig:PGC0011670_r_kur}, and the 
		same for the remaining figures. Region 1 is the bulge (high and large peak in 
		$\tilk$\ ). Region 2 is the  disk with spiral arms and the outer disk is region 3 with
		low values in curvature. Again, the dashed vertical line ($\mbox{SNR=2}$) 
		is our limit of confidence. See the text for further details of this galaxy.}
	\label{fig:ngc1357}
\end{figure}

We trace the behaviour of $\widetilde{\kappa}(R)$ for galaxies 
presented in \tb{tab:data_sample} with the aim to distinguish galaxy 
multicomponents and decipher different morphologies. Each galaxy in our sample 
will be discussed individually in separated sections, divided by their 
global morphologies, e.g. lenticulars (Section \ref{lenticulars}), 
spirals (Section \ref{spirals}) and ellipticals/spheroidal components (Section \ref{spheroidal}). 
We begin by discussing $\tilk$\ for two cases in detail: NGC\,1211 and NGC\,1357. 
The full data sample is discussed below. The complete set of figures for the 
curvature are given in the Appendix \ref{app:kurvature_sersic}.

For each galaxy (see the model reference \fg{fig:PGC0011670_r_kur}), 
we present the broad band image of the galaxy {in the top panel}. 
In the middle panel we show the residual map from a single Sérsic 
fit to the broad band image  made with \textsc{Morfometryka} 
-- for the structures are easily visualized on it. For both, yellow 
dotted lines correspond to regions identified in the curvature plot 
and yellow dot-dashed line mark the $2R_p$ region. The bottom panel 
shows the normalized brightness profile $\nu(R)$ (red curve -- scale 
in the axis at right) and curvature  $\tilk(R)$\ calculated from it 
(green curve -- scale on axis at left). The solid vertical lines 
delimit regions of different components (discussed below for each 
galaxies); they  correspond to the yellow ellipses in the images, but 
are inferred from the $\tilk$ behaviour. The vertical dashed line found in more external 
regions is the limit of confidence of $\mbox{SNR}=2$.

\subsection{Case Study 1: Curvature of   NGC\,1211}
\label{example_kur_ngc1211}
We begin by analysing $\tilk$\ and components from NGC 1211 
(PGC\,11670) galaxy of the EFIGI sample. It is classified as
(R)SB(r)0/a \citep{RC3_1991}, i.e. barred spiral/lenticular 
with rings.
\fg{fig:PGC0011670_r_kur} shows the SDSS galaxy $r$ 
image (top), the residual from a single Sérsic fit by {\sc Morfometryka} 
(middle) and the curvature $\tilk$ (bottom).  Changes in 
$\tilk$\ are related to the transition between regions dominated by 
different components. 
In the galaxy image on the top panel, the size of the yellow dotted ellipses 
overlaid  were determined by the local peaks in the curvature plot 
(bottom), which are signalled as vertical black solid lines in the curvature plot. 
They delimit regions (identified by numerical labels) that are dominated by different
components. 

Starting from the central region, we have two local peaks in $\tilk$ 
at $R/R_p\sim 0.1$ and $R/R_p\sim 0.3$. The first region
(label 1) seems to be a small structure inside the bulge which 
is represented by region 2. However, both components have small values in 
curvature compared to the maximum amplitude ($\tilk \sim 4$ at $R/R_p\sim 0.4$), 
the bulge in region 2 reaches values close to zero, therefore 
this indicates that the Sérsic index of this component is close to 1 
(this can also be seen by the straight brightness profile in the region 
-- red curve in bottom plot). This may indicate that the bulge of  
NGC\,1211 is a pseudobulge. Note also that the innermost component also has a small $\tilk$. 
In \cite{abreu2018} they comment that NGC\,1211 has an internal 
structure called ``barlenses'', which is a component different of a bulge 
contained inside the bar \citep{laurikainen2010}. \cite{gadotti2007} also 
observed a nuclear structure in this galaxy. Therefore, the signature in 
$\tilk$ in regions 1 and 2 are not of a classical bulge, but indicate the 
presence of the nucleus and a pseudobulge. 

The second peak at $R/R_p\sim 0.3$ is the transition between the bulge 
and the bar+inner ring -- a narrow and negative valley in $\tilk$. 
Usually, bars and inner rings are associated with narrow valleys in 
$\tilk$. For inner rings they are negative (as will be seen further) 
and for bars they can be negative or not, but in general both have narrow valleys.
The third peak at $R/R_p \sim 0.75$  defines the end of the bar and the 
start of the outer region of the galaxy, label 4 and 5. The transition 
$4\to 5$ is indicated by the local peak at 
$R/R_p\sim 1.2$. Intermediate regions in 4 shows $\tilk \sim 0$ 
regarding a disk like structure. As indicated by \cite{buta2013}, 
NGC\,1211 has two outer rings: i) an inner outer ring identified to be red; 
ii) an outer outer ring identified to be blue (see his Figure 2.30).

\subsection{Case Study 2: Curvature of   NGC\,1357}
\label{example_kur_ngc1357}
NGC 1357 is a non-barred spiral galaxy classified as SA(s)ab 
\citep{RC3_1991}. \cite{gao2017}  considered that the galaxy 
has two disks: an inner blue disk which contains well defined spiral arms,
and an outer red disk with no spiral arms. 
\fg{fig:ngc1357} shows the curvature for NGC\,1357. 

The bulge is indicated by the region of high curvature (1); the valley at (2)
is related to the tight shape of the spiral arms, having a ring-like shape,
pointed out by \cite{gao2017}. We found that a high and narrow negative
gradient in curvature is characteristic of ring and bar components -- 
which corresponds to Sérsic index lower than unity. The transition of this 
ring with the spiral structure corresponds to the middle inwards part of
region 2. However, all spiral structure is contained inside region 2
which is the inner disk.

The transition between the inner and the outer disk is 
underpinned by the 
decrease in the oscillations after the local peak at $R/R_p\sim 0.65$, 
therefore region 3 corresponds to the outer disk with no spiral arms.
As mentioned before, a disk has values of $\tilk$ close to zero. 
In the following Section we extend the same analysis for the rest of the 
sample, separating them according to each morphological type. 
The remaining the figures are in Appendix \ref{app:k_plots}.

\subsection{Lenticular Galaxies}
{Lenticular galaxies are at first order composed by a
	bulge+disk structure and sometimes having a bar, and the bulge 
	can be often classical or pseudo. These kind of galaxies are a 
	suitable tool to use the curvature in order to unveil the nature 
	of the bulge, since the the slope of the light profile is related  with such morphologies} \citep{laurikainen2016}{. 
For classical bulges, in principle the higher the slope (high S\'ersic index $n$) the higher the curvature. 
On the other hand, for pseudo,  lower slopes (and $n$) may imply lower curvatures. Therefore,
analysing the bulge region with $\tilk$ enables one more way to characterize its morphology.  }

\label{lenticulars}
\subsubsection*{NGC\,936}
NGC936 is classified as barred lenticular galaxy SB0 
\citep{wozniak1991a}, formed by a structure of bulge+bar+disk. 
In \fg{fig:pgc0009359rkur}  we infer from $\tilk$ that  the bulge dominates region 
1, while the bar becomes dominant in region 2 -- highlighted by the straight
valley and negative $\tilk$. The bar ends close to $R/R_p \sim 0.5$. Its
properties were also extracted by \cite{mateos2013}, regarding the bar
length,{ they provide $~50$arcsec}.The peak in $\tilk$ at $R/R_p \sim 0.5$ ($\sim 54$arcsec) indicating the end of the 
bar seems to be in {agreement}\footnote{{The pixel scale of EFIGI (SDSS DR4) is $0.396"/$pix and our $R_p = 273$px, therefore 
$R_p[{\rm arcsec}] = 108$''. Then  our bar length value gives $0.5Rp \sim 54$arcsec, 
in agreement with }\cite{mateos2013}.} with the outer limit found by the authors
(see their figure 3). The bar of NGC\,936 was also studied by 
\cite{erwin2003} together with a nuclear ring inside the bar region. 
Returning back to the inner component, $\tilk$ is nearly 
constant and $\tilk \sim 1$ (small value compared to the maximum amplitude of $\tilk \sim 10$). 
This suggest that the inner region is not a classical 
bulge, but rather being a pseudobulge -- which also may contain a 
nuclear ring.

In region 3 (around 
$R/R_p\sim 0.8$) we see a broad and smooth valley of negative 
curvature, and subsequently being close to zero for $R/R_p> 1.0$ 
-- this pattern appears because the disk is of type II 
\citep{erwin2008,mateos2013} which explains the broad valley, since it comprises
larger extension in radius. 
It is possible to discriminate a valley in 
$\tilk$ entailed by the disk of Type II morphology
and one related to a ring or a bar. Valleys for bars and 
rings have high negative gradient in $\tilk$ and are thin in width
(see additional examples in Figs.\ref{fig:PGC0011670_r_kur}, \ref{fig:ngc1357},
\ref{fig:pgc0009359rkur}, \ref{cut_NGC2767_f160w_kur}, \ref{fig:ngc1512ird2009kur},
\ref{fig:ngc2273rkur} 
and \ref{fig:ngc7723kur}).

For completeness on the analysis of NGC\,936, \fg{fig:pgc0009359rkurcomps2} shows 
each fundamental region of the profile represented in terms of shaded ellipses, gathered from the behaviour of the curvature. 
See labels in the plot. Also, the smooth grey circles shows a practical
example of \fg{fig:kurcircles1} for different osculating circles drawn 
in the galaxy profile. 

\begin{figure}
	\centering
	\includegraphics[width=0.9\linewidth]{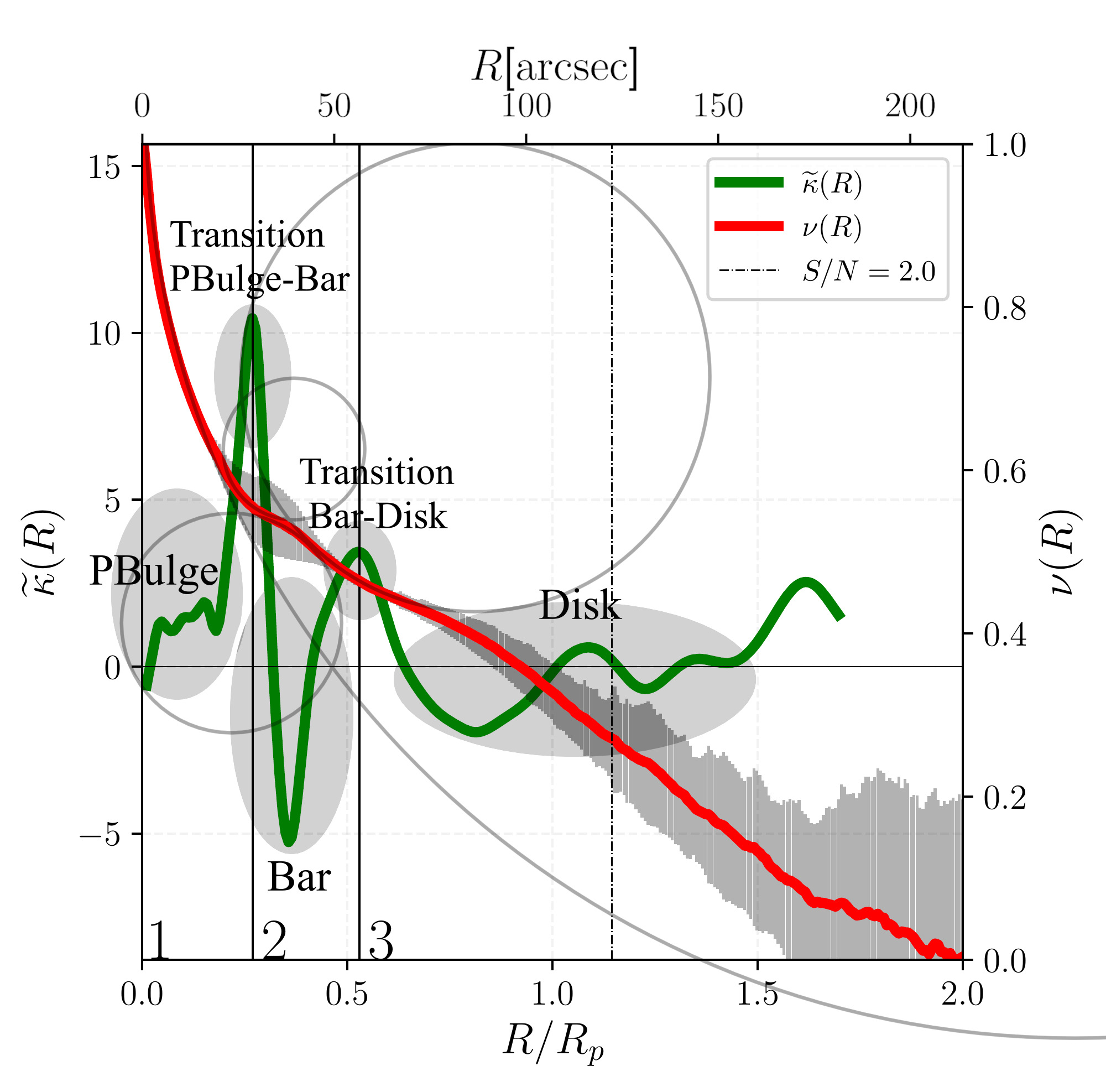}
	\caption{
			Extra representation of the curvature measured on NGC\,936.
			The shaded ellipses indicates, according to the curvature behaviour, each fundamental region of the profile (see the
			labels). The four smooth grey circles represents the osculating circles, as in \fg{fig:kurcircles1}, in some regions of the profile.
	}
	\label{fig:pgc0009359rkurcomps2}
\end{figure}

\subsubsection*{NGC\,2767}
NGC 2767 is classified as an elliptical \citep{RC3_1991} and as a S0
\citep{nair2010}. However, the structure of this galaxy 
is not trivial to be resolved. \cite{yildrim2017} comment that the PA 
twist in the centre of the galaxy may imply the existence of a bar 
and a dust disk. Examining  $\widetilde{\kappa}$ in \fg{cut_NGC2767_f160w_kur}
shows the existence of two peaks and a narrow deep valley,  
revealing that the galaxy has at least three components (regions 1, 2 and 3).
The bulged central part is demarcated by 1. The valley in 2 ($R/R_p \sim 
0.4$), even not being negative (but $\sim 0$), is much smaller than the 
neighbouring peaks in terms of amplitude and also is thin, therefore it points to the bar 
component. The region 3 seems to be of a disk structure of type II,
very similar to NGC\,936 (note that after $R/R_p\sim 1.0$, the 
curvature is very small). A final note is that, photometrically, we suggest that the bulge 
of this galaxy is indicated to be pseudobulge due to the behaviour of $\tilk$ in 
the inner part of the profile -- it takes some distance from the centre to
increase significantly until the transition region at $R/R_p\sim 0.4$. 

Summarizing, in agreement with \cite{yildrim2017}, the curvature entail
that the galaxy may have a bulge+bar+disk structure. The residual image 
in the middle panel of \fg{cut_NGC2767_f160w_kur} demonstrates 
these structures well within the delimited regions gathered from the 
curvature in the bottom plot.

\subsubsection*{NGC\,4267}
In earlier catalogues \citep{nilson1973} NGC\,4267 was classified as S0, but the 
presence of a bar is also suggested 
\citep{sandage1981,wozniak1991a,jungwiert1997,gadotti2006,erwin2008}
-- SB0.
The shape of $\tilkR$ is shown in 
\fg{fig:pgc0039710rcomponentsgray}.  In comparison to NGC\,2767, 
NGC\,4267 exhibit an akin radial profile, however $\tilk$ is considerably 
different. Region 1 around $R/R_p\sim 0.3$  refers to the bulge while the disk 
takes part in the remaining of the profile in region 2 (lower $\tilk$ below 
zero). In respect to the bar, the curvature does not trace a sign of
it because the pattern of a narrow valley is not exhibited.  Also combining 
together the curvature and the residual image (middle panel of 
\fg{fig:pgc0039710rcomponentsgray}), shows no evidence of a 
photometric bar component. In relation the bulge and the disk, the 
transition of both is highlighted by the peak in $\tilk$ at $R/R_p \sim 
0.3$. The inner region shows an abrupt 
increase in $\tilk$ from zero to the transition region with the disk, 
therefore this points that the bulge is a classical bulge. This is in 
agreement with \citep[][see their Table 2]{fisher2010}.

\subsubsection*{NGC\,4417}
NGC4417 is classified as SB0:edge-on \citep{RC3_1991,hinz2003} and also 
S0:edge-on  \citep{sandage1981,nilson1973}. In the curvature plot of 
\fg{fig:NGC4417rkur}, two peaks stand out in the first half of the $R_p$ 
region: 1 and 2 seems to have a bulge morphology, while 3 is a disk. 
Analysing $\tilk$, it seems plausible that it distinguishes two components 
inside $R/R_p\sim 1.0$, the first part identified by the first peak and 
the second by the second peak. 
\cite{kormendy2012} argued that the inner part (region 1) of this galaxy 
is an inner disk while the outer region (region 2) is the bulge
\footnote{\cite{kormendy2012} found a prominent bulge ($B/T\sim 0.88)$ to this galaxy.} 
with a boxy shape due to a faint bar. This seems reasonable because 
the second component has a small $\tilk$ in region 2 resembling the behaviour of a bar 
as in other galaxies. However, in this case we could not differentiate the bulge 
from the bar. In the residual image it is clear that there are signs of two
components, in region 1 a more oval shape while in region 2 a boxy one.
Region 3 is related to the edge-on disk, since decreases after $R/R_p \sim 1.0$. 
But, a final note is related to the peak in $\tilk$ close to 	
$R/R_p\sim 1.5$. Due to the signal-to-noise limit and considering that 
this region is close to the edge of the galaxy, we do not treat it 
as a transition of components. 

\subsection{Spirals}
\label{spirals}
\subsubsection*{NGC1512}
NGC 1512 is classified as a barred spiral galaxy SBa
\citep{RC3_1991}, however  there are many references bringing up 
other components in its morphology: rings, pseudorings, nuclear ring and 
pseudobulge \citep[see for example][]{laurikainen2006,fisher2008}.
According to \cite{compere2014}, who  conducted  a detailed study on this 
galaxy,  it  has a bulge, a bar and a disk.
\fg{fig:ngc1512ird2009kur} shows the results for $\tilk$. The region 
inside $R/R_p<0.5$ shows two components delimited by the local peak at 
$R/R_p\sim 0.15$, and both shows small $\tilk$ compared with the 
overall curvature profile. From 
\citep{laurikainen2006,compere2014}, the innermost region (1) is a nuclear 
ring and region 2 is dominated by the bulge. By the shape of $\tilk$ 
one must consider a pseudobulge component, which is in agreement 
with \citep{fisher2009,fisher2010}.

Regarding the bar, it seems that pseudobulge and bar coexist 
(the change in the position angle and ellipticity between the bulge and the 
bar is very smooth \cite[see figure 4 of ][]{jungwiert1997}), with the bar 
connecting the pseudobulge and the outer ring -- according to 
\cite{fisher2008}, a pseudoring (region 3). 
With this configuration, the distinction of the bar and the pseudobulge 
with $\tilkR$ is not clear, since region 2 encloses both components. 
\cite{fisher2008} also did 
not fit a bar component separately to NGC\,1512 and considered only a 
Sérsic+Exp fit to it. 
What is noticeably is the outer ring, the peak at $R/R_p \sim 0.5$ points 
to the transition between the pseudobulge+bar with this ring, which is
dominant in region 3. The last smooth peak at $R/R_p \sim 1.1$ 
binds the dominance of the outer ring with a faint disk ($\tilk$ approaches
to zero).

\subsubsection*{NGC\,2273}
NGC\,2273 is classified as SBa \citep{nilson1973} and SB(r)a 
\citep{RC3_1991}. In \fg{fig:ngc2273rkur} the curvature indicates a small 
bulge contained inside $R/R_p \sim 0.1$ (region 1). Considering the valley in 
region 2, it retain a gradient in $\tilk$, considerable lower compared to the 
first and second local peaks, suggesting a bar component. However 
this behaviour is not well pronounced here, but still indicates correctly 
the existence of different structural components. 
The next valley in region 3 is formed by two tightly wound spiral arms 
\citep{laurikainen2017}, which behaves like a ring. The fourth peak at 
$R/R_p\sim 0.5$ makes the transition of this component with the outermost 
region of the galaxy. 

\cite{lavers2004} worked on the details considering the aforementioned 
structures, adding a lens component and assuming that the outer region  is 
a disk. Also, they argue that the bulge is prominent, nevertheless $\tilk$  
indicates that the bulge is not too prominent.  This can be evidenced by 
the fact that the inner region can be formed by other components:
\cite{moiseev2004A} noted that the bar is of large scale; 
\cite{erwin2003,moiseev2004A} also assert that the innermost region can
be built by a nuclear spiral while \cite{mulchaey1997} says that it is a 
secondary bar. Also, \cite{erwin2003} says that this galaxy is four-ringed, 
two of them outer rings  (outer region). {And additionally,
NGC\,2273 is also characterized with a barlens component by }\cite{laurikainen2017}.

The curvature give a picture of 
these rings in region 4 due to small perturbations.
In summary we find that $\tilk$ gives and acceptable constrain that 
NGC\,2273 has a bulge+bar/ring+disk structure. 

\subsubsection*{NGC\,7723}
NGC 7723 is classified as SB(r)b/SB(rs) \citep{RC3_1991,comeron2013}.
\fg{fig:ngc7723kur} shows the result of the curvature
for NGC\,7723. The bulge is contained inside the region 1 delimited
by $R/R_p \sim 0.2$. \cite{eskridge2002} argue that the central part 
of this galaxy is composed by a boxy bulge with a symmetric nucleus 
embedded in. 

Analysing the valley in region 2 together with the residual image 
we see that it is related to a bar and a ring. However, the 
bar is much more prominent \citep{prieto2001b} than the later and 
it becomes non-trivial to demonstrate the existence of a ring.
In fact, this component is considered to be a pseudoring 
formed by the way the spiral arms emerges from the bar 
\citep{delrio1998,delrio1999,prieto2001b,eskridge2002}.
In \cite{aguerri2000,prieto2001b} they mentioned the ring structure but 
in their structural decomposition they have not taken it into account. 
The spiral arms begin  at $R/R_p\sim 0.35$ -- and manifests in oscillations of $\tilk$.
\cite{delrio1998} made another commentary arguing that there is a disk 
outside the spiral arms, however due to our confidence limit $S/N =2 $ 
it is not possible to conclude that this disk exists using $\tilk$.
Still a disk feature can be noted between $0.35\lesssim R/R_p \lesssim 1.0$ because 
the mean value of $\tilk$ is very low.

\subsubsection*{NGC 6056} 
This galaxy is considered to be an barred 
spiral with an internal bulge (SB(s)0) \citep{RC3_1991,prieto2001b}, and 
also non-barred (S0/a) \citep{tarenghi1994}. 
In $\tilk$ of \fg{fig:ngc6056rkur}, the bulge is within $R/R_p \lesssim 0.18$ 
(first peak). The valley in 2 is the bar (the same behaviour in $\tilk$ as 
the other galaxies), and confirmed by \cite{prieto2001b}. The transition of the
bar with the outer disk (spiral) is indicated by the local peak at $R/R_p \sim 0.6$.

A pseudobulge component was pointed by \cite{prieto2001b} and they found 
a Sérsic fit of $n=1.1$ in the $r$ band.
But the pseudobulge part does not shows the same behaviour in $\tilkR$ as the other galaxies. 
This may be attributed to the low resolution of the galaxy image 
-- note that there are few data points in the region related to the bulge.

\subsection{Spheroidal components}
\label{spheroidal}
Unlike spiral and lenticular galaxies, whose components are structurally 
very diverse (bulge, bar, spiral arms and disk, for example), in the case 
of spheroidal galaxies (the various kind of ellipticals, bulges and some 
lenticular) the differences in the components are much more subtle. 
Consider, for instance, that many elliptical galaxies substructures cannot 
be seen in raw images, but rather, reveal themselves after image manipulation 
techniques, such as unsharp masking, brightness profile modelling and so on. 
In this way, we may regard that the effects that the different 
component of elliptical galaxies have on the curvature are more faint than for 
spiral galaxies. 

\subsubsection*{NGC\,1270}
NGC\,1270 is classified  as an elliptical galaxy \citep{RC3_1991} and further 
as a compact elliptical galaxy \citep{yildrim2017}. The curvature 
$\widetilde{\kappa}(R)$ is displayed in 
\fg{fig:cut_NGC1270_f160w_kur_new}, its shape is not continuous (as 
expected) and having small oscillations of low scale. However, the curvature 
has a smooth overall shape, is high in inner regions (around $R/R_p\sim 
0.2$) and drops to near zero at large radius. 
As can be seen in the residual image, the central region ($R/R_p 
\lesssim 0.20$) indicates an inner spheroidal component with different 
properties compared to the main body of the galaxy (missing light regarding 
the model for the main body). There is a sign of a small disk within $R/R_p 
\lesssim 0.1 $ aligned with the apparent main axis of the galaxy. 

\subsubsection*{NGC\,472}
NGC\,472 is a compact elliptical galaxy \citep{yildrim2017}. Similar to 
NGC\,1270, it does show some perturbations in the inner regions -- the 
two local peaks are attributed to inhomogeneities -- but 
the shape of $\tilk$ follows an overall shape resembling a pure component, 
which is smooth and decreases from inner to outer regions. The great difference between
NGC\,472 and NGC\,1270 is that for NGC\,1270 the increasing of $\tilk$ in the  innermost 
part is more subtle than NGC\,472. Lastly, a single Sérsic fit $n=5.8$ returned a robust 
representative model for NGC\,472, since there is no residual structures in the outer regions. 

\subsubsection*{NGC\,384}
NGC\,384 is classified in literature as E \citep{RC3_1991,yildrim2017} but 
also S0 \citep{nilson1973}. In comparison to NGC\,472 and NGC\,1270, 
$\widetilde{\kappa}(R)$ is quite different 
(\fg{fig:cut_NGC0384_f160w_kur}) due to the emergence of the substantial 
oscillations in the region $R/R_p \lesssim 0.5$. These delineates some 
internal and small structure. In the residual image of
\fg{fig:cut_NGC0384_f160w_kur} it is readily seem such structure with a 
disk oval shape.

\subsubsection*{NGC 1052}
NGC1052 is classified as an elliptical galaxy \citep{RC3_1991} and as E/S0
\citep{sandage1981}.  \cite{lauer2007} indicates the presence of a core. In 
\fg{fig:pgc0010175rkur} the curvature suggests some possible transition
around $R/R_p\sim 0.15$, in which, therefore, might be between 
the core structure and the remaining structure of an elliptical. 
Furthermore, 
\cite[][and references therein]{milone2007} conclude that NGC\,1052 
has a rotating disk. The curvature may indicate such kind of component 
because the mean small values after $R/R_p\sim 0.6$, however such disk seems not 
to be exponential.   The disk is also seen in the 
residual image. 

\section{Discussion}\label{sec:discussion}
Galaxy structural analysis has been traditionally tackled with parametric 
methods of model fitting. Currently, given the amount of data available,  
there is a growing need of using non-parametric methods that do not have 
an underlying model and that require less user intervention. The curvature, 
being non-parametrical, can provide us with a framework that allow us to 
automate the galaxy structural analysis and thus  its appliance to large datasets.

With our set of 14 galaxies, distinct behaviours appeared 
to relate to the curvature. The basics is that regions dominated 
by different structural components have its own shape in the 
surface brightness and we can use $\tilk$  to unveil 
the difference between each one. The curvature is  sensitive 
 to smooth (e.g NGC\,384, NGC\,6056) or abrupt variations (e.g. 
NGC\,1512, NGC\,7723) of the light profile. High concentrated regions 
have high values of $\tilk$ (in relation to the overall of each curvature profile), 
but transitions regions also reveal high $\tilk$. This happens  because 
along the radius, a transition is a smooth discontinuity of the inner part, 
which gives rise to the next component -- i.e. the point where the 
profile changes from one component to another  --  and frequently has a different 
slope (higher) than the slope characteristic of each profile alone.

Below we discuss the signature of each component in the curvature profile related to the individual galaxies. 
At this stage of the technique,  there is no clear discrimination between bars and rings from the 1D curvature profile.  
For these cases there may be a limit of the method when signatures of bulges, bars, and rings overlap in the 1D profile.
We  will explore this and other issues in a forthcoming  paper.

\subsection*{Bars}
Bars and rings show a  depression  pattern in $\tilk$ where these 
components dominates{.}  For 
NGC\,936 (\fg{fig:pgc0009359rkur}), 
NGC\,2767 (\fg{cut_NGC2767_f160w_kur}) and 
NGC\,7723 (\fg{fig:ngc7723kur})
the bars have a distinct valley between the two local peaks, 
region $2$ of each galaxy. These bars are represented 
by the narrows valleys in $\tilk$
(note the difference between a narrow valley in $\tilk$ related to a bar and a 
broad valley related to a disk type II in region 3 of NCG\,936). 
For the faint bar of NGC\,2273 (\fg{fig:ngc2273rkur}), 
the curvature exhibit a deep but non-negative valley in region 2
(between the two local peaks), and the same applies to NGC\,6056 (\fg{fig:ngc6056rkur}).
We found that the depression in $\tilk$ for bars is not always negative ,
but considerably lower than the local peaks. 
{For NGC\,1512 (}\fg{fig:ngc1512ird2009kur}{) the curvature does not 
	indicate well the bar in region 2 because bulge and bar appear 
	to show a subtle transition. }

\subsection*{Rings}
Galaxies NGC\,1512  (\fg{fig:ngc1512ird2009kur}), NGC\,2273 (\fg{fig:ngc2273rkur}) 
and NGC\,7723 (\fg{fig:ngc7723kur}) have rings. They exhibit a high and narrow
negative gradient in region 3 (first two) and region 2 (last one). These ring structures are 
also likely to be formed by the tight morphology of the spiral arms after the end of the 
bar, for example NGC\,2273 and NGC\,7723. 
One issue in 1D curvature profile at the current 
stage is that it may not be possible unravel the differences between bars and rings.

\subsection*{Disks and spiral arms}
The most notable galactic disk absent of spiral arms that shows values 
of $\tilk$ close to zero is that of NGC\,4267 (\fg{fig:pgc0039710rcomponentsgray}). 
For NGC\,936 (\fg{fig:pgc0009359rkur}) the disk 
is of Type II, therefore it shows a broad valley in region 3 
(a kind of transition between inner and outer parts of type-II disks). In the outer 
part of the disk, the curvature approaches zero.  
A similar result is found for NGC\,2767 (\fg{cut_NGC2767_f160w_kur}).
Another case is NGC\,4417 (\fg{fig:NGC4417rkur}) which in region 3 shows
a decreasing  $\tilk$ and reach small values, regarding that not all galaxy disks 
follows a $n=1$ Sérsic law, therefore there are disk components that exhibit 
$\tilk >0$ or $\tilk <0$.

NGC\,1357 (\fg{fig:ngc1357}), NGC\,2273 (\fg{fig:ngc2273rkur}), 
NGC\,7723 (\fg{fig:ngc7723kur}) and NGC\,6056 (\fg{fig:ngc6056rkur}) 
have spiral arms.  For NGC\,7723 the spirals arm are  
well identified by $\tilk$ at the inner to middle part of region 3, due to 
the homogeneous oscillations.
Galaxy NGC\,1357 also shows two oscillations of this kind 
from the middle onwards part of region 3. 
For NGC\,2273, the pattern in $\tilk$ is not well defined, 
but we still get an indication of a perturbed disk in region 4.
The spiral galaxy in which the spiral structure does not appear 
clearly is NGC\,6056.

\subsection*{Spheroidal}
For our elliptical galaxies the curvature behaviour also revealed 
interesting results. For NGC\,384 (\fg{fig:cut_NGC0384_f160w_kur})
$\tilk$ revealed an inner
component due to the perturbations in region $R/R_p<0.5$, which is a inner 
dust disk. For NGC\,472 (\fg{fig:cutngc0472f160wkur}),  
NGC\,1052 (\fg{fig:pgc0010175rkur}) and 
NGC\,1270 (\fg{fig:cut_NGC1270_f160w_kur_new})
$\tilk$ also indicates perturbations of $I(R)$.
This behaviour indicates that these galaxies are 
not a complete homogeneous system but are formed 
by heterogeneous light distribution. This means that even 
in systems with subtle variations of the brightness profile, we can 
unveil the existence of perturbations in the galaxy's structure. 

\subsection*{Pseudobulges}
Another important result we found on this work
is the behaviour of $\tilk$ for pseudobulges. Inner regions 
that shows small values of curvature compared to the values of posterior 
regions{,} and a subtle increasing in the values,  might infer pseudobulge 
components. In our sample the cases are 
NGC\,1211 (\fg{fig:PGC0011670_r_kur}),
NGC\,936 (\fg{fig:pgc0009359rkur}), 
NGC\,2767 (\fg{cut_NGC2767_f160w_kur}), 
NGC\,1512 (\fg{fig:ngc1512ird2009kur}), 
NGC\,2273 (\fg{fig:ngc2273rkur}) 
and 
NGC\,7723 (\fg{fig:ngc7723kur}) but for NGC\,2273 {and NGC\,7723} this pattern is not too clear. 
For NGC\,2767 there is no reference in literature indicating that 
the bulge is pseudo.

In summary, pseudobulges curvatures display a distinct pattern. Since 
pseudobulges are in some cases similar to disks -- having surface 
brightness profiles similar to exponentials -- the curvature might in general 
shows a behaviour that resembles a disk, we mean this by having small 
values in inner regions and no abrupt variations (fast increasing) 
in some range of $R$. 

\section{Summary}\label{sec:conlcusions}
We summarise bellow our conclusions in using the curvature 
of the brightness profile for structural analysis.
We have introduced the curvature of the brightness profile of 
galaxies as a tool to identify and study their different 
structural components, most notably  bulges, bars, discs, rings 
and spiral arms. The underlying argument for such is that these 
components have distinct Sérsic index (or concentration) which 
directly impact the curvature. But, unlike standard multicomponent 
modelling of light profiles, the curvature is non parametric and 
does not depend on model parameters.

We measured the curvature profile for structural analysis in 14 galaxies 
(\tb{tab:data_sample}) comprising different  morphologies. For these galaxies, 
we identified their	structural components and inferred their domain regions  
in terms of the local peaks in $\tilkR$; following the peaks in the 
measured curvature profile we spotted the matching regions in the image 
and residual maps where the different components can be identified and 
related to the curvature behaviour. Thus regarding the curvature curve 
(in terms of logarithm normalized scale of intensity): 
\begin{enumerate*} 
	\item  pseudobulges have small values (compared to the highest absolute 
	values) or near zero curvature in their regions;
	\item {Bars and rings have similar narrow valleys in $\tilk$ (negative or small values) although 
		they are inextricable regarding the profile;} 
	\item Disks have a broad profile in the curvature with near zero (Type I) 
	or negative (Type II)  values. 
	\item the curvature in elliptical galaxies shows that they also are not 
	completely homogeneous systems.
\end{enumerate*}

\section*{Acknowledgments}
We would like to thank Horácio Dottori and Leonardo de Albernaz Ferreira for 
useful comments in the original manuscript. Val\'erie de Lapparent for kindly 
providing the original high resolution EFIGI stamps used in this work. 
The anonymous reviewer for many useful questions and suggestions that helped improve the 
manuscript. Matheus
Jatkoske Lazo and Lucas Bonifácio Selbach for useful discussions. The Programa 
de Pós Graduação em Física of Universidade Federal do Rio Grande for all
the infrastructure and PROPESP-FURG for part of financial support. We also thank 
Coordenação de Aperfeiçoamento de Pessoal de Nível Superior - Brasil (CAPES)
for the research grant funding, supporting the development of this work.

\bibliographystyle{mnras}

\bibliography{ references_rev}

\appendix 

\section{Curvature for a Sérsic profile}
\label{app:kurvature_sersic}
\label{sec:sersic_law}

A good description of brightness profile of different components of galaxies is given by the 
 Sérsic law \citep{sersic1963,Sersic1968,ciotti1999} 
\begin{align}\label{sersic_law}
I(R) = I_n \exp\left\{
-b_n \left[\left(\frac{R}{R_n}\right)^{1/n}-1\right]
\right\}
\end{align}
where $R_n$ (effective radius) is the radii  that contain half of the
total luminosity of the galaxy integrated to infinity $L(R = R_n) = 0.5 L_T^\infty$, 
and $I_n$ is the effective surface brightness, i.e.  the 
value of $I(R)$ at $R=R_n$, and $n$ controls the concentration of the profile.
The term $b_n$ is defined  to make the above definitions hold \citep{ciotti1999}  
\begin{align}
b_n =  2n-\frac{1}{3} + \frac{4}{405n^2} + \frac{46}{25515 n^2}.
\end{align}

\begin{figure}
	\centering
	\includegraphics[width=0.99\linewidth]{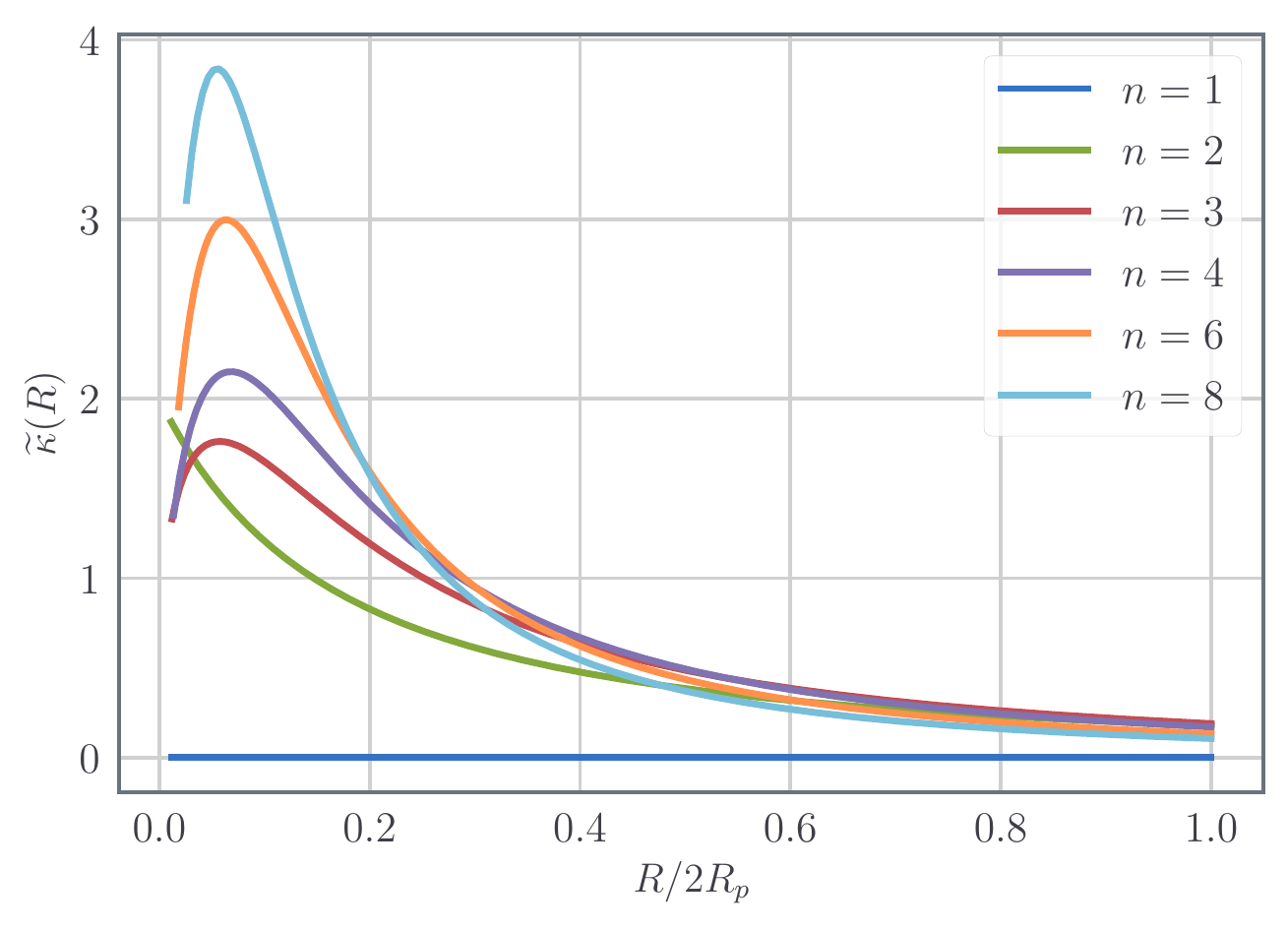}
	\caption{Plots of Equation (\ref{normalized_single_sersic_curvature_nu}) for 
		different Sérsic indexes. The higher is $n$, the higher is $\tilk$ in the 
		inner regions. Note that for large $R$, the $\tilk$ profiles are similar,
		which is reasonable because high $n$ represent more changes in inner part 
		of the profile.}
	\label{fig:kurmunsersic2}
\end{figure}

Here we derive the curvature of a Sérscic law for a single component. 
As commented, to remove the scale on $I$, we use normalization 
given by (\ref{normalization_IR}). 
Taking the logarithm of equation (\ref{sersic_law}) we have
\begin{align}
\nu(R) \equiv \log I(R) = \log I_n + b(n)\left[ 
1-\left(\frac{R}{R_n}\right)^{1/n}
\right].
\end{align}
Since we consider galaxies extending up to $2R_p$, the maximum and minimum are
at $R=0$ and $R=2R_p$, respectively, hence 
\begin{align}
& \max \log I = I(0) = \log I_n +b(n)\\
& \min \log I = I(2R_p) = \log I_n + b_n\left[ 
1 - \left( \frac{2R_p}{R_n}\right)^{1/n}
\right]
\end{align}
thus, inserting these into (\ref{normalization_IR}) gives
\begin{align}
\nu(R) = 1 - \left( \frac{R}{2R_p}\right)^{1/n}.
\end{align}
This result is independent of $I_n$ and $R_n$ because of the normalization on $I$ and $R$.
The first and second derivative are
\begin{align}
\frac{\dd \nu(R) }{\dd R} &= -\frac{1}{2R_p n} \left( \frac{R}{2R_p}\right)^{\frac{1-n}{n}}\\
\frac{\dd^2 \nu(R) }{\dd R^2} & = \frac{n-1}{4R_p^2 n^2} \left( \frac{R}{2R_p}\right)^{\frac{1-2n}{n}}.
\end{align}
Using these two result in equation (\ref{final_normalized_curvature}) we obtain 
the normalized curvature for a single Sérsic profile: 
\begin{align}
\label{normalized_single_sersic_curvature_nu}
\tilkR = 
\frac{ 
\frac{n-1}{n^2}\left(\frac{R}{2R_p} \right)^{\frac{1-2n}{n}}
}
{
\left[ 1+\frac{1}{n^2}\left( \frac{R}{2R_p}\right)^{\frac{2-2n}{n}}
\right]^{3/2}
}
\end{align}

\fg{fig:kurmunsersic2} shows 
plots of (\ref{normalized_single_sersic_curvature_nu}) for different 
values of $n$ against the normalized variable 
$\chi = R/(2R_p)$. 
For a disk, which frequently follows a $n\sim1$ law, we obtain that 
$\tilk \sim 0$ because the numerator vanishes. For $n>1$, $\tilk$ 
is correlated to $n$, the higher the $n$ the higher the curvature 
(in inner regions), for outer regions the Sérsic law becomes more flat and 
$\tilk$ behaves a disk.

\section{Additional Curvature Plots}
\label{app:k_plots}

\begin{figure}
	\centering
	\includegraphics[width=0.9\linewidth]{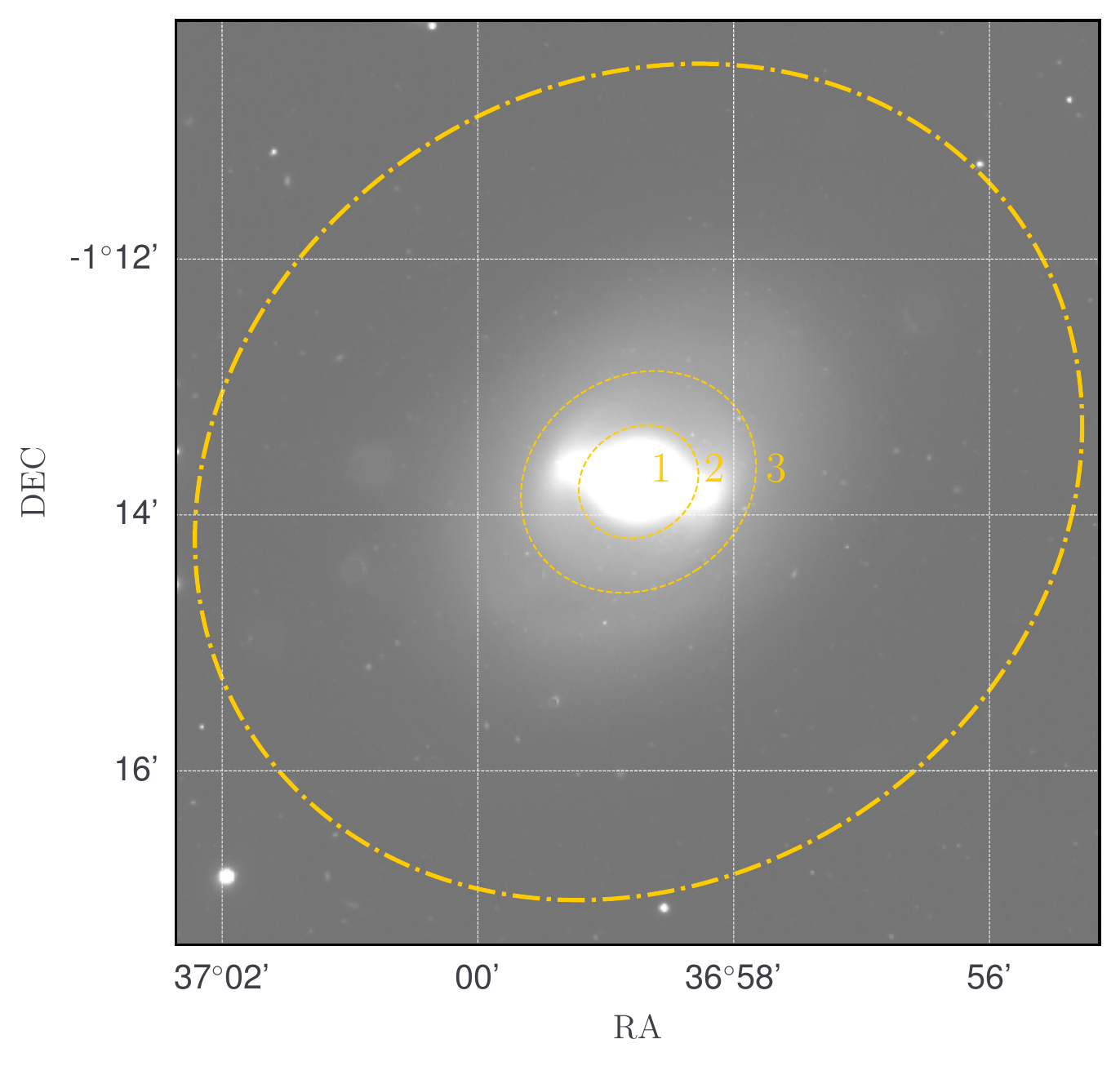}
	\includegraphics[width=0.9\linewidth]{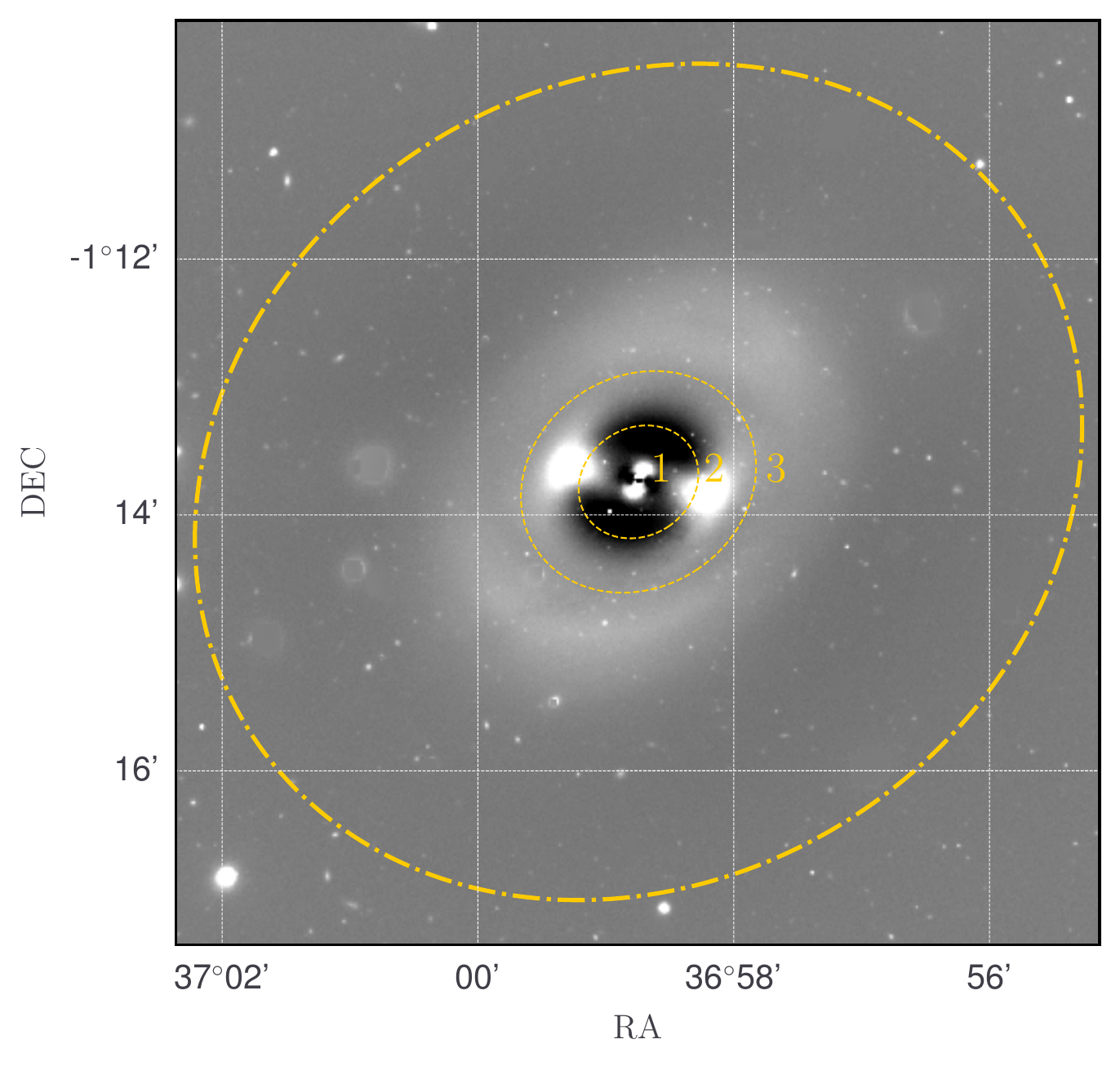}
	\includegraphics[width=0.99\linewidth]{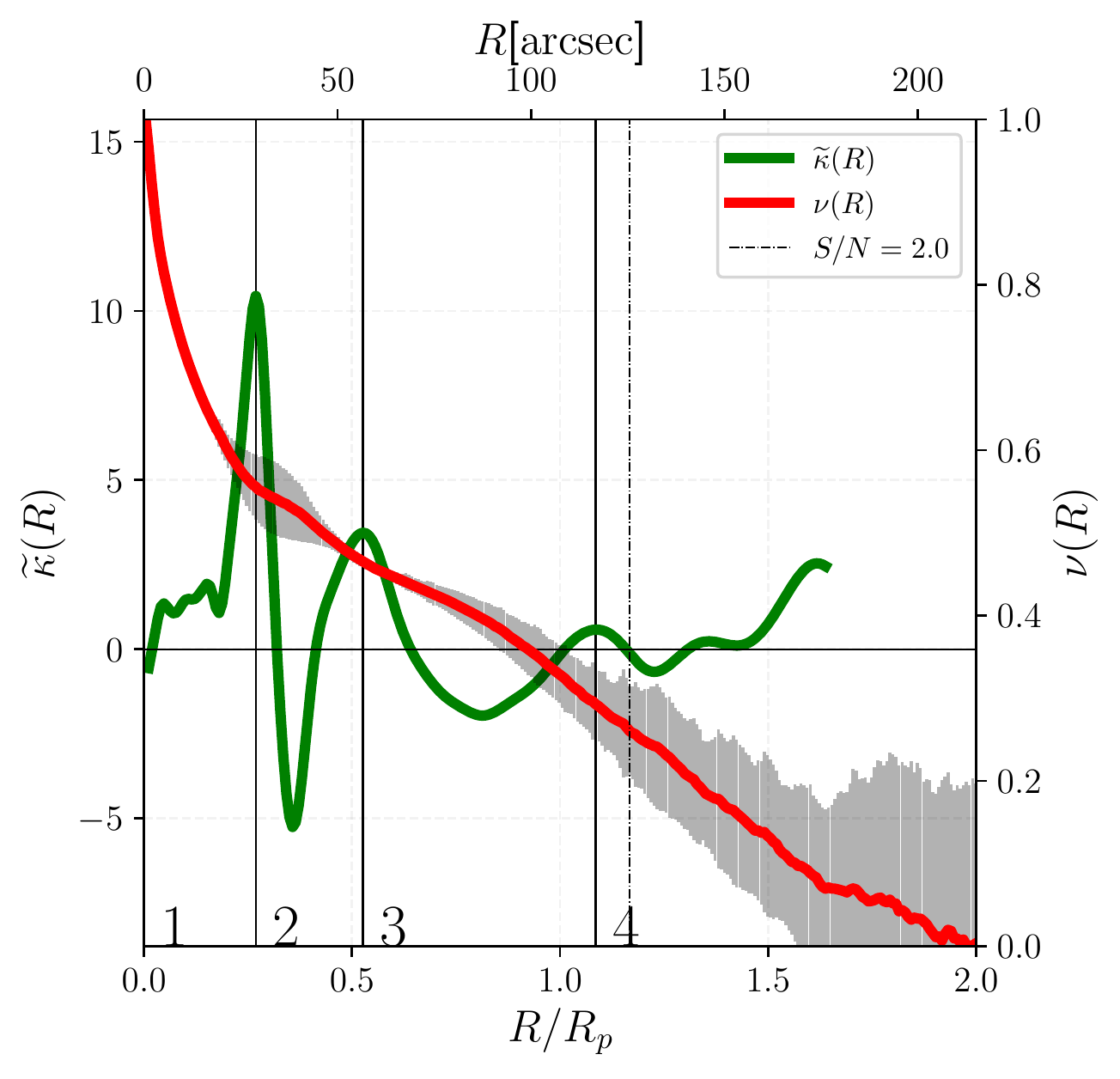}
	\caption{The same as \fg{fig:PGC0011670_r_kur}, but for $r$ band of NGC\,936 from EFIGI. See text for details.}
	\label{fig:pgc0009359rkur}
\end{figure}

\begin{figure}
	\centering
	\includegraphics[width=0.9\linewidth]{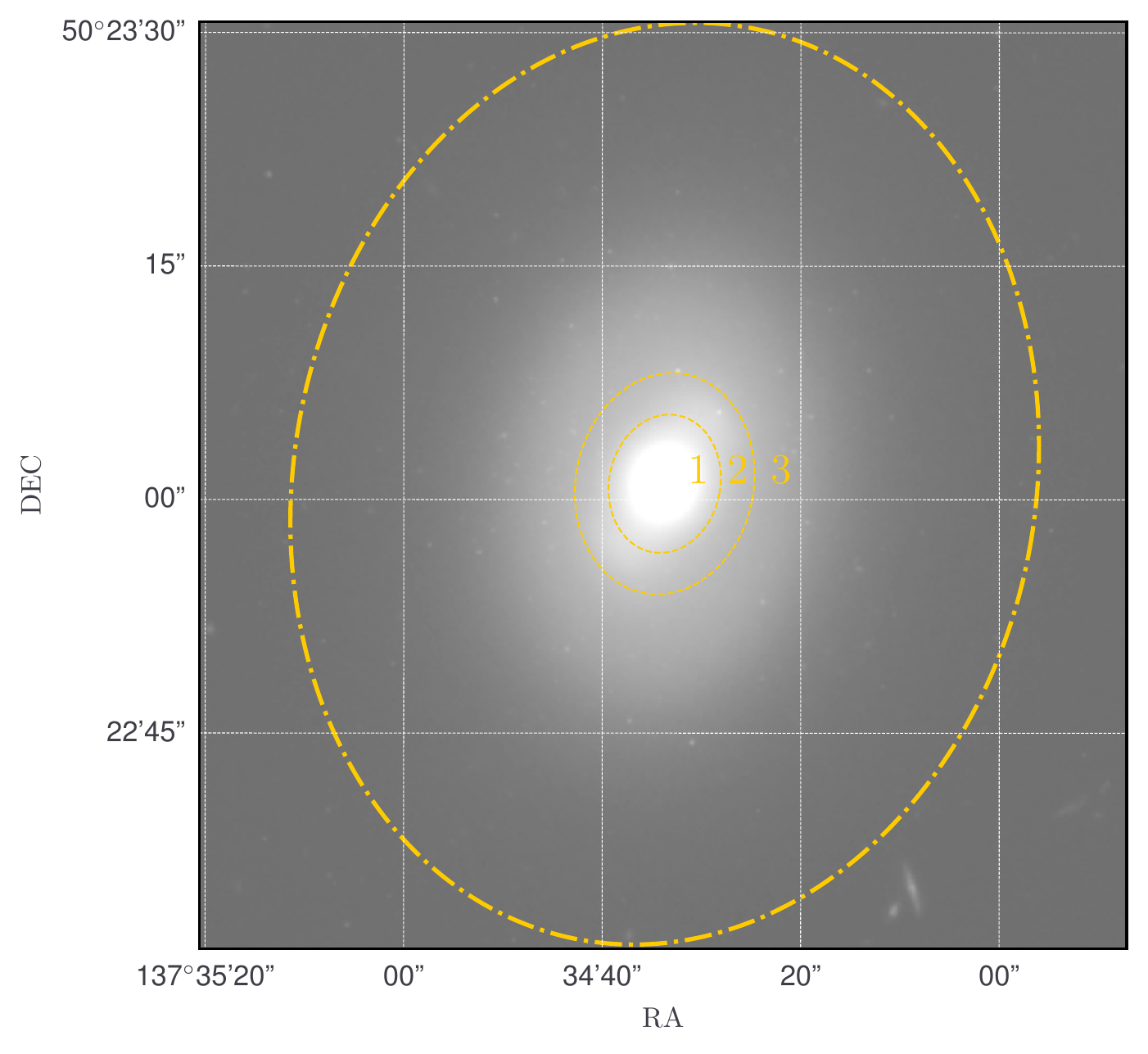}
	\includegraphics[width=0.9\linewidth]{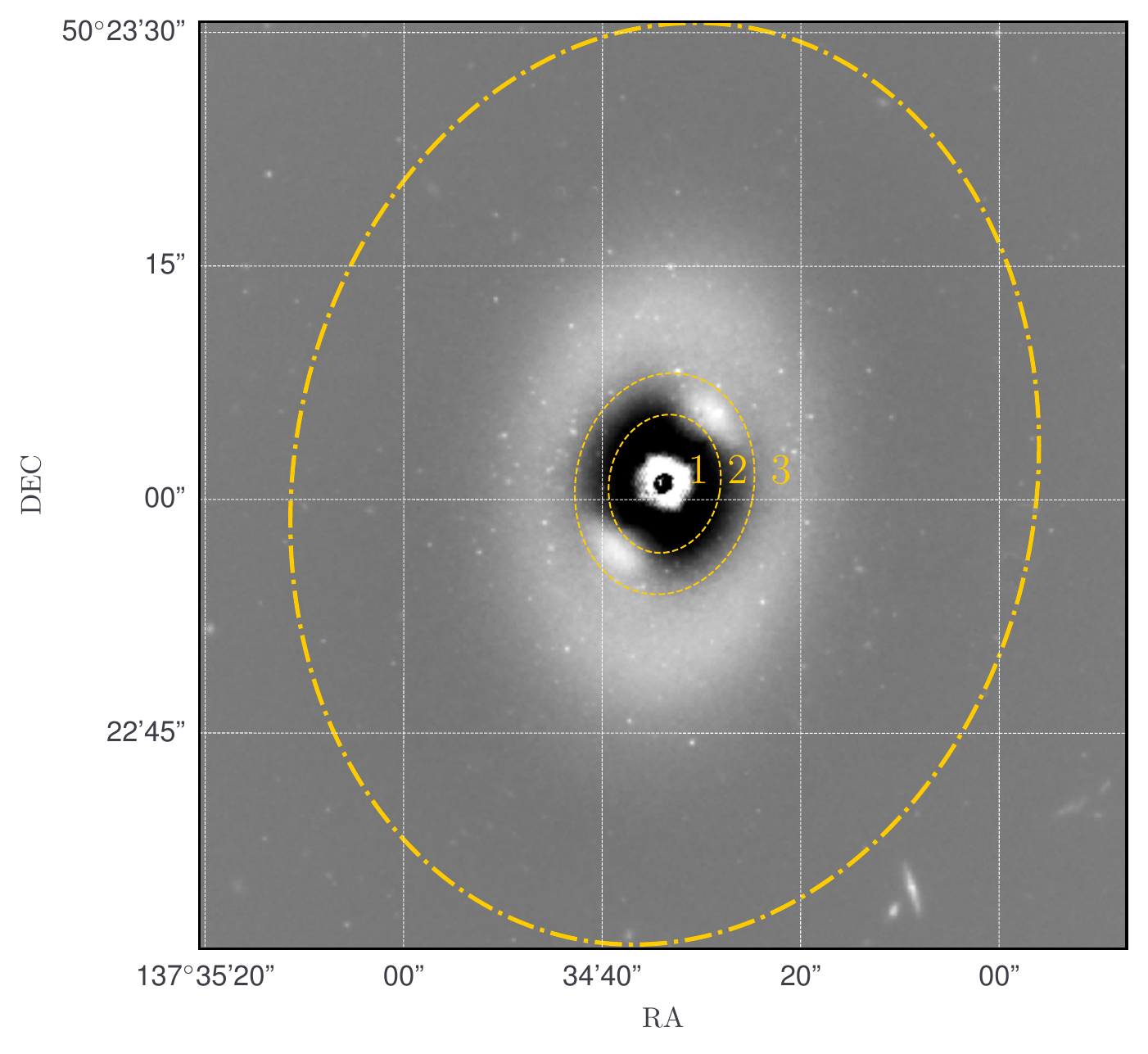}
	\includegraphics[width=0.99\linewidth]{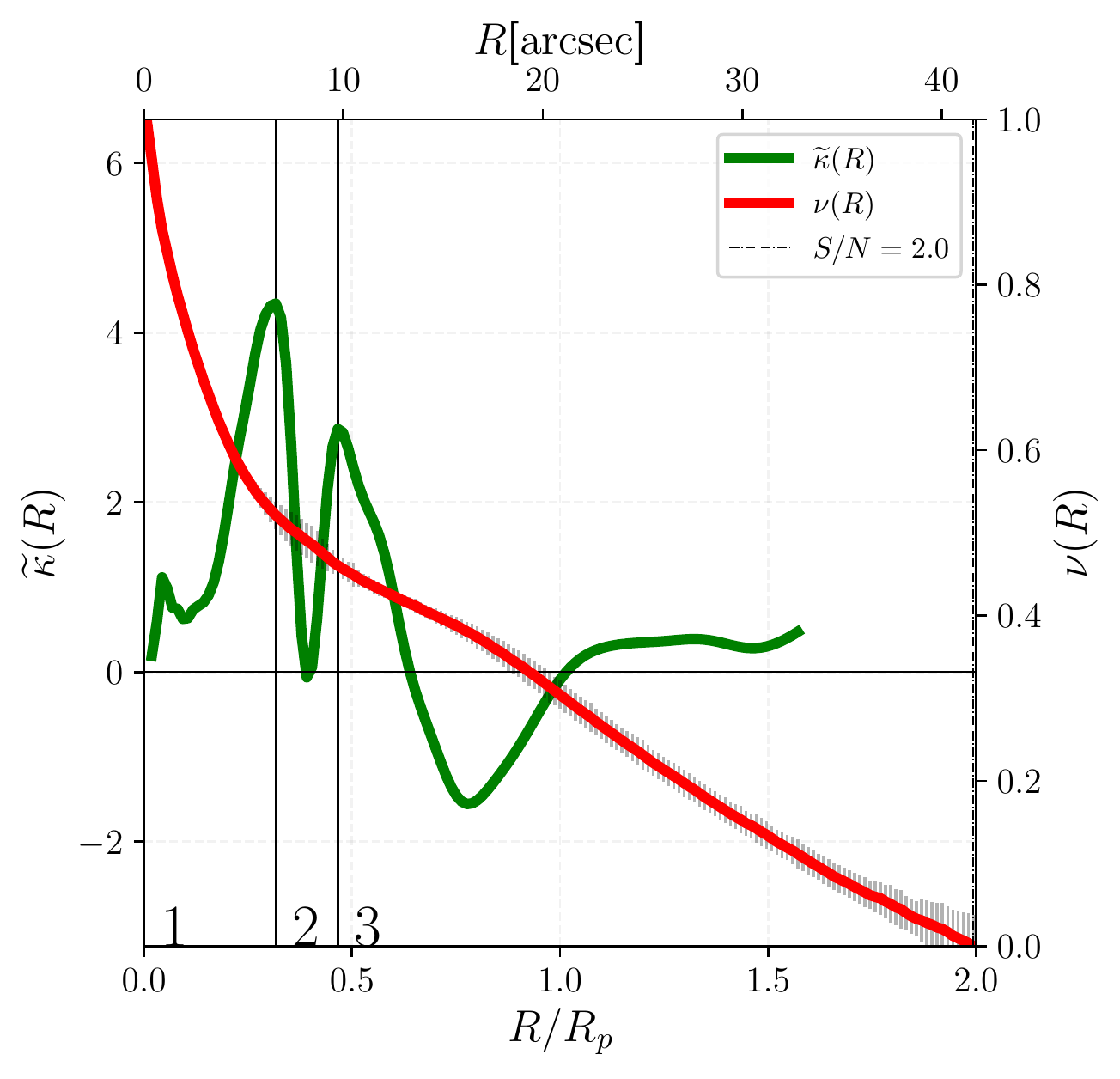}
	\caption{The same as \fg{fig:PGC0011670_r_kur}, but for NGC\,2767 from HST (f160w). }
	\label{cut_NGC2767_f160w_kur}
\end{figure}

\begin{figure}
	\centering
	\includegraphics[width=0.9\linewidth]{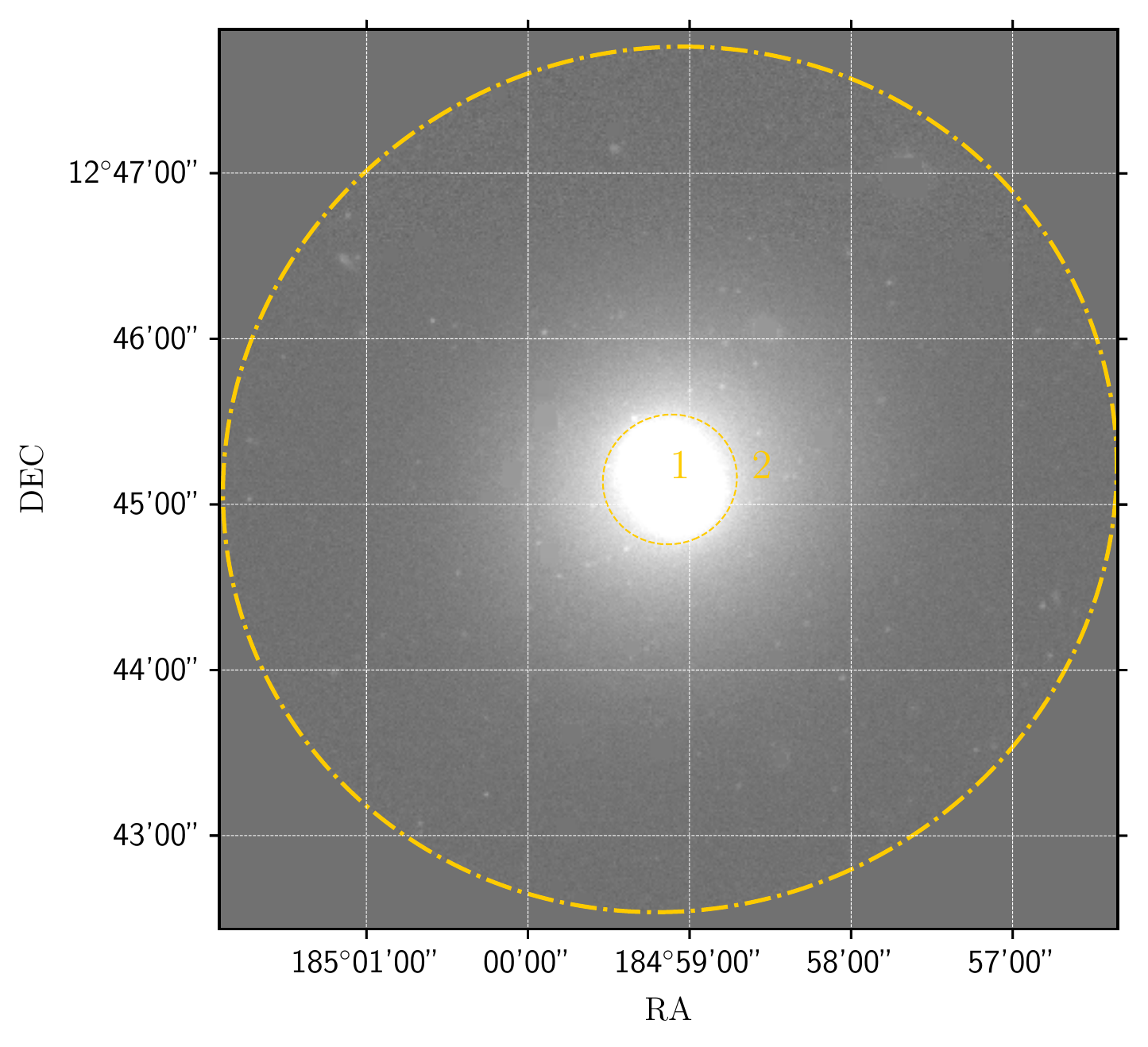}
	\includegraphics[width=0.9\linewidth]{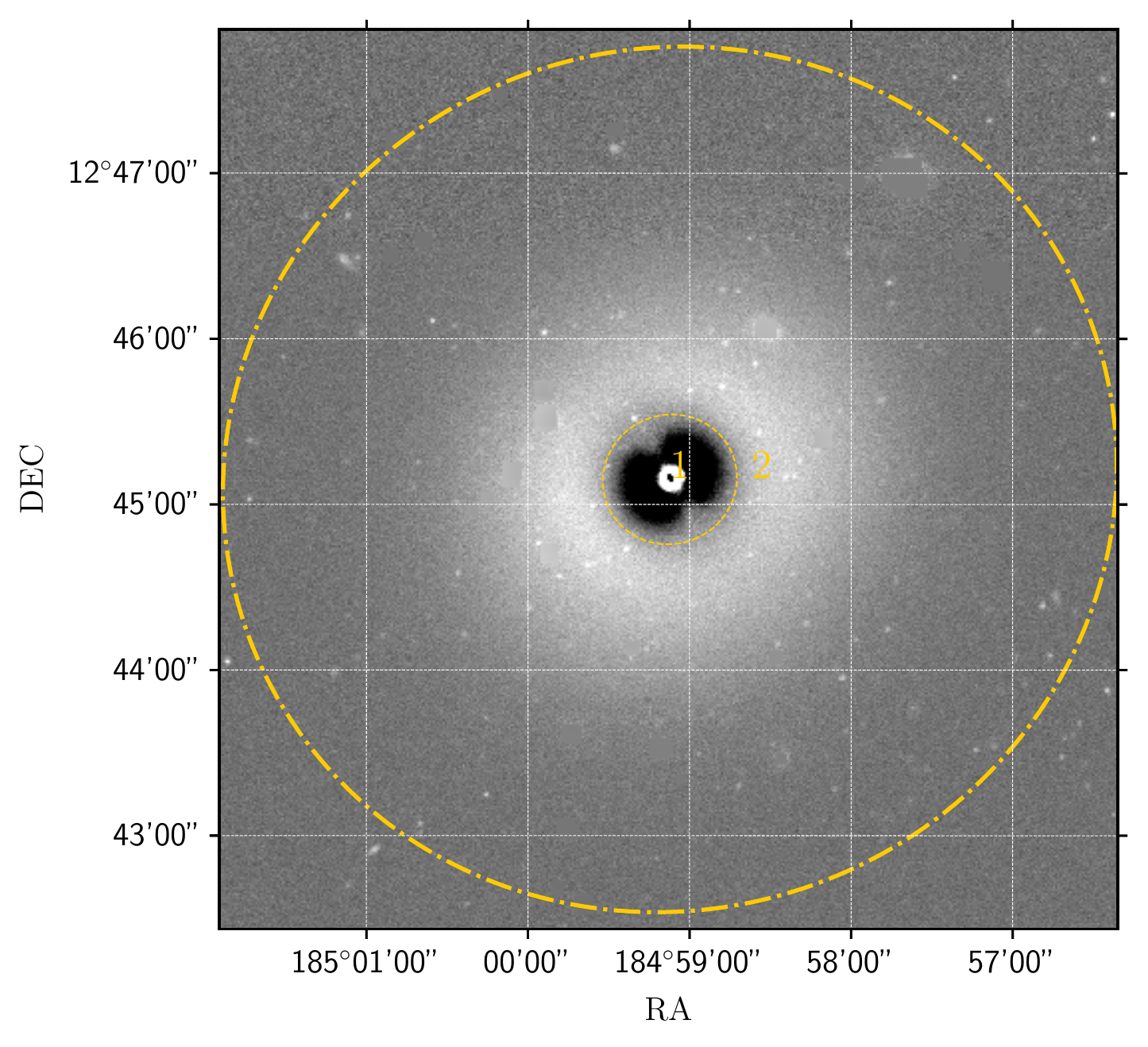}
	\includegraphics[width=0.99\linewidth]{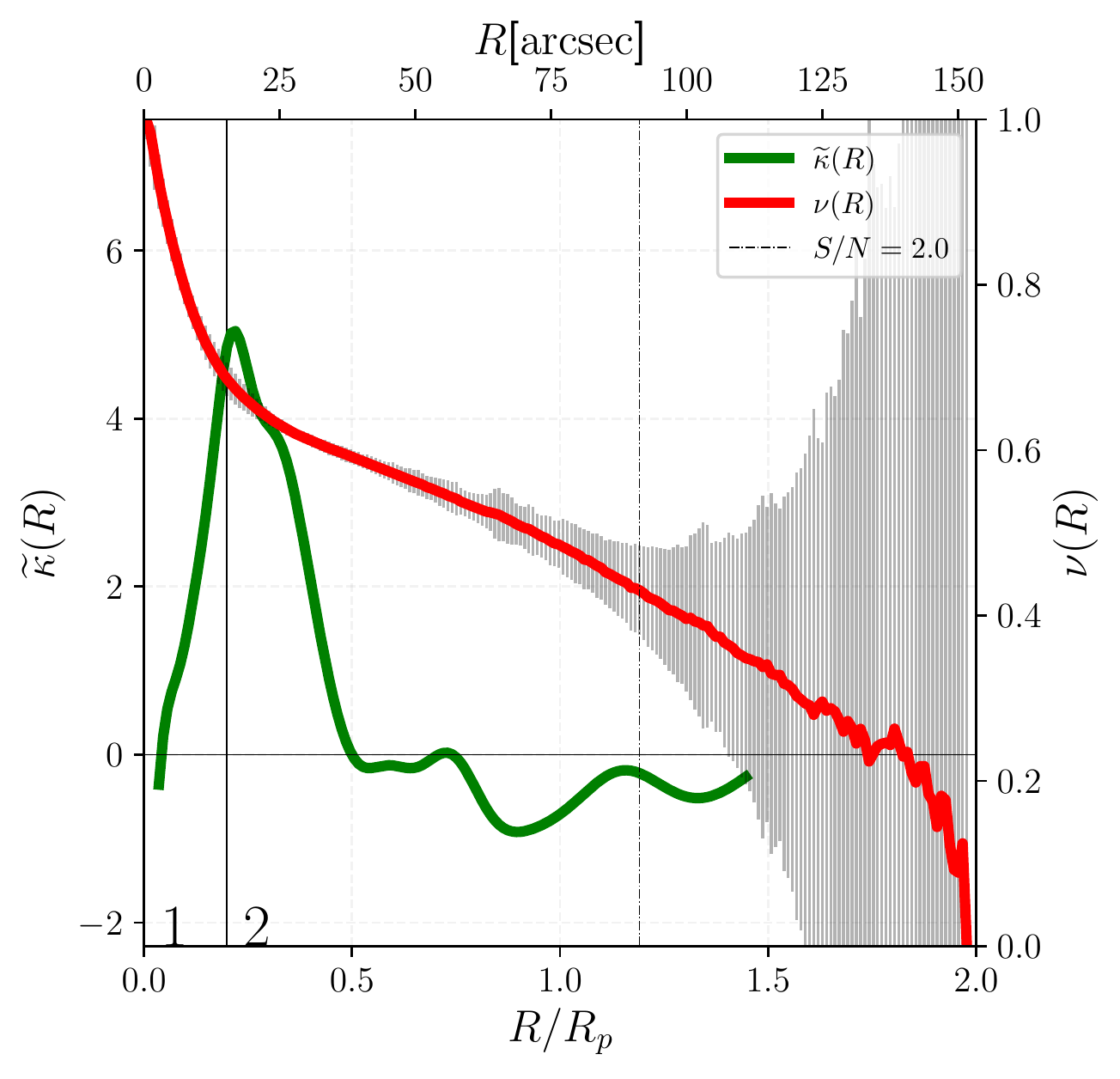}
	\caption{The same as \fg{fig:PGC0011670_r_kur}, but for $r$ band of NGC\,4267 from EFIGI. }
	\label{fig:pgc0039710rcomponentsgray}
\end{figure}

\begin{figure}
	\centering
	\includegraphics[width=0.9\linewidth]{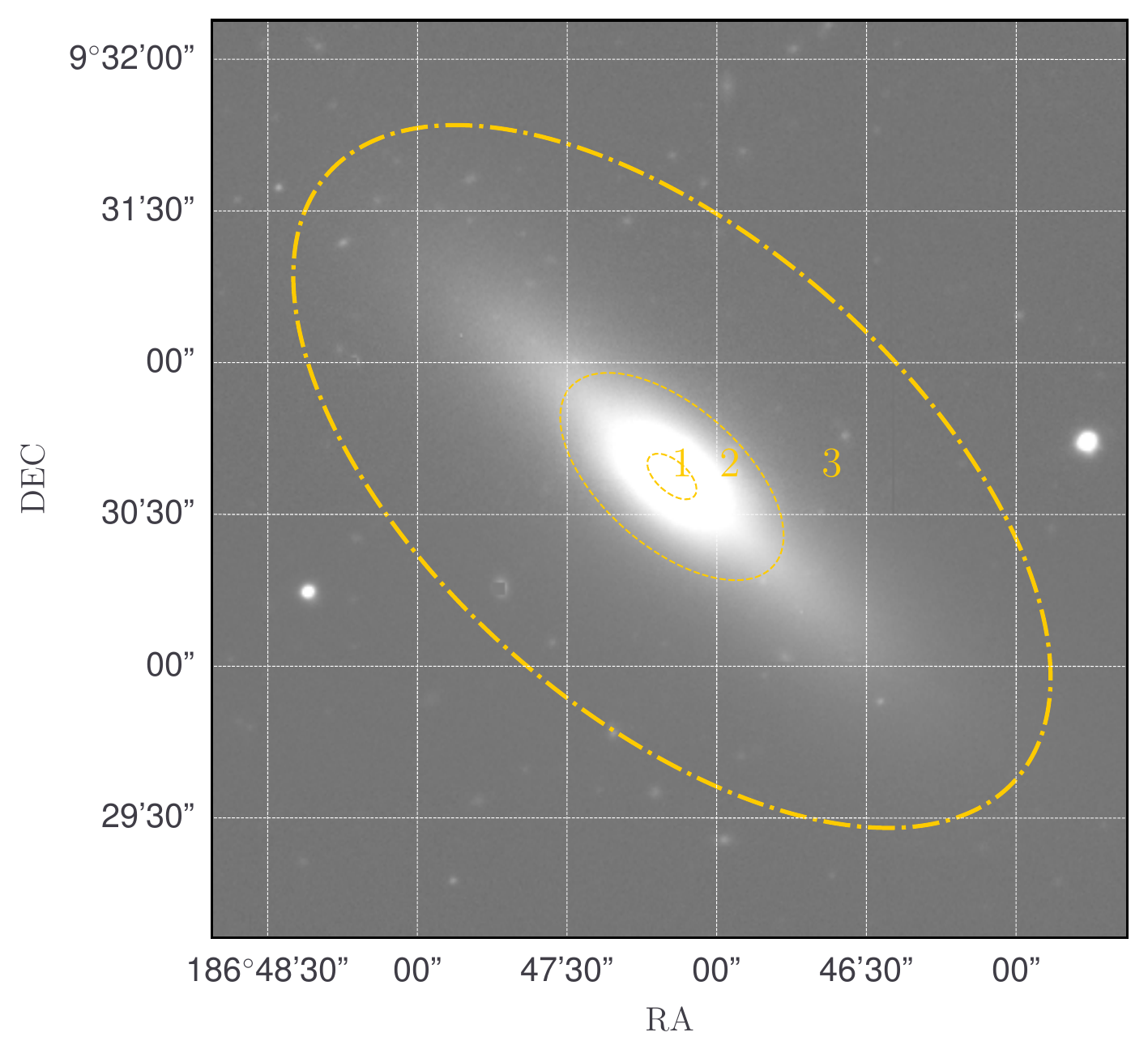}
	\includegraphics[width=0.9\linewidth]{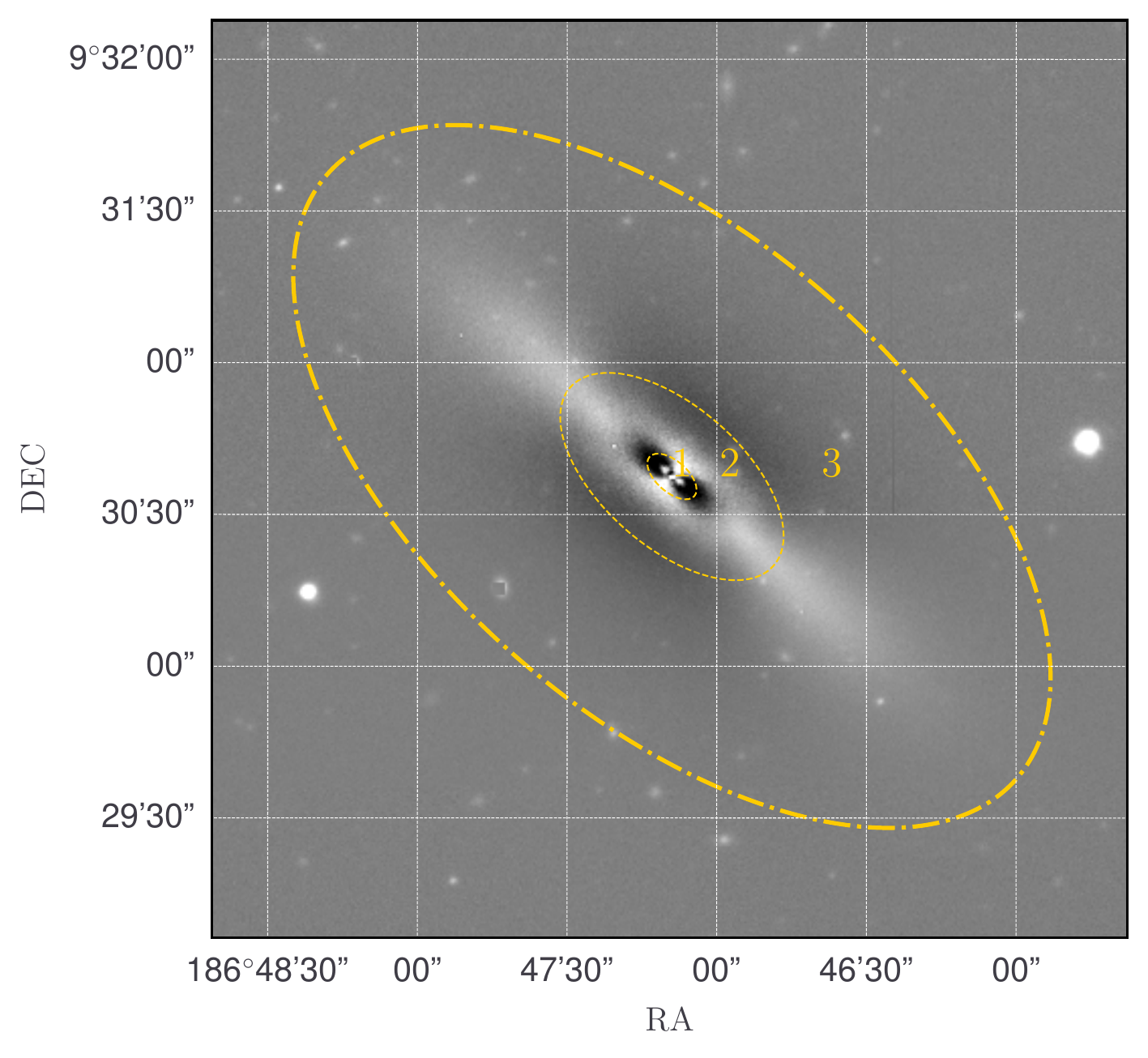}
	\includegraphics[width=0.99\linewidth]{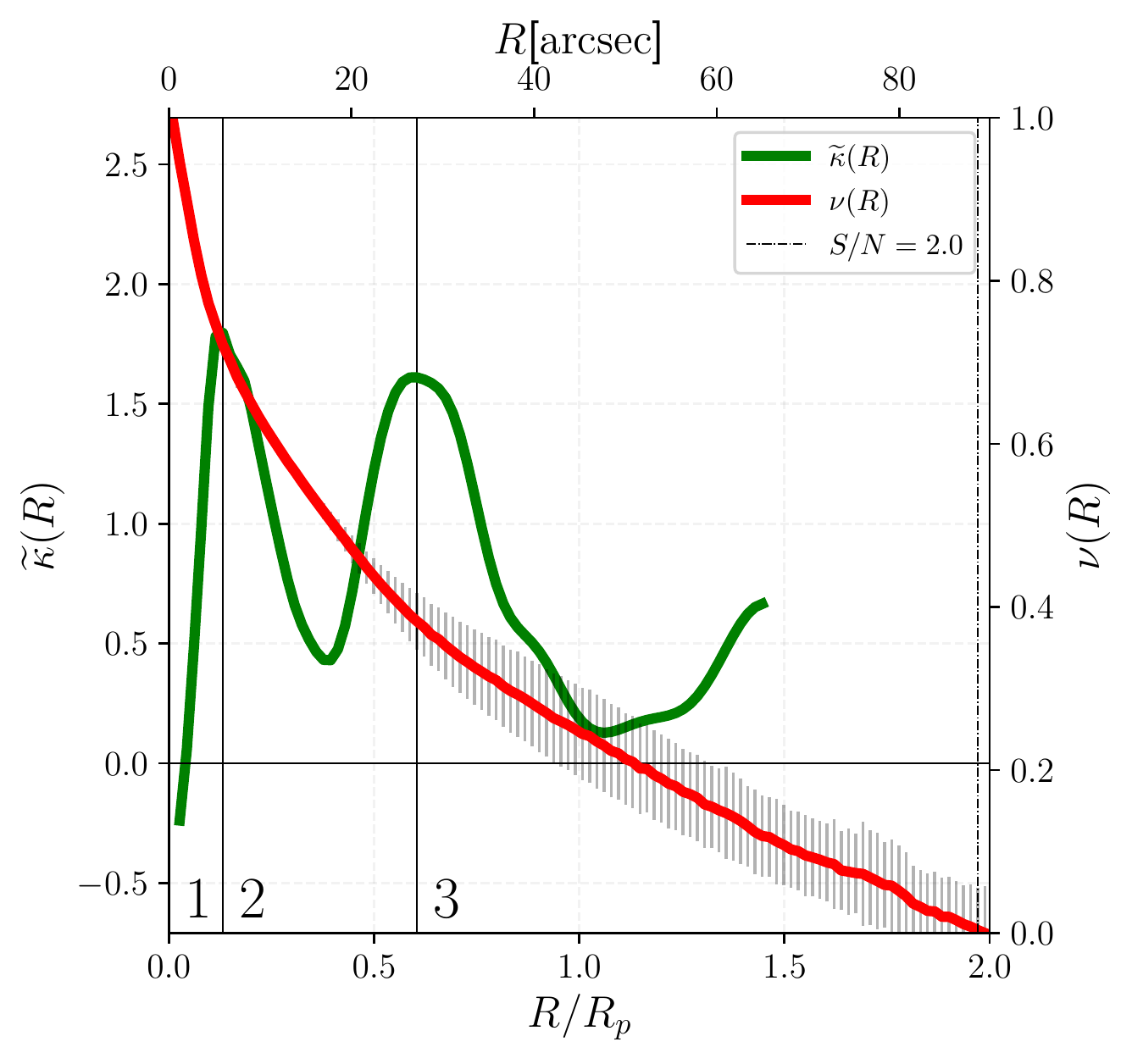}
	\caption{The same as \fg{fig:PGC0011670_r_kur}, but for $r$ band of NGC\,4417 from EFIGI.}
	\label{fig:NGC4417rkur}
\end{figure}

\begin{figure}
	\centering
	\includegraphics[width=0.9\linewidth]{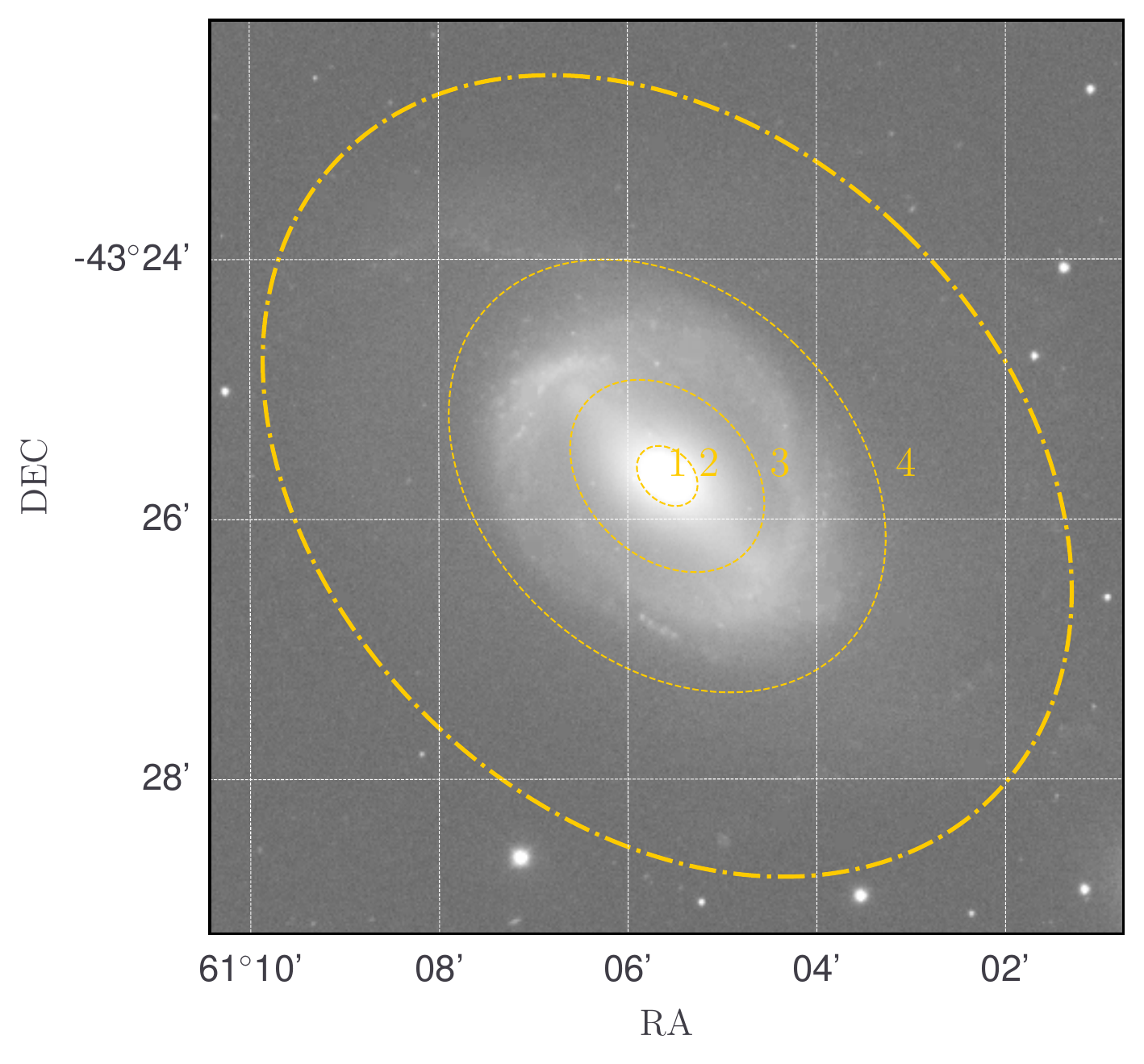}
	\includegraphics[width=0.9\linewidth]{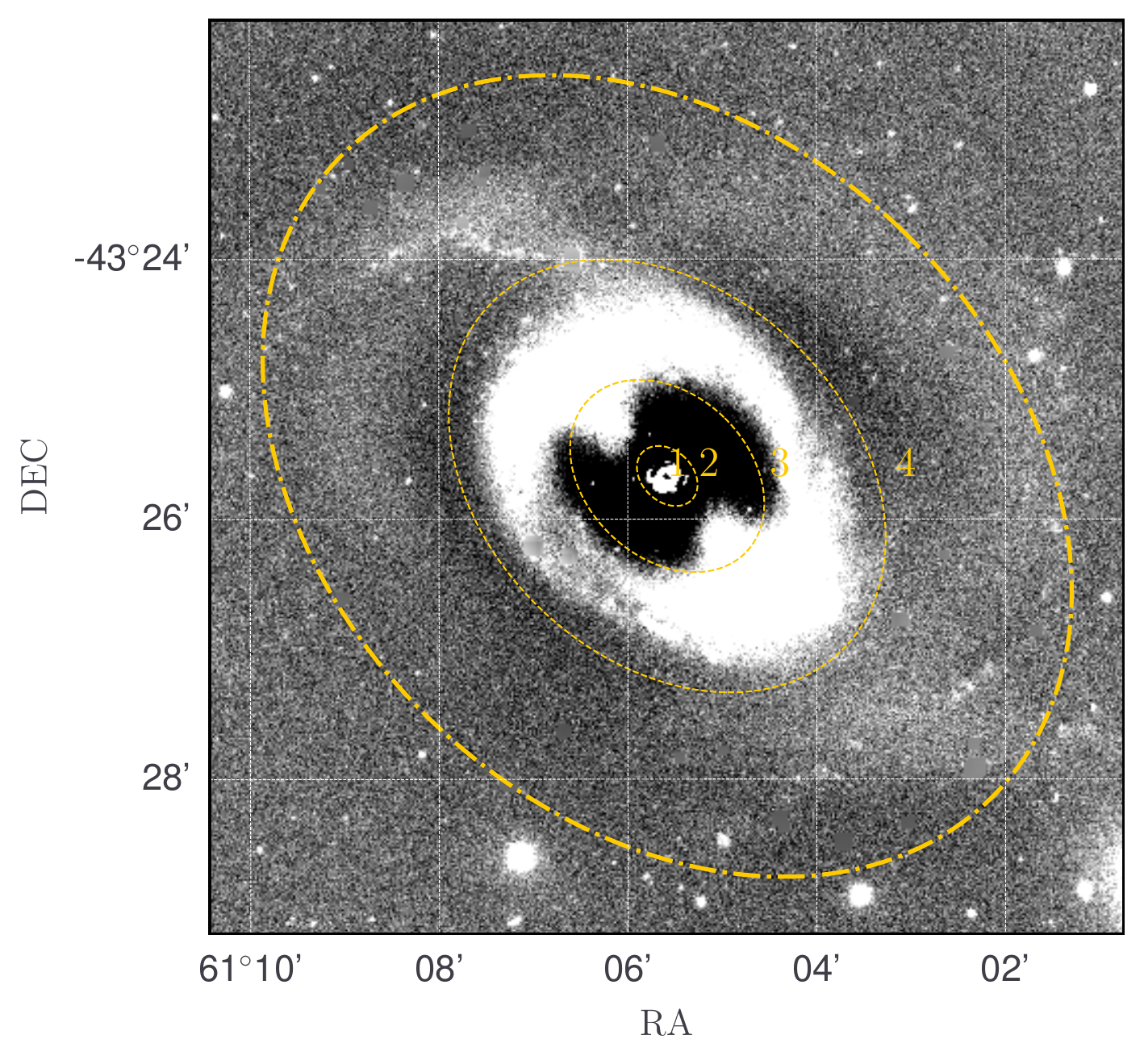}
	\includegraphics[width=0.99\linewidth]{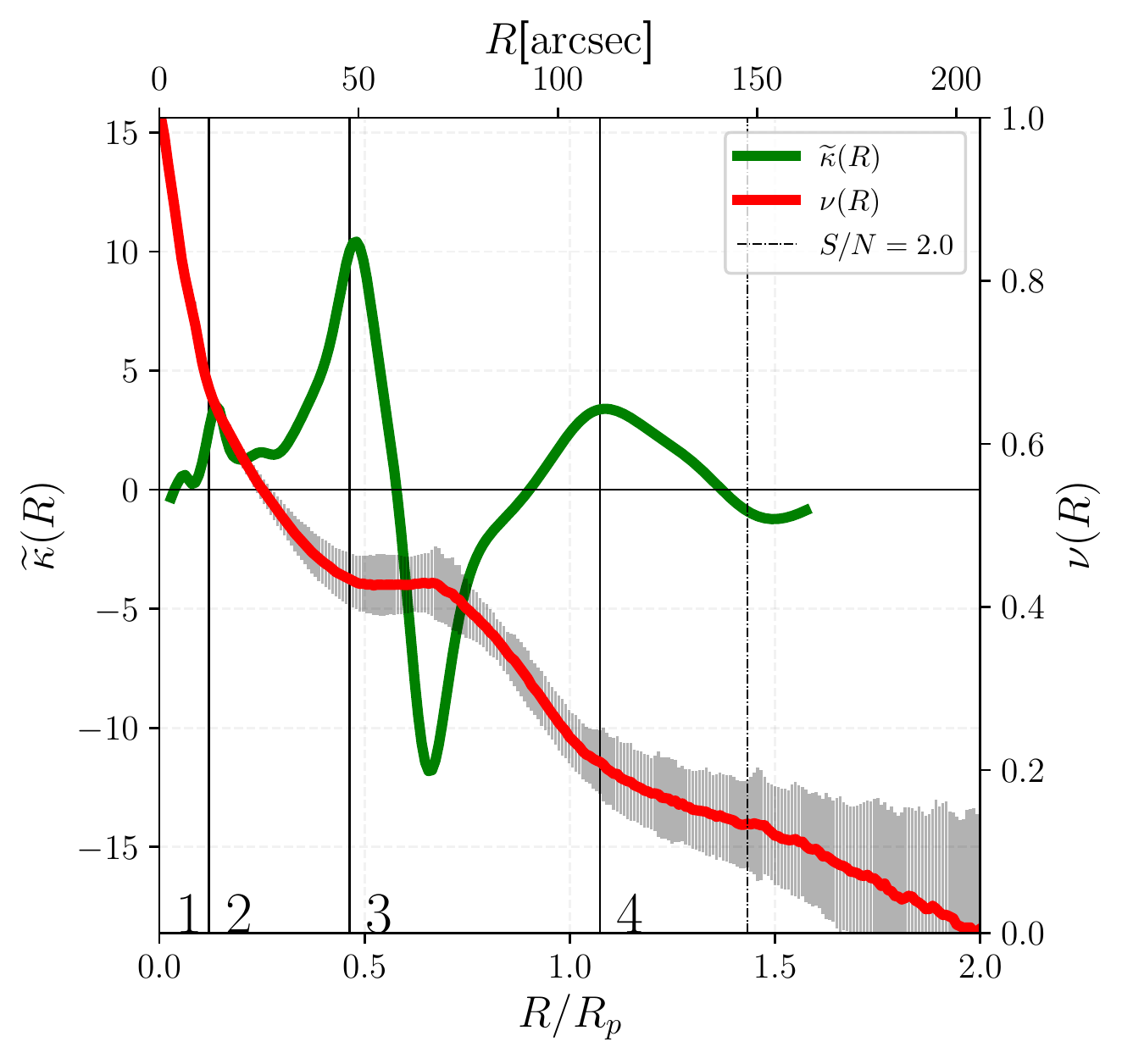}
	\caption{The same as \fg{fig:PGC0011670_r_kur}, but for $r$ band of NGC\,1512 from CTIO telescope.} 
	\label{fig:ngc1512ird2009kur}
\end{figure}

\begin{figure}
	\centering
	\includegraphics[width=0.9\linewidth]{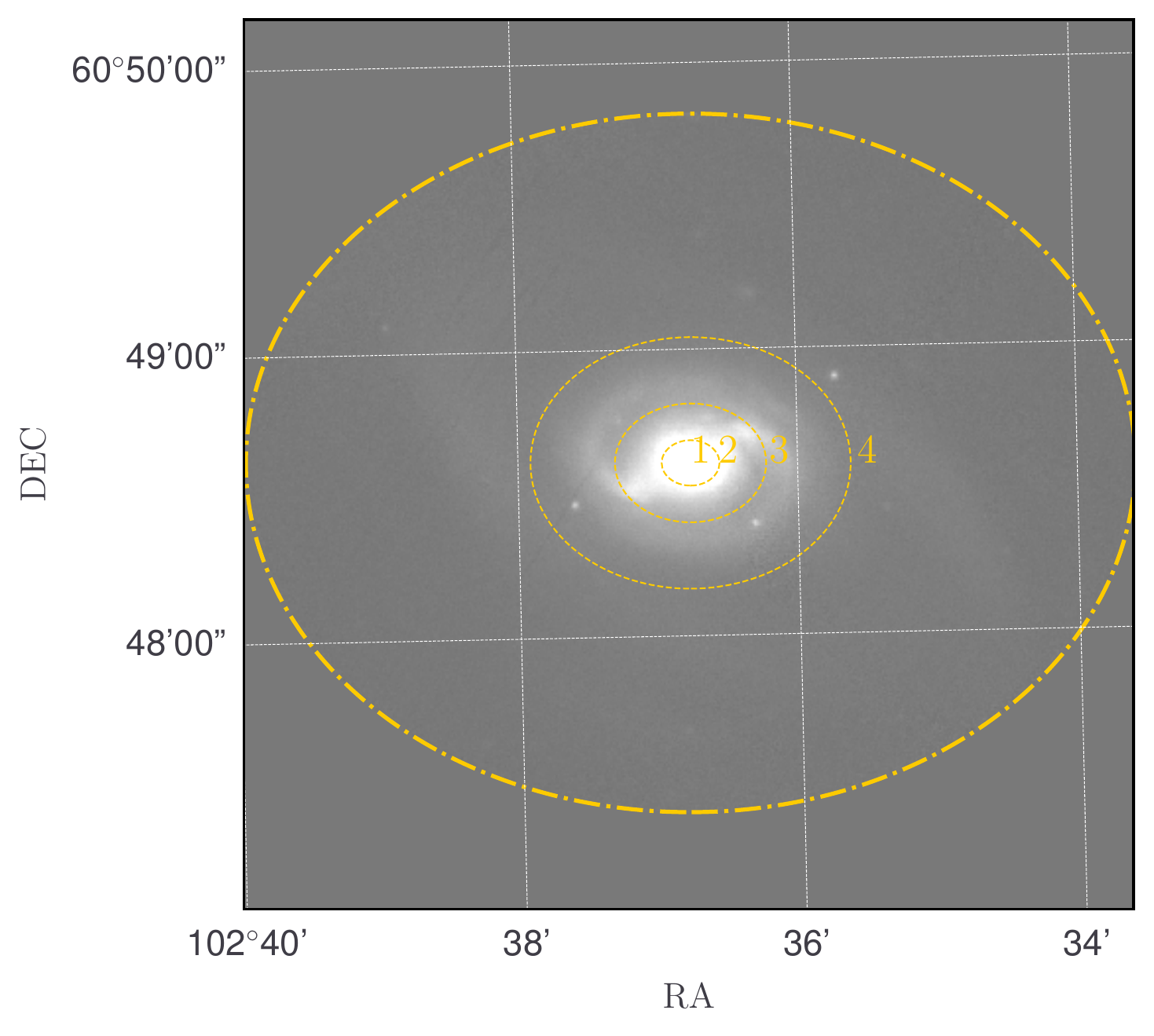}
	\includegraphics[width=0.9\linewidth]{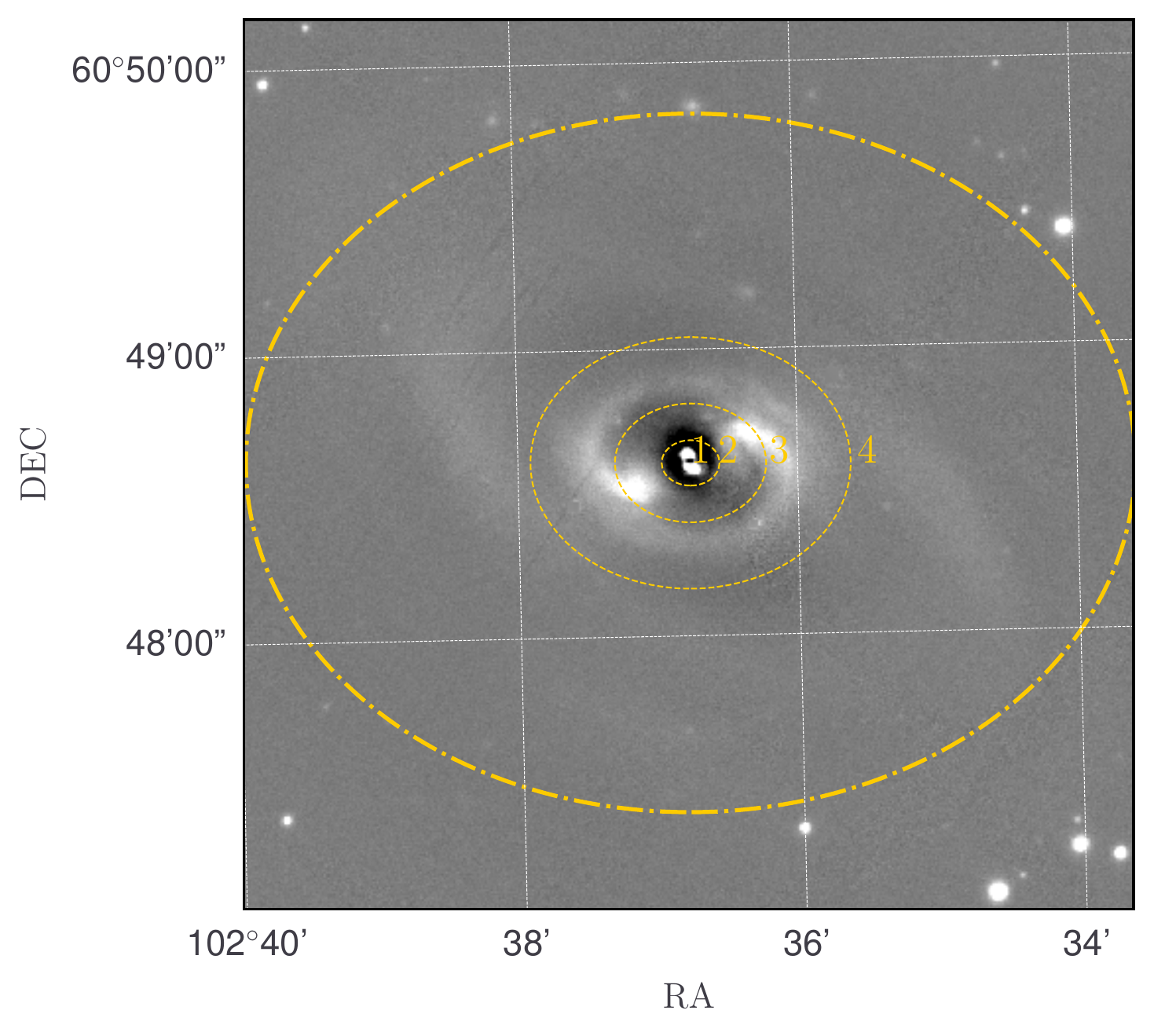}
	\includegraphics[width=0.99\linewidth]{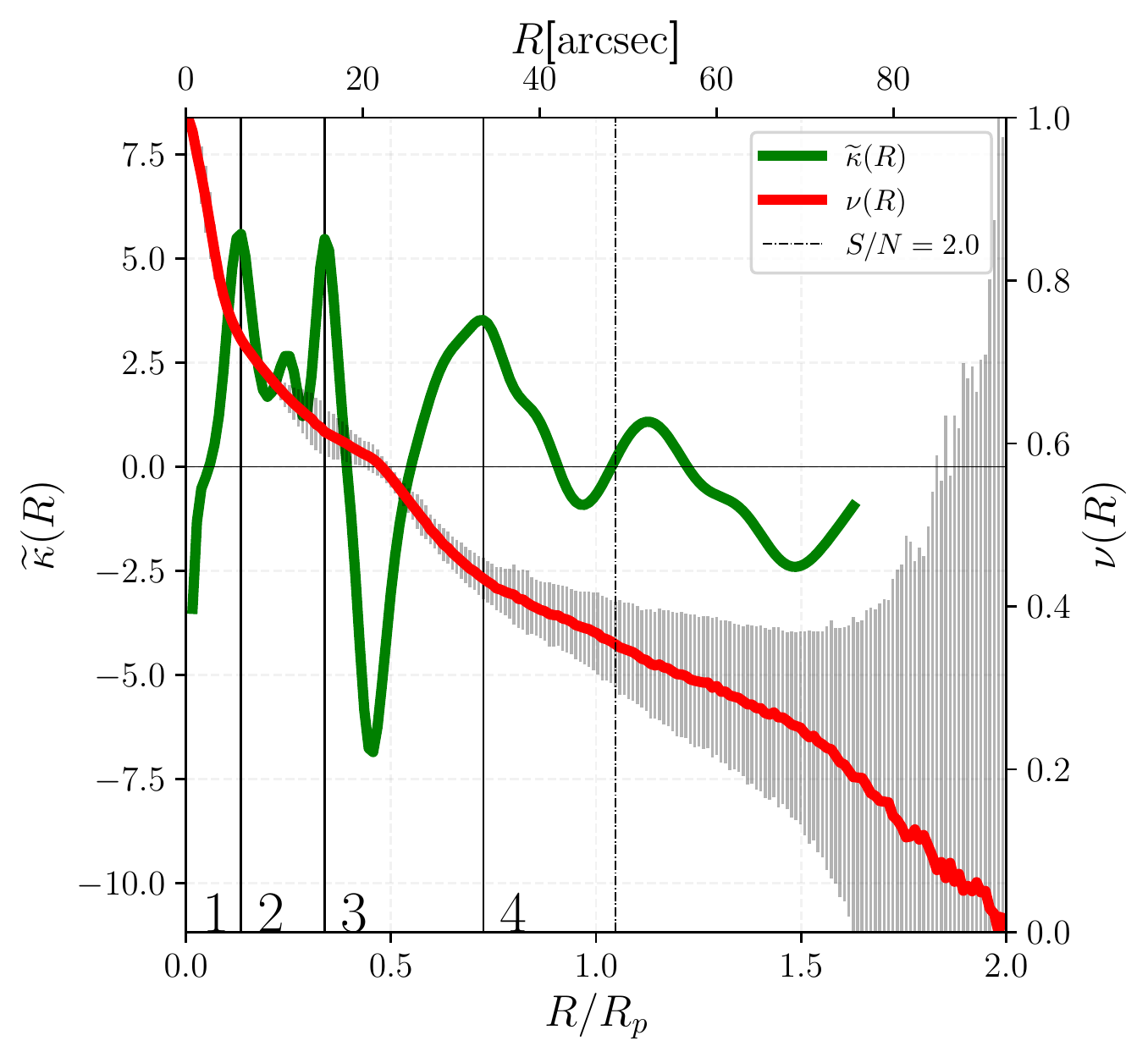}
	\caption{The same as \fg{fig:PGC0011670_r_kur}, but for $r$ band of NGC 2273 from Pan-STARRS.}
	\label{fig:ngc2273rkur}
\end{figure}

\begin{figure}
	\centering
	\includegraphics[width=0.9\linewidth]{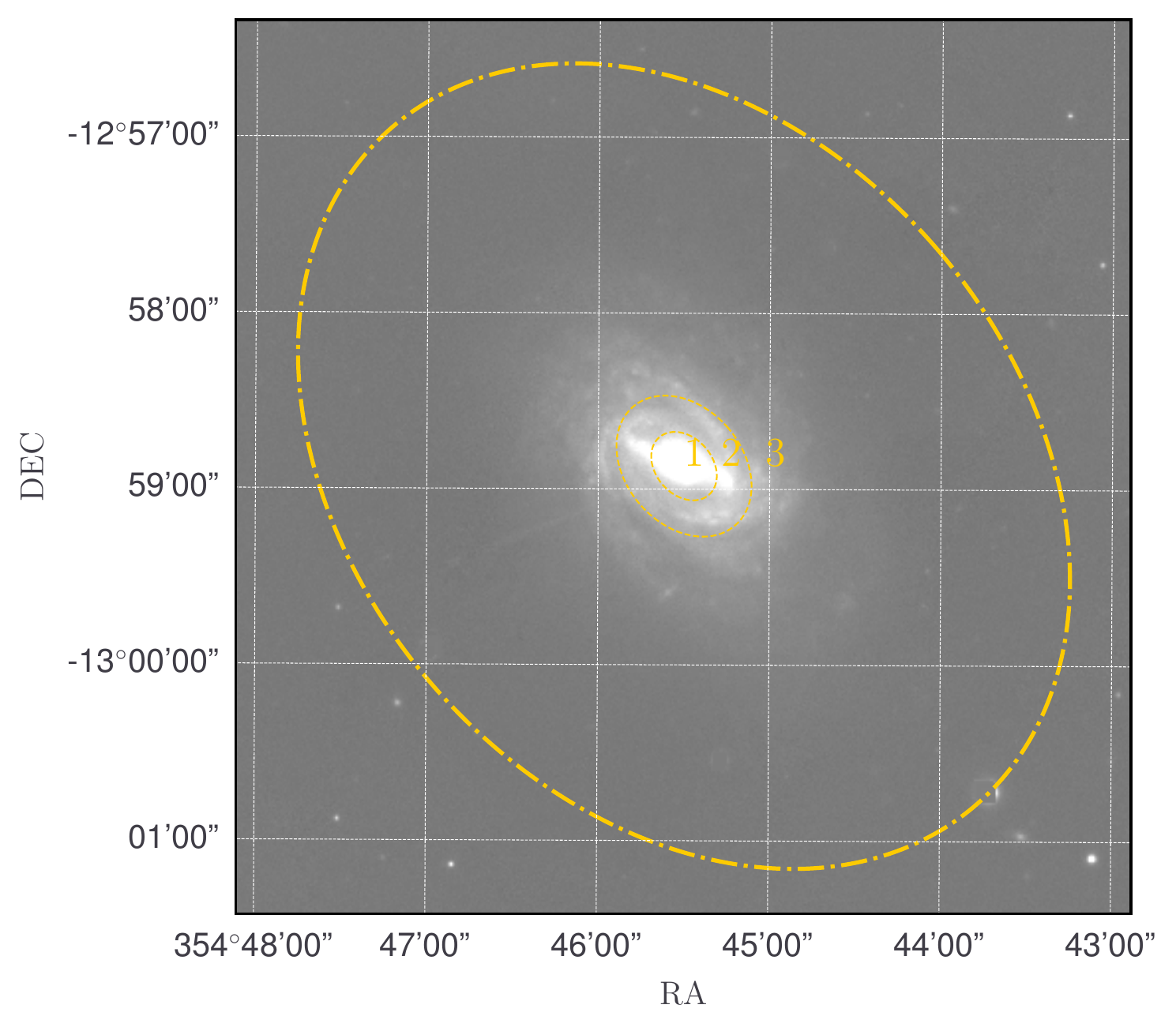}
	\includegraphics[width=0.9\linewidth]{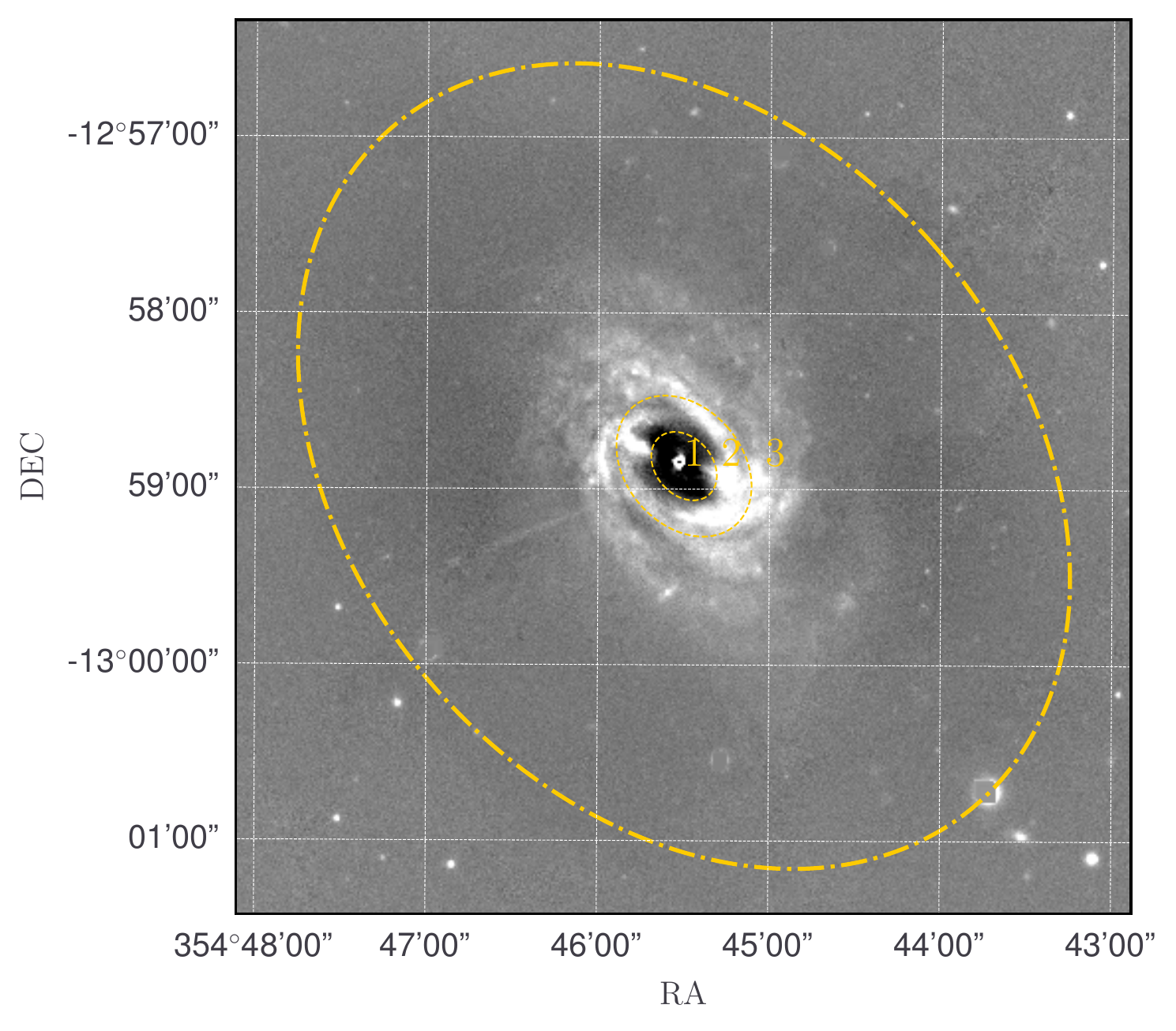}
	\includegraphics[width=0.99\linewidth]{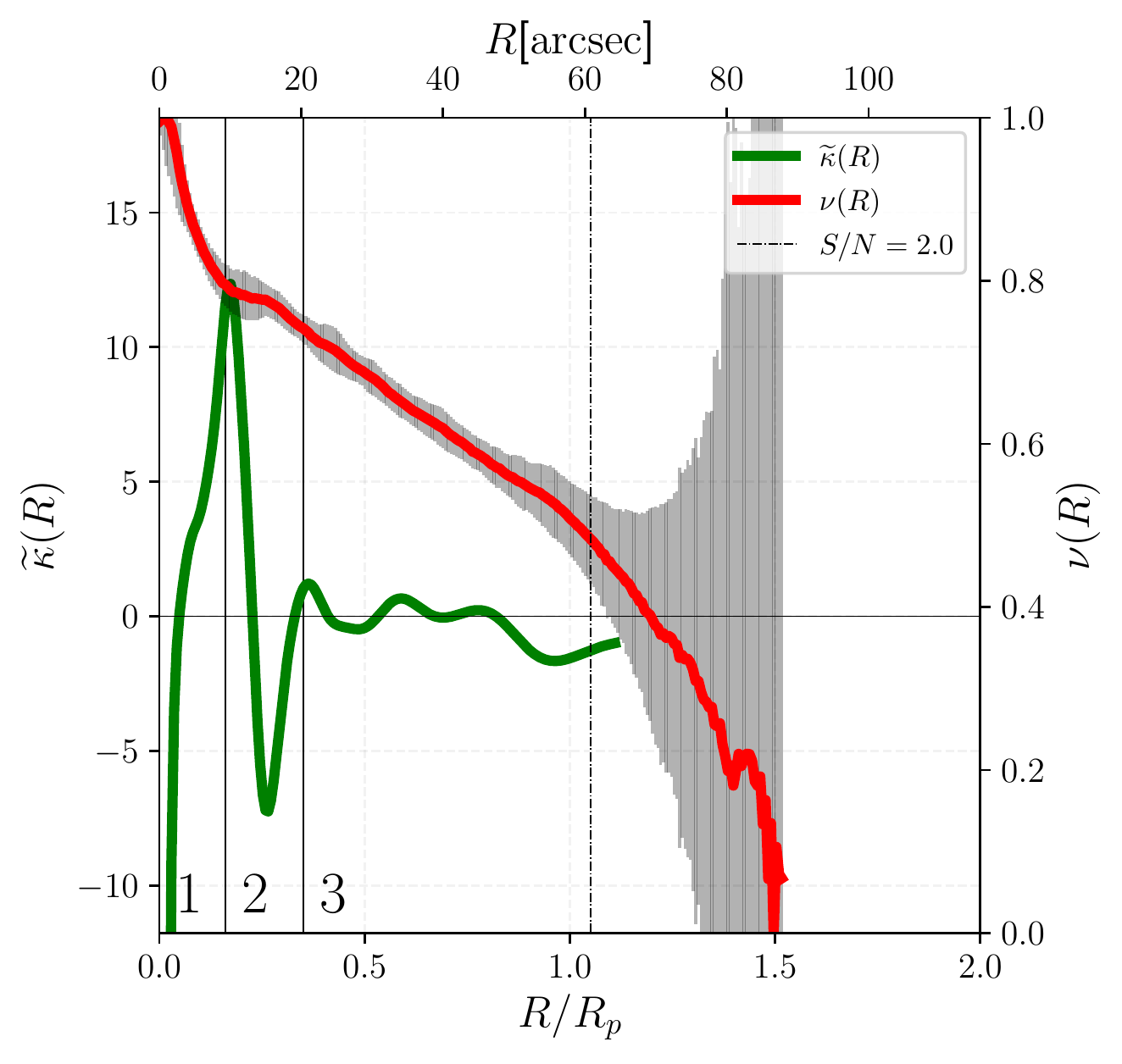}
	\caption{The same as \fg{fig:PGC0011670_r_kur}, but for $r$ band of NGC\,7723 from Pan-STARRS.}
	\label{fig:ngc7723kur}
\end{figure}

\begin{figure}
	\centering
	\includegraphics[width=0.9\linewidth]{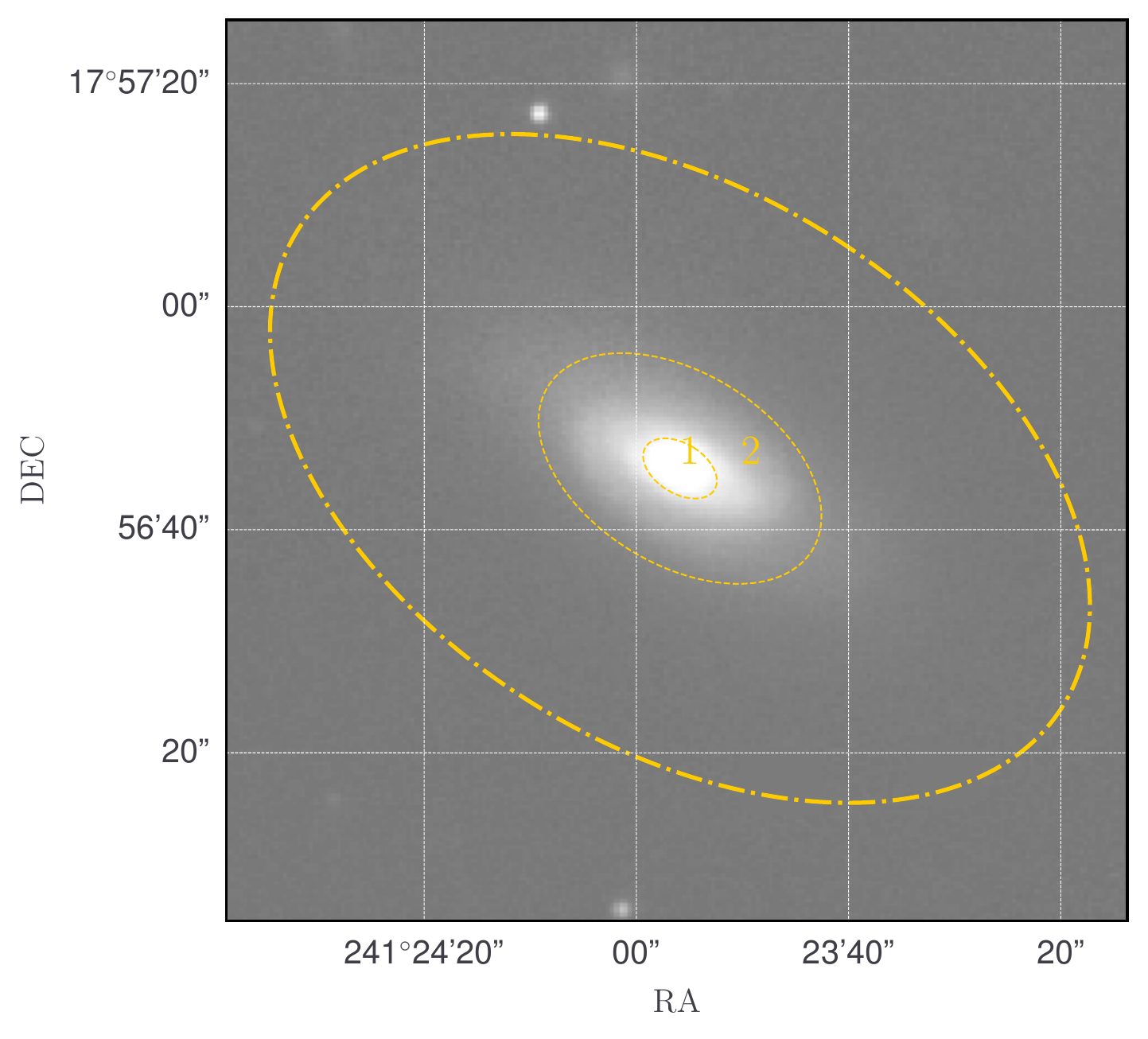}
	\includegraphics[width=0.9\linewidth]{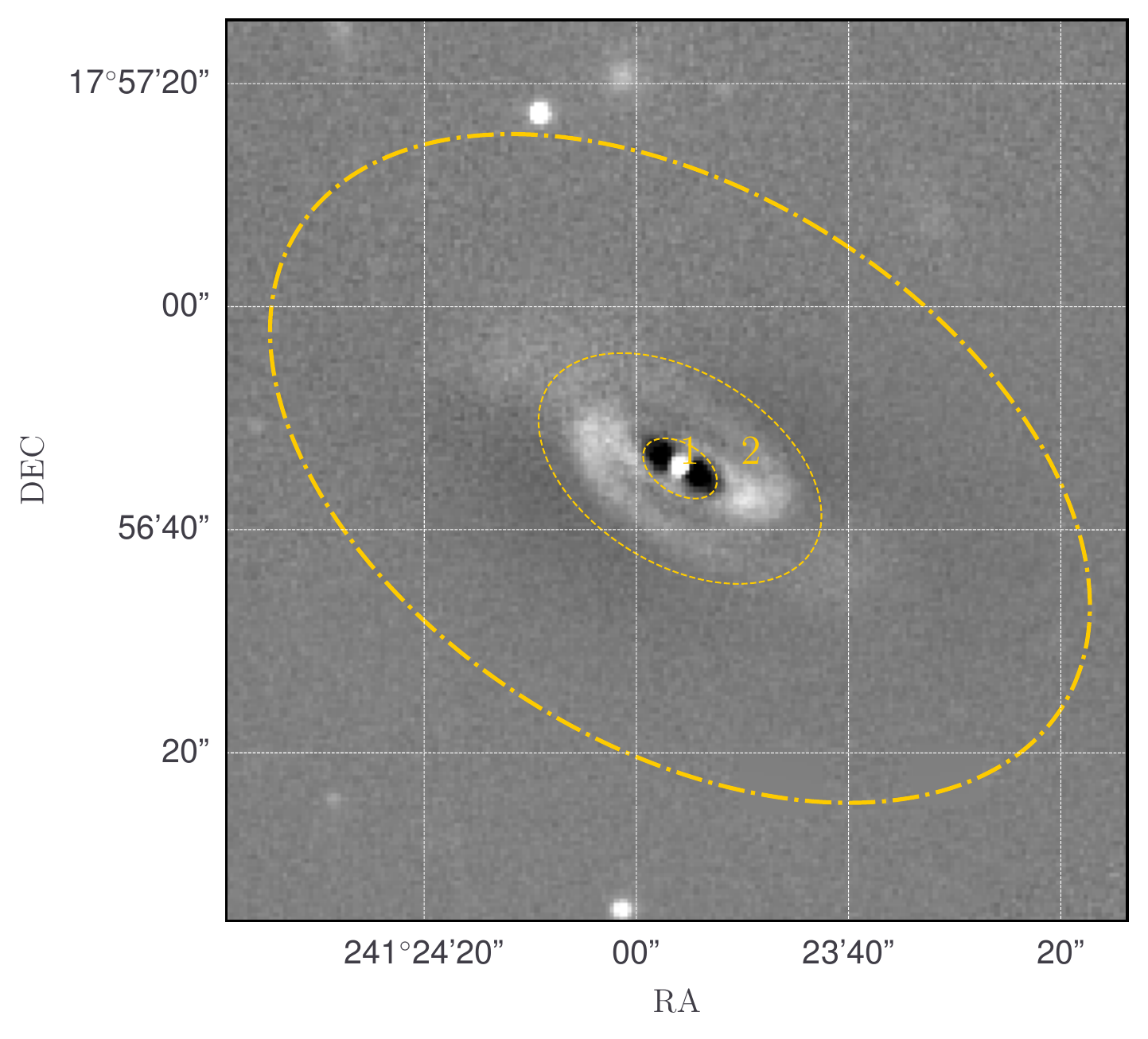}
	\includegraphics[width=0.99\linewidth]{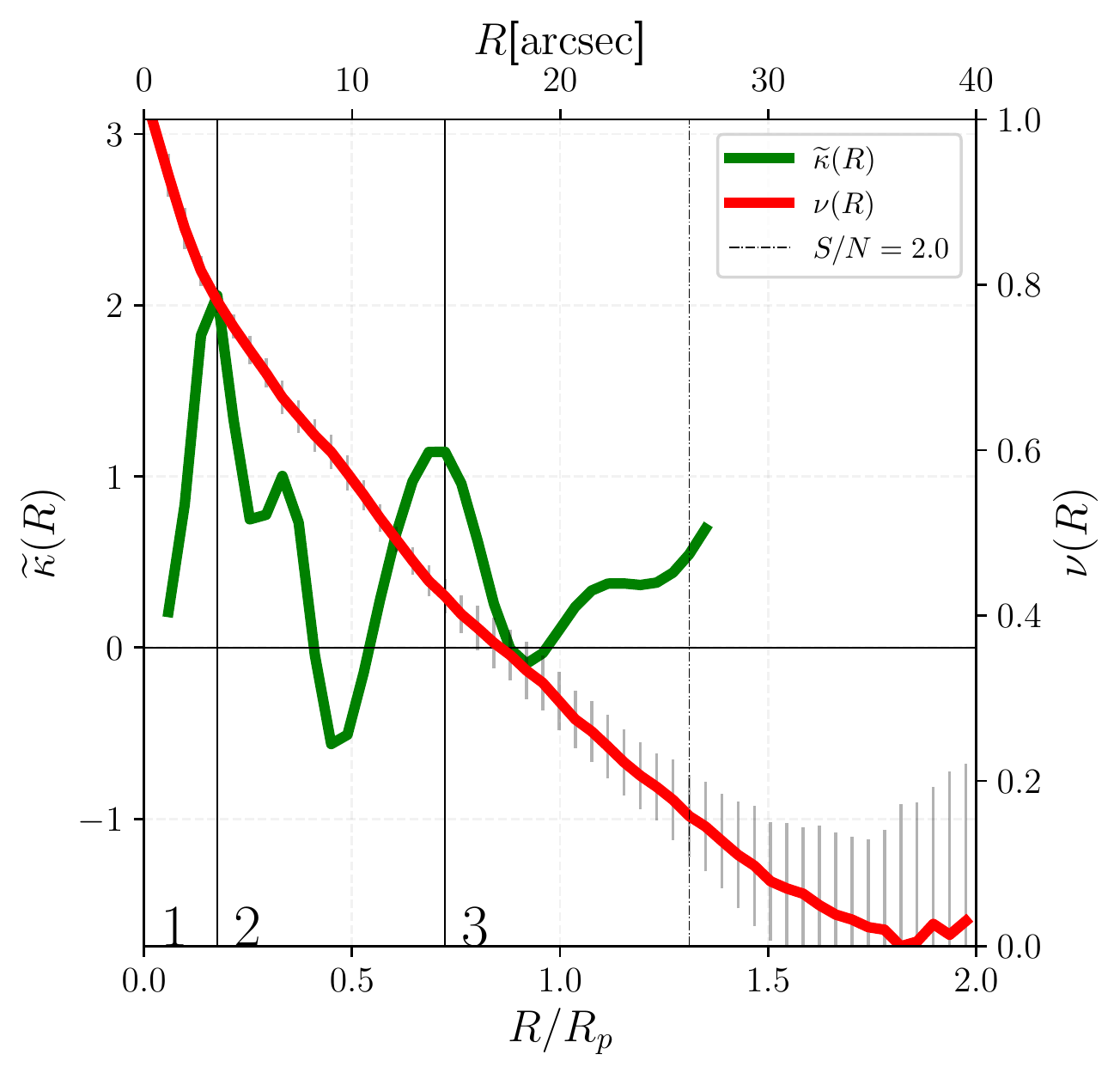}
	\caption{The same as \fg{fig:PGC0011670_r_kur}, but for $r$ band of NGC\,6056.}
	\label{fig:ngc6056rkur}
\end{figure}

\begin{figure}
	\centering
	\includegraphics[width=0.9\linewidth]{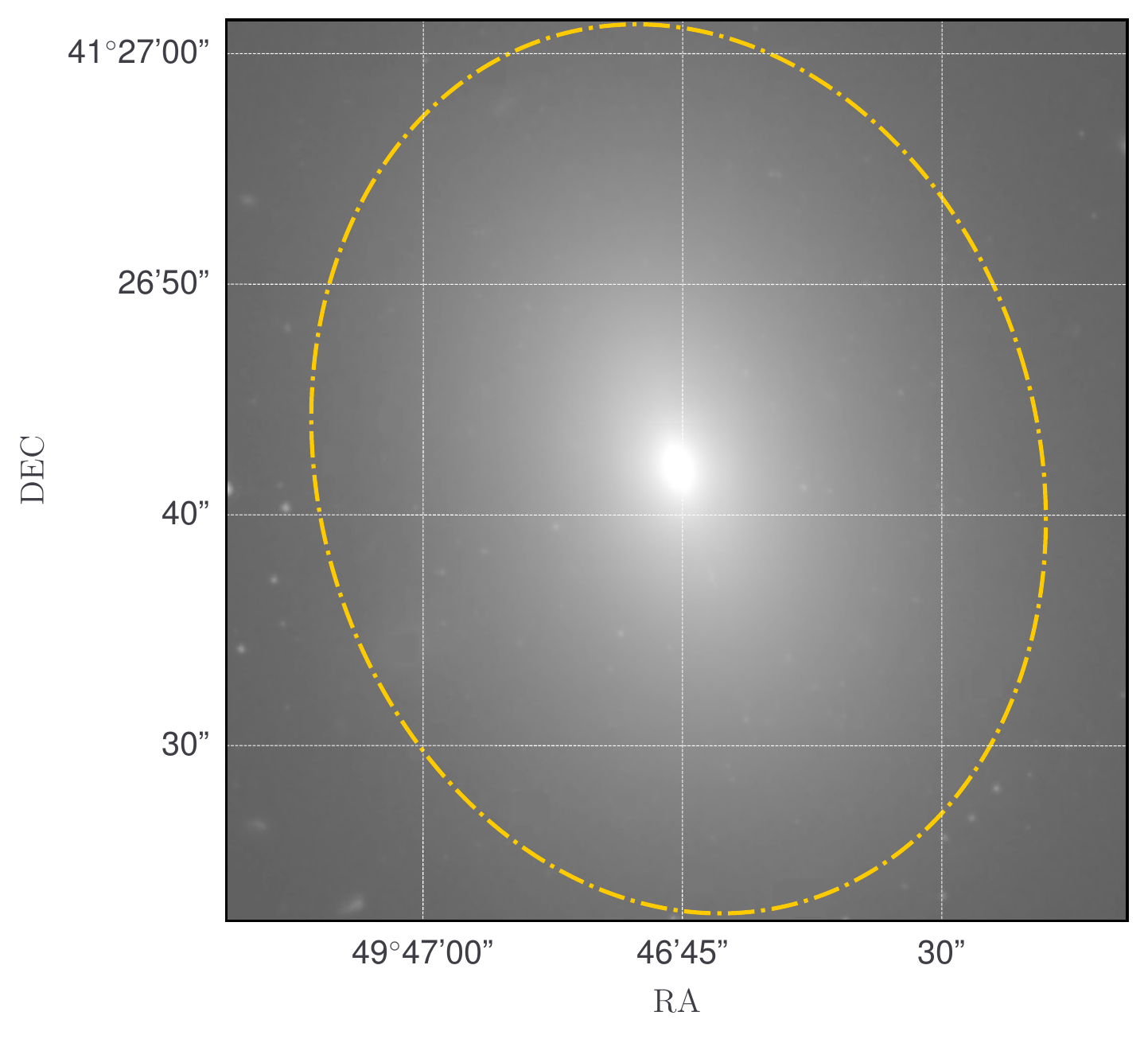}
	\includegraphics[width=0.9\linewidth]{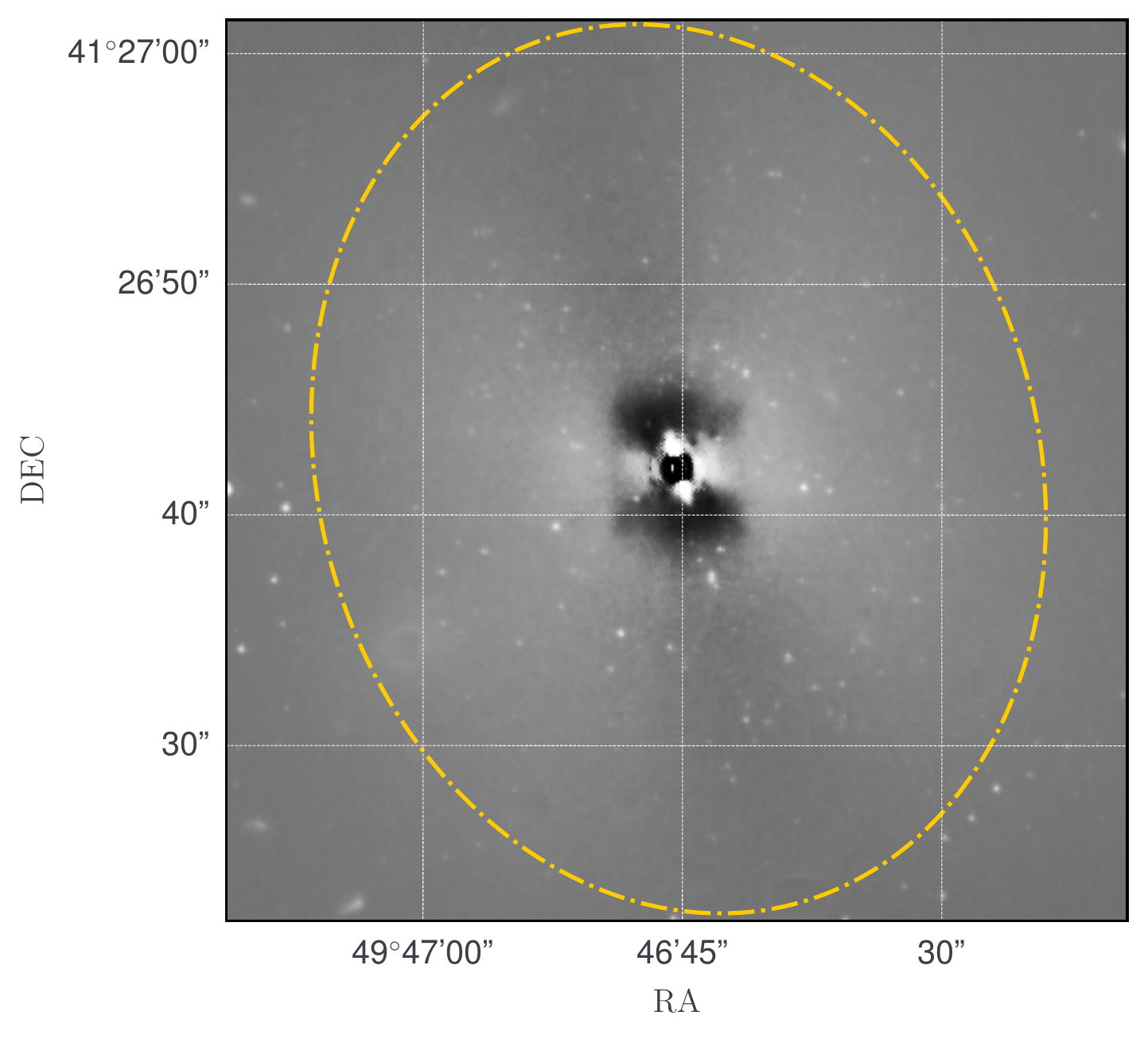}
	\includegraphics[width=0.99\linewidth]{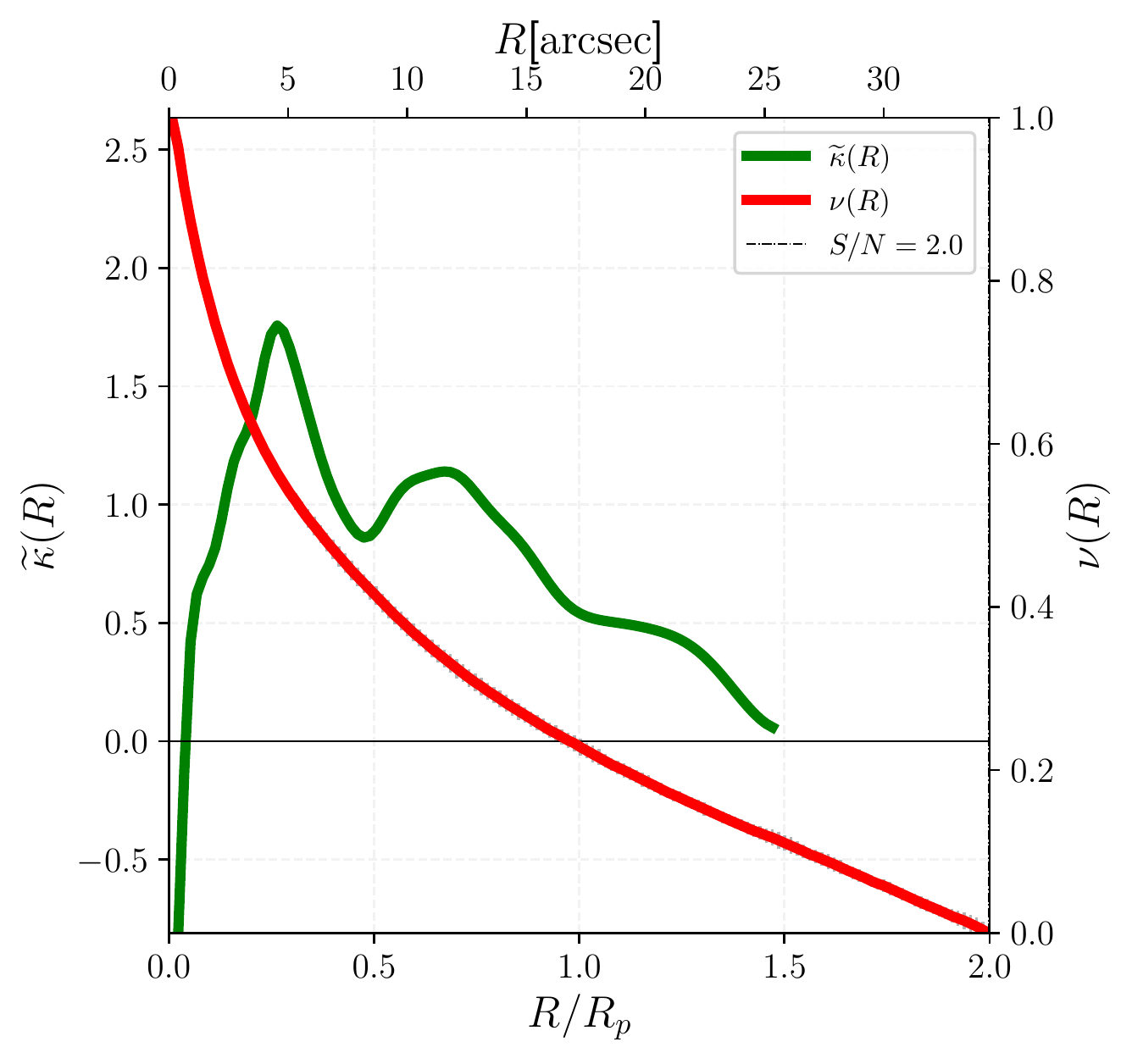}
	\caption{The same as \fg{fig:PGC0011670_r_kur}, but for NGC\,1270 from HST (f160w).}
	\label{fig:cut_NGC1270_f160w_kur_new}
\end{figure}

\begin{figure}
	\centering
	\includegraphics[width=0.9\linewidth]{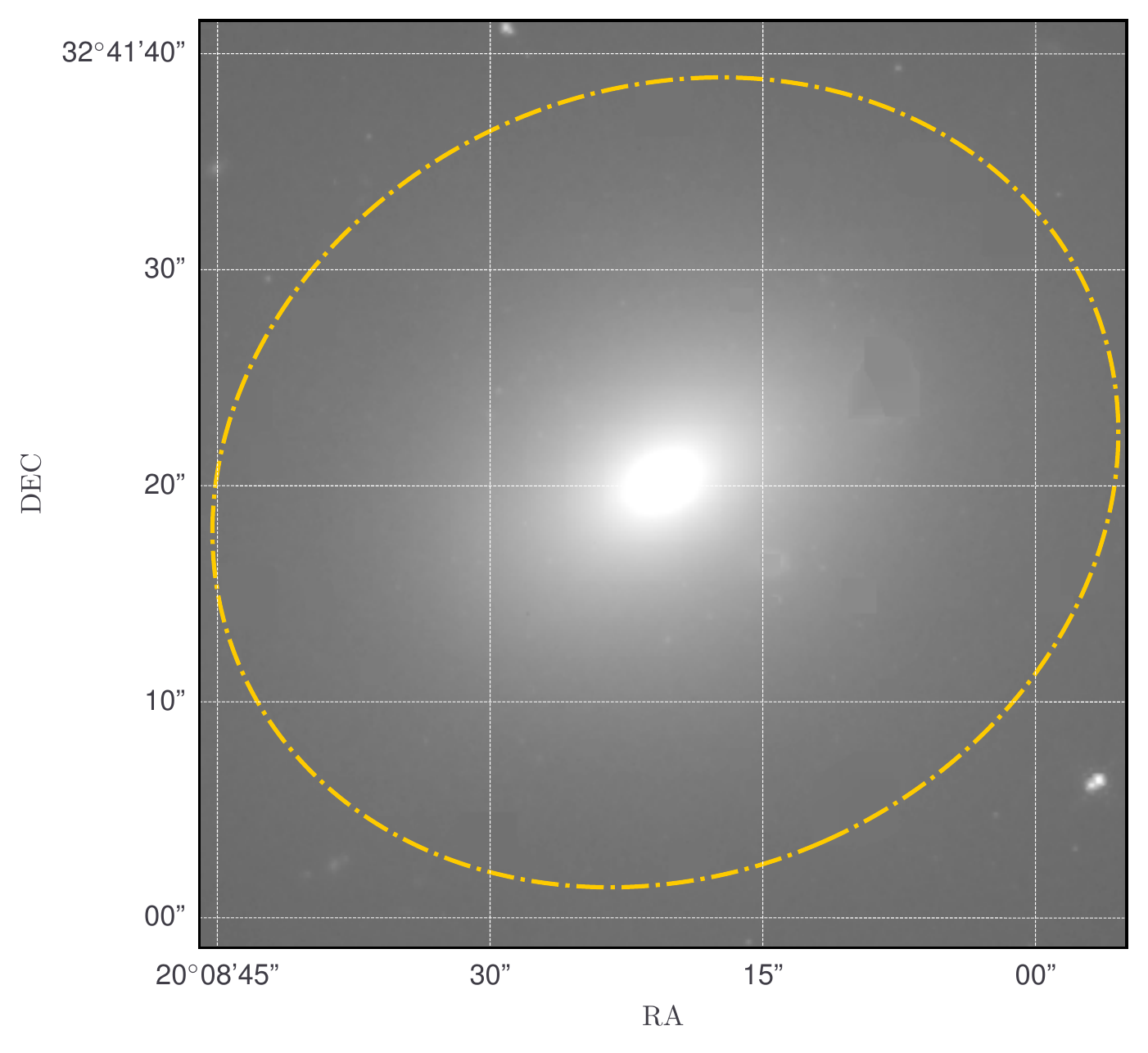}
	\includegraphics[width=0.9\linewidth]{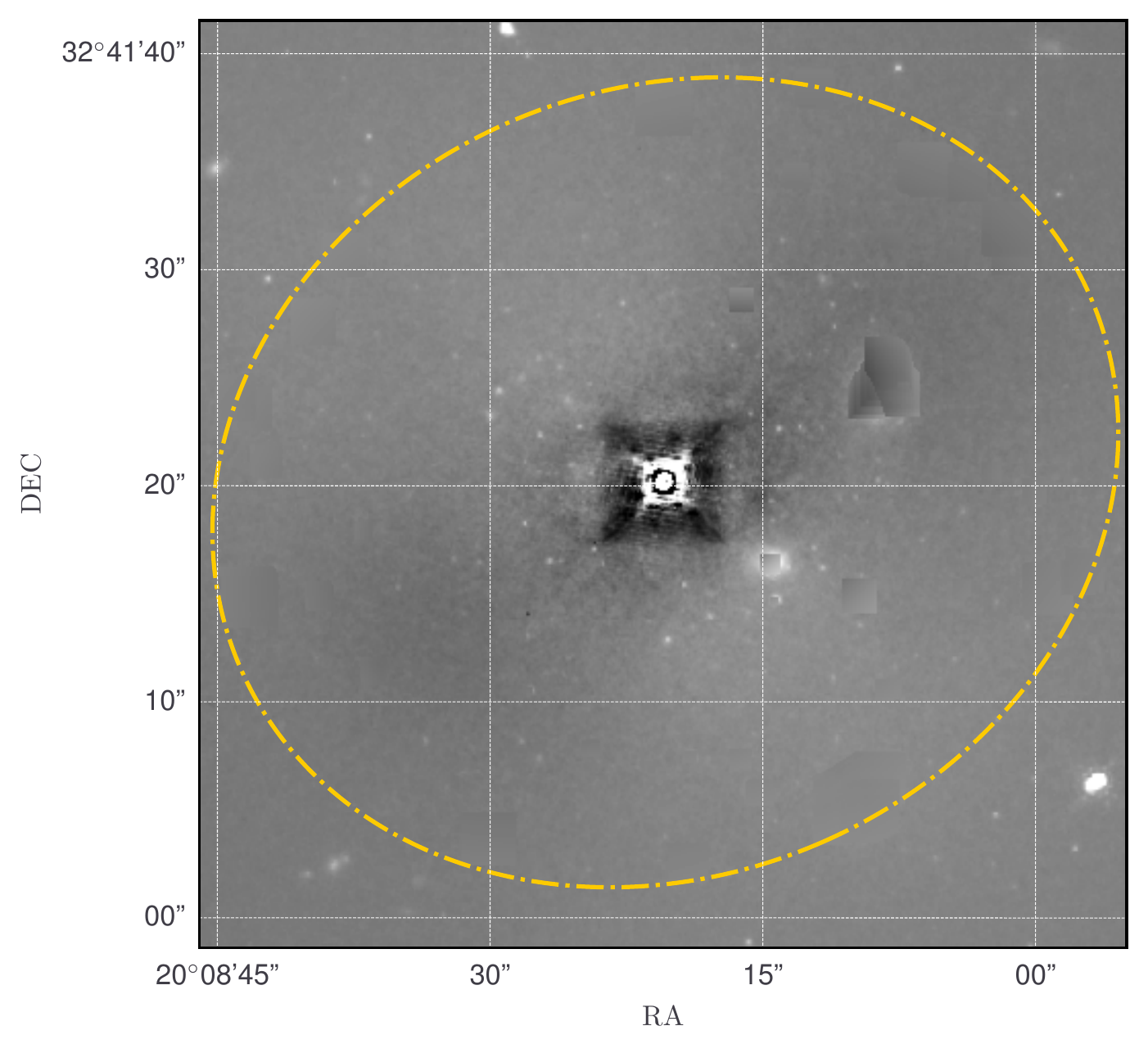}
	\includegraphics[width=0.99\linewidth]{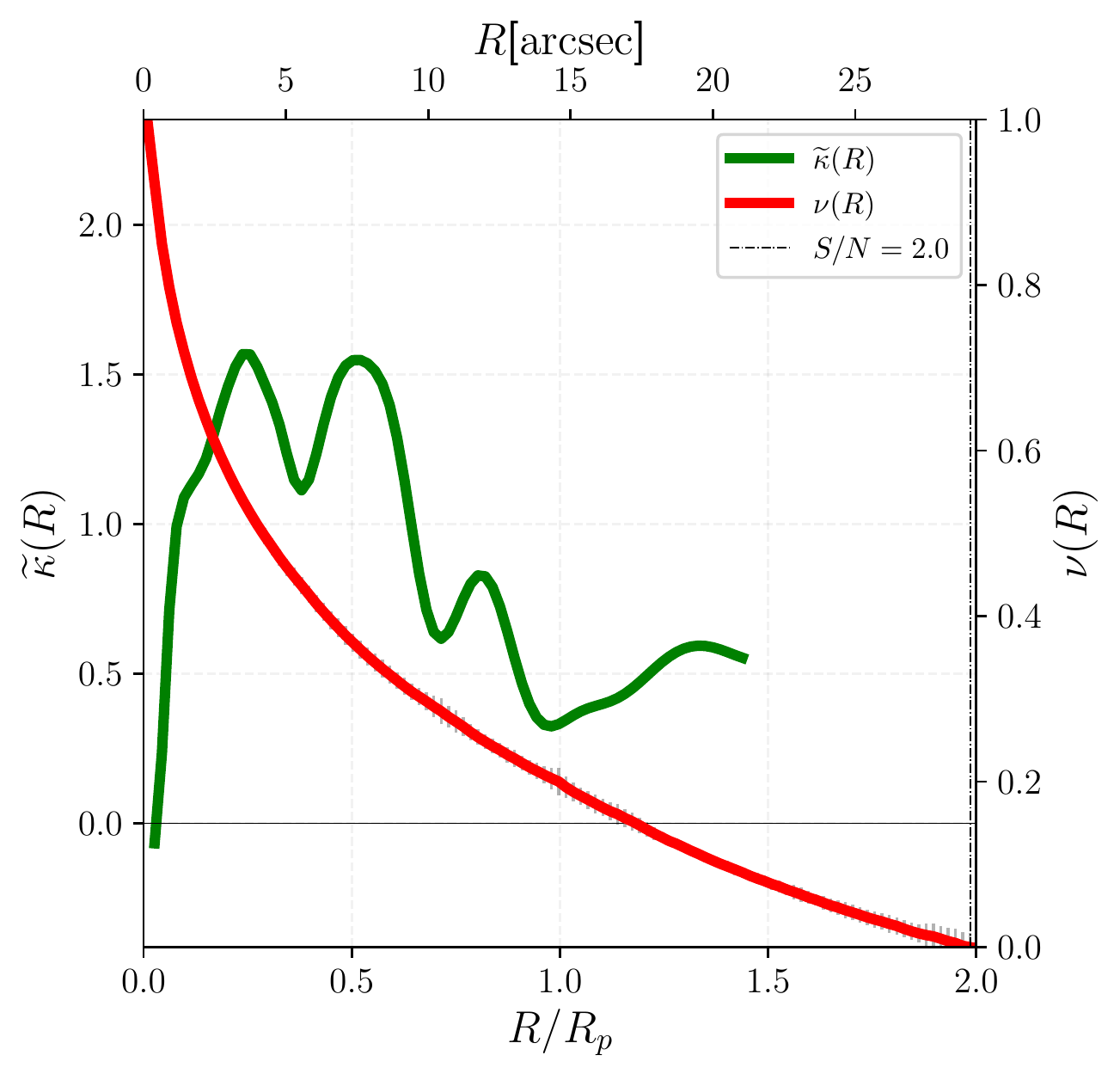}
	\caption{The same as \fg{fig:PGC0011670_r_kur}, but for NGC\,472 from HST (f160w).}
	\label{fig:cutngc0472f160wkur}
\end{figure}

\begin{figure}
	\centering
	\includegraphics[width=0.9\linewidth]{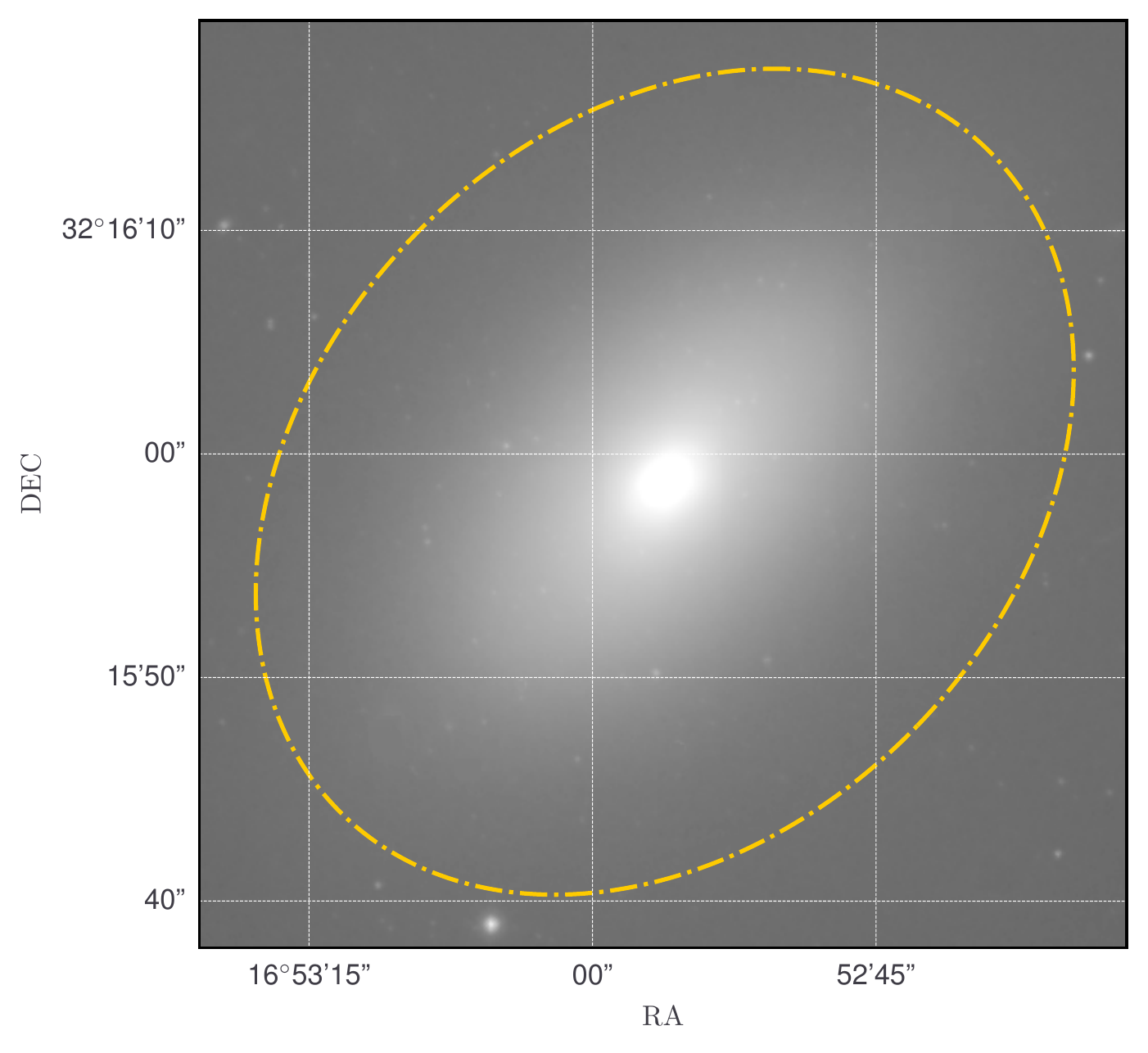}
	\includegraphics[width=0.9\linewidth]{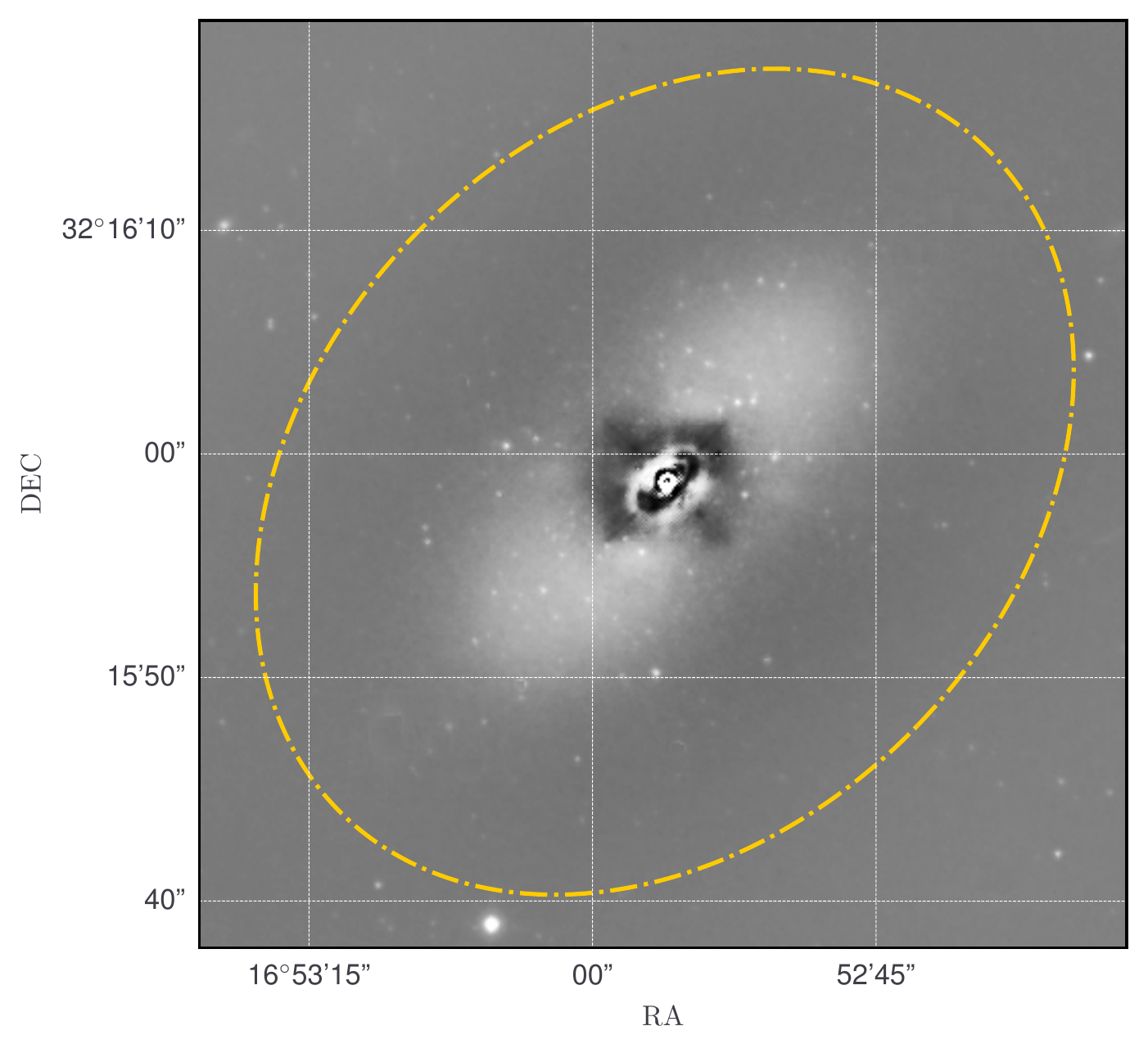}
	\includegraphics[width=0.99\linewidth]{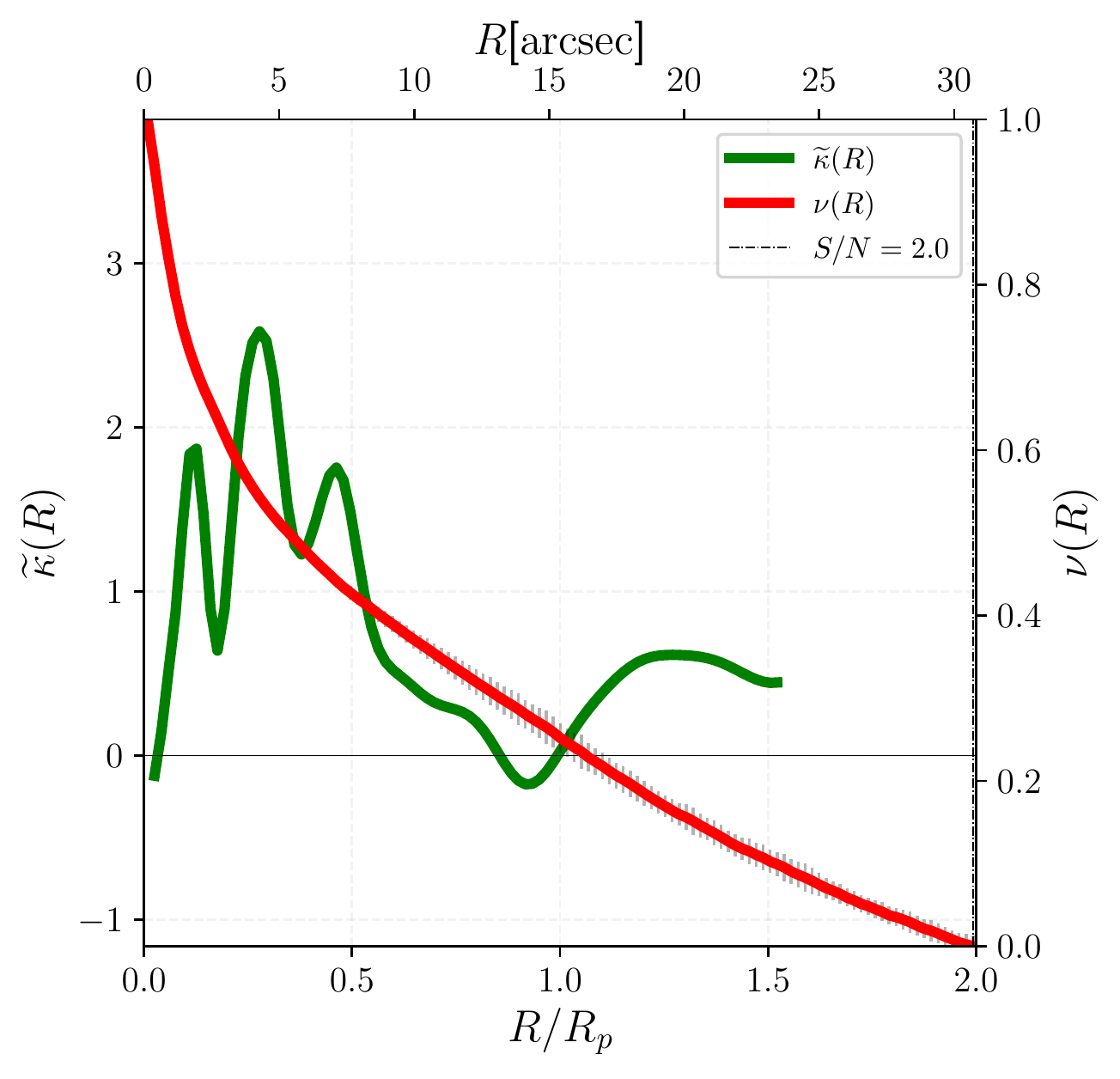}
	\caption{The same as \fg{fig:PGC0011670_r_kur}, but for NGC\,384 from HST (f160w). }
	\label{fig:cut_NGC0384_f160w_kur}
\end{figure}

\begin{figure}
	\centering
	\includegraphics[width=0.9\linewidth]{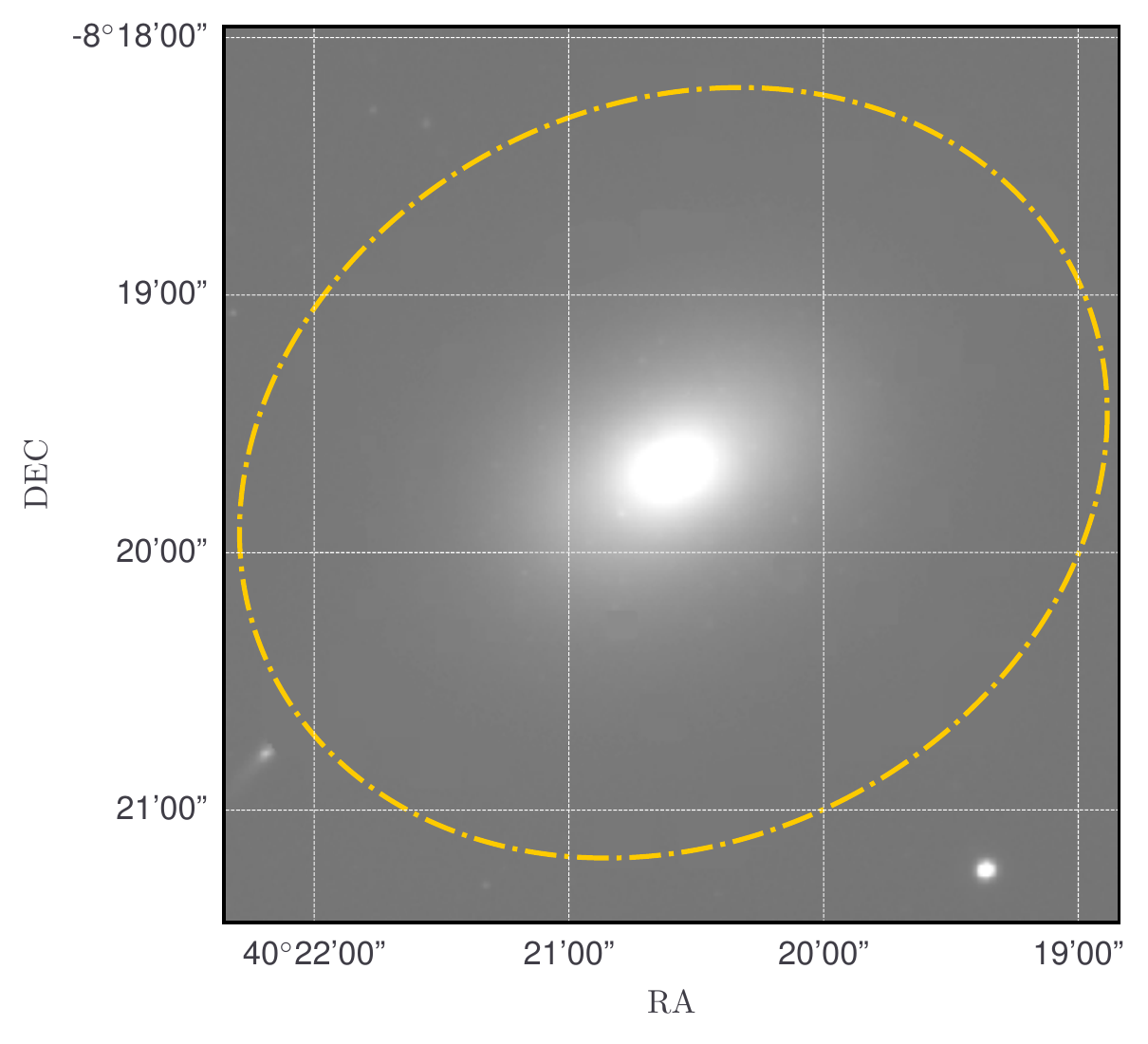}
	\includegraphics[width=0.9\linewidth]{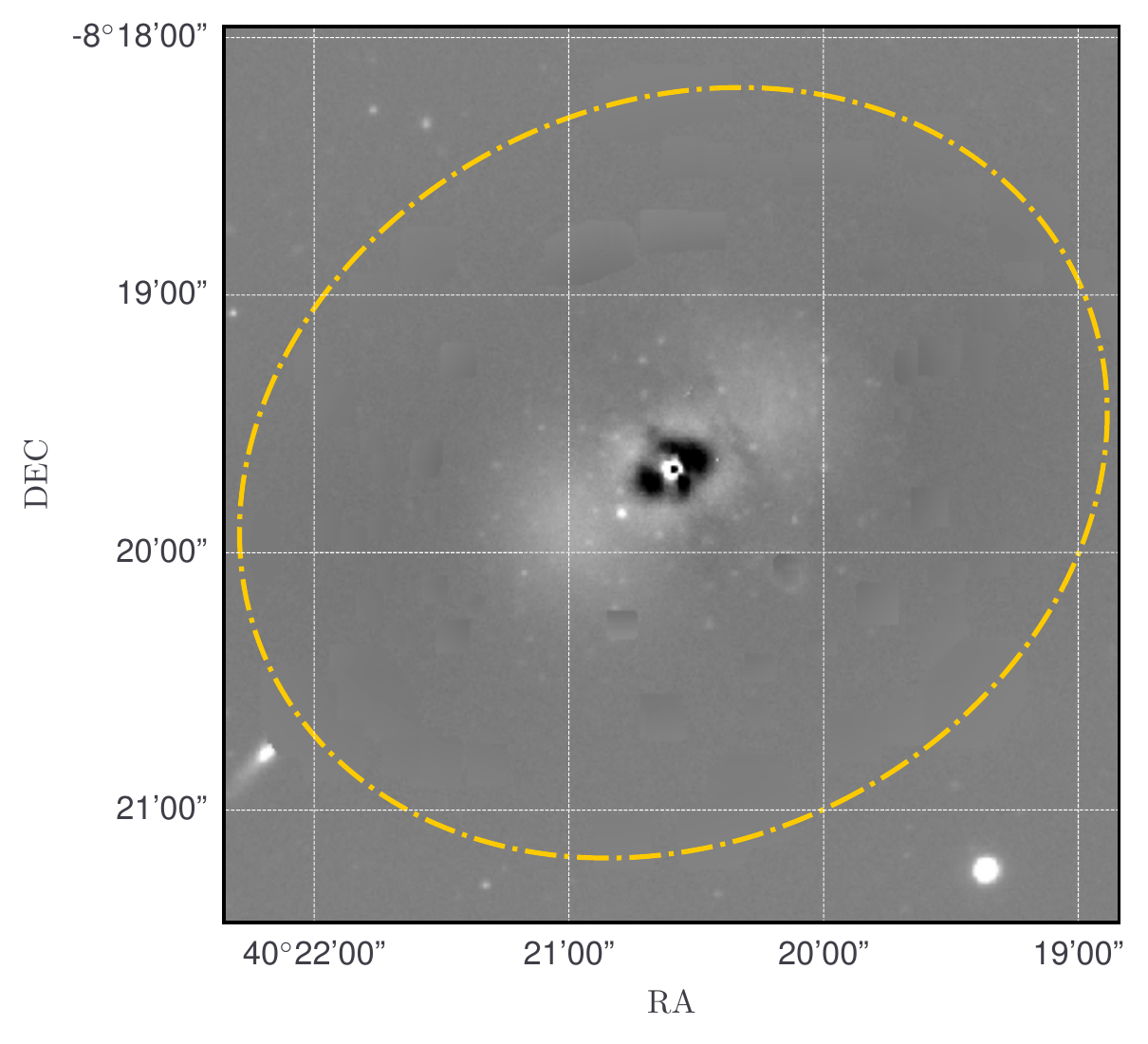}
	\includegraphics[width=0.99\linewidth]{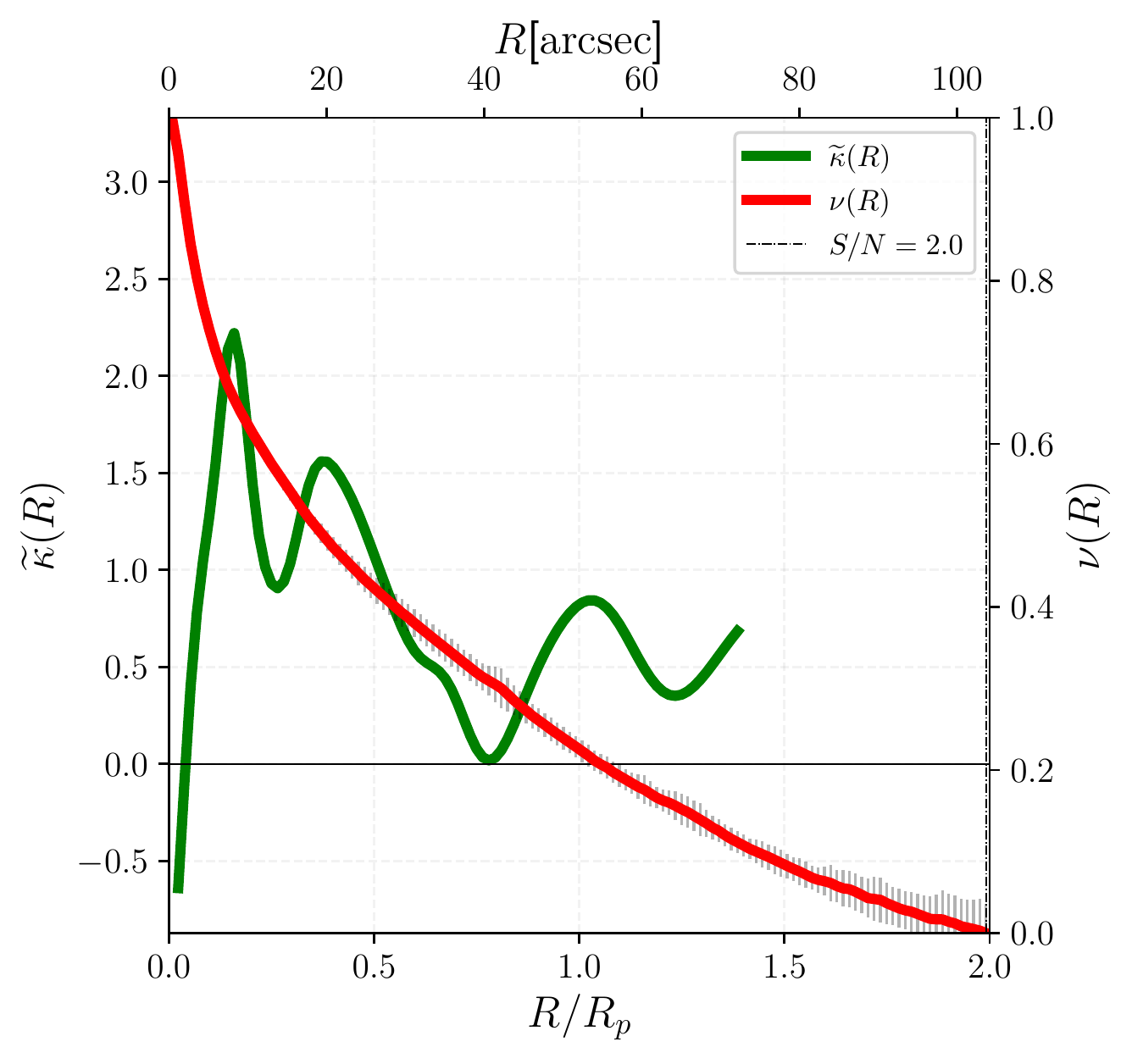}
	\caption{The same as \fg{fig:PGC0011670_r_kur}, but for $r$-band of NGC\,1052 from EFIGI.}
	\label{fig:pgc0010175rkur}
\end{figure}

\label{lastpage}
\end{document}